%% file: aos-sample.tex
\documentclass[aos]{imsart}
\RequirePackage{amsthm,amsmath,amsfonts,amssymb}
\makeatletter
\@imsart@seceqntrue
\numberwithin{equation}{section}  
\makeatother
\usepackage{bm}
\AtBeginDocument{\renewcommand{\bm}[1]{\boldsymbol{#1}}}
\usepackage[linesnumbered,ruled,vlined]{algorithm2e}
\RequirePackage[authoryear]{natbib}
\RequirePackage[colorlinks,citecolor=blue,urlcolor=blue]{hyperref}
\RequirePackage{graphicx}

\hypersetup{
    linkbordercolor={0 0 0}, 
    colorlinks=true,         
    allcolors=black,          
    linkcolor=blue,         
    citecolor=blue,
    urlcolor=blue
}

\startlocaldefs
\theoremstyle{plain}

\newtheorem{theorem}{Theorem}
\newtheorem{lemma}{Lemma}
\newtheorem{condition}{Condition}
\newtheorem{remark}{Remark}
\newtheorem{corollary}{Corollary}
\newtheorem{proposition}{Proposition}
\theoremstyle{definition}

\allowdisplaybreaks

\DeclareMathOperator*{\argmin}{arg\,min}
\DeclareMathOperator*{\argmax}{arg\,max}

\newcommand{\iid}{\textit{i.i.d.}\ }

\DeclareMathOperator*{\E}{\mathbb{E}} 

\newcommand{\ie}{\emph{i.e.},\ }
\newcommand{\eg}{\emph{e.g.},\ }
\newcommand*{\tr}{^{\mkern-1.5mu\mathsf{T}}}

\newcommand*{\T}{\tr}

\newcommand{\bw}{\bm{w}}
\newcommand{\bmu}{\bm{\mu}}
\newcommand{\bX}{\bm{X}}
\newcommand{\by}{\bm{y}}

\newcommand{\bV}{\bm{V}}

\newcommand{\bx}{\bm{x}}
\newcommand{\bbeta}{\bm{\beta}}
\newcommand{\bdelta}{\bm{\delta}}

\newcommand{\bG}{\bm{G}}
\newcommand{\bSigma}{\bm{\Sigma}}

\newcommand{\brho}{\bm{\rho}}
\newcommand{\bvarphi}{\bm{\varphi}}

\newcommand{\bA}{\bm{A}}
\newcommand{\bpsi}{\bm{\psi}}

\newcommand{\bOmega}{\bm{\Omega}}

\newcommand{\un}{\underline{n}}
\newcommand{\ueta}{\underline{\eta}}
\newcommand{\tA}{\tilde{A}}

\newcommand{\note}[1]{\textcolor{ForestGreen}{#1}}

\endlocaldefs

\begin{document}

\begin{frontmatter}
\title{Sufficiency-principled Transfer Learning via Model Averaging}
\runtitle{Sufficiency-principled Transfer Learning via Model Averaging}

\begin{aug}
\author[A]{\fnms{Xiyuan}~\snm{Zhang}\ead[label=e1]{zhangxiyuan@amss.ac.cn}\orcid{0009-0001-4814-1430}}
\author[B]{\fnms{Huihang}~\snm{Liu}\ead[label=e2]{huihang@mail.ustc.edu.cn}}
\and
\author[A]{\fnms{Xinyu}~\snm{Zhang}\ead[label=e3]{xinyu@amss.ac.cn}\thanks{\textbf{Corresponding author.}}}
\address[A]{
Academy of Mathematics and Systems Science, Chinese Academy of Sciences\printead[presep={ ,\ }]{e1,e3}}

\address[B]{
International Institute of Finance, School of Management, University of Science of Technology of China\printead[presep={,\ }]{e2}}
\end{aug}

\begin{abstract}
When the transferable set is unknowable, transfering informative knowledge as much as possible\textemdash a principle we refer to as \emph{sufficiency}, becomes crucial for enhancing transfer learning effectiveness.
However, existing transfer learning methods not only overlook the sufficiency principle, but also rely on restrictive single-similarity assumptions (\eg individual or combinatorial similarity), leading to suboptimal performance.
To address these limitations, we propose a sufficiency-principled transfer learning framework via unified model averaging algorithms, accommodating both individual and combinatorial similarities.
Theoretically, we establish the asymptotic/high-probability optimality, enhanced convergence rate and asymptotic normality for multi-source linear regression models with a diverging number of parameters, 
achieving  sufficiency, robustness to negative transfer, privacy protection and feasible statistical inference. 
Extensive simulations and an empirical data analysis of Beijing housing rental data demonstrate the promising superiority of our framework over conventional alternatives. 
\end{abstract}



\end{frontmatter}

\input{body/intro}

\input{body/model}

\input{body/theory}

\input{body/sim}

\input{body/realdata}

\input{body/discuss}


\begin{supplement}
    \stitle{Supplement A: Technical Details}
    \sdescription{This supplement provides formal mathematical proofs and theoretical analysis supporting the main results.}
\end{supplement}

\begin{supplement}
    \stitle{Supplement B: Additional Numerical Results}
    \sdescription{This supplement provides additional numerical experiment results and implementation specifics.}
\end{supplement}

\begin{acks}[Acknowledgments]
The authors would like to thank the anonymous referees, an Associate
Editor and the Editor for their constructive comments that improved the
quality of this paper.
\end{acks}

\bibliographystyle{imsart-nameyear}
\bibliography{bibliography}       





\end{document}

%% file: body/intro.tex
\section{Introduction} \label{intro}

\renewcommand{\note}[1]{} 

\note{迁移学习的重要性}
Transfer learning improves the performance of target models by leveraging knowledge from related auxiliary domains \citep{pan2009survey, weiss2016survey, zhuang2020comprehensive, bao2023survey}. 
It has achieved significant success in various fields, including medical and biological research \citep{deepak2019brain,theodoris2023transfer}, computer vision \citep{gopalakrishnan2017deep}, natural language processing \citep{ruder2019transfer}, and recommendation systems \citep{zhao2017unified}, among others. 
\note{统计迁移学习的发展}
In the community of statistics, transfer learning has garnered increasing attention, starting with single-source transfer learning for linear regression \citep{bastani2021predicting}, and has been extended to various statistical tasks involving multiple sources, such as
high-dimensional linear models \citep{li2022transfer,  zhao2023residual,lin2024profiled}, 
high-dimensional generalized linear models \citep{tian2023transfer,li2024estimation}, 
semi-parametric models \citep{hu2023optimal}, 
nonparametric models \citep{lin2022transfer, cai2024transfer,auddy2024minimax}, 
quantile regression models \citep{huang2022transfer, zhang2025transfer}, 
panel regression models \citep{duan2024target}, 
spatial autoregressive models \citep{zeng2024transfer}, 
and others. 

\note{源域集成方法是关键技术, 它的作用}
\note{
这些方法依赖于源数据集的两个关键因素: 
(1)源数据集和目标数据集具有某种相似相似性
(2)源数据集的大量样本(它可以带来的估计或预测精度的提高)
}
\note{详细列举: 多模型、数据源集成的文献.}
\note{集成相比于其他方法的优势: 在xx方面优于选择方法}
\note{有很多文献基于这些平均方法提出了迁移学习方法.列举文献.}
\note{转折: while, these aggregation methods suffer from some critical issues. 然后接后文的详细描述}
Given the transferable set, multi-source transfer learning has shown promising success \citep{li2022transfer, li2024estimation, tian2023transfer,lin2024profiled}, depending on two primary factors: 
(i) the similarity between the transferable sources and the target, and 
(ii) the enhanced estimation or prediction accuracy enabled by the abundance of auxiliary samples from transferable sources.
However, the transferable set is always unknowable in practice and transfering some sources far away from the target can arise harmful negative transfer phenomena \citep{tian2023transfer}, 
arising a significant challenge in multi-source transfer learning.
To address this gap, various strategies have been proposed, including 
model selection aggregation \citep{li2022transfer}, 
transferable source detection \citep{tian2023transfer}, 
and model averaging approaches \citep{hu2023optimal,zhang2024prediction,lin2024profiled}.
The current article also utilizes model-averaging based transfer learning approaches when the transferable set is unknown, 
since model averaging incorporates all available information and constructs a weighted average of the individual prediction from all potential models, 
which provides a kind of insurance against selecting a very poor model, and tends to perform better than the model selection estimator in finite samples \citep{raftery1997bayesian, hansen2007least, claeskens2008model, hansen2012jackknife, moral2015model, steel2020model, zhang2023model}.
For the two aforementioned factors, we have the following specific thoughts to motivate our transfer learning methods.

\note{(对于前者, 个体相似性要求target和source的参数足够接近, 比如代表性的工作lisai, 她考虑了高维线性模型要求target参数和source参数在0-1范数的意义下足够小, 且把这样的source称为informative.)}

As the primary factor (i), the similarities in the literature are typically categorized into two types: individual-similarity and combinatorial-similarity. 
Individual-similarity considers sources whose parameters are close enough to the target parameter to be informative, which has been applied by \cite{li2022transfer} in high-dimensional linear models, \cite{tian2023transfer} and \cite{li2024estimation} in high-dimensional generalized linear models, \cite{lin2022transfer} in functional linear regression, \cite{li2023transfer} in high-dimensional Gaussian graphical models, and \cite{cai2024transfer} in nonparametric regression, among others. 
\note{对于后者, 组合相似性要求target参数和source的某种组合足够接近, 比如lin(2024)考虑了source参数的线性组合, zhang2024和hu(2024)考虑了target和source参数的凸组合.} 
Combinatorial-similarity requires that the distance between the target model parameter and a combination of source parameters be sufficiently small, such as linear combinations \cite{lin2024profiled} or convex combinations \citep{hu2023optimal, xiong2023distributionally, zhang2024prediction}.
\note{虽然直观上, 组合相似性是个体相似性的延伸, 因为它允许source和target参数的距离任意大, 但是其实两种相似性各有好处}
Combinatorial-similarity extends individual-similarity by allowing a larger distance between source and target parameters \citep{lin2024profiled}. 
However, determining which type of similarity is superior in practice is still complex. 
\note{1.如果信息集确实很多, 关注个体相似性可以更充分地利用信息集的信息, 而此时若考虑组合相似性, 则可能因非信息集的估计预测精度较低而拖慢迁移学习的效率, 甚至产生负迁移.}
On one hand, if there are indeed many informative sources, focusing on individual-similarity may be more effective in utilizing the information from these sources,
while focusing combinatorial-similarity may slow down the efficiency of transfer learning, or even arise negative transfer; see numerical examples in Section~\ref{sim}. 
\note{如果信息集比较少, 而含有信息的组合较多时, 显然组合相似性有更强的潜在迁移学习能力, 而个体相似性则可能因为过于严格而使迁移学习失效.}
On the other hand, in scenarios where few or no sources are informative while there actually exist combinations of sources close to the target, combinatorial-similarity holds greater potential than individual-similarity, as highlighted in \citet[Corollary 2]{lin2024profiled}. 
\note{(然而, 目前的文章都只侧重于某一种相似性, 我们的问题是: 能否设计算法更充分地同时利用个体相似性和组合相似性的知识, 进行更有效地迁移学习?)}
Notice that existing works always focus on only one type of similarity. 
Thus, we aim to address the following question. 
\begin{itemize}
  \item[(Q1)] \textit{How to design a paradigm/framework that concurrently leverages both individual-similarity and combinatorial-similarity to enhance transfer learning effectiveness?} \label{Q1}
\end{itemize}

As the primary factor (ii), given a predefined transferable set, utilizing a large number of auxiliary samples from informative sources can substantially improve estimation and prediction performance \citep{li2022transfer,li2023transfer}.
Moreover, more informative auxiliary samples imply better transfer learning performance, as shown in \citet[Theorem 1]{li2022transfer}.
Thus, when the truly transferable set is unknown, maximizing the transfer of informative knowledge, \ie ensuring \emph{sufficiency}, is important for preserving transfer learning effectiveness with the prior knowledge of the transferable set.
However, sufficiency is always overlooked in existing transfer learning methods, resulting in suboptimal transfer learning performance.
For instance, when sufficiency is neglected, the enhanced convergence rate achieved by Oracle Trans-Lasso (\citet[Theorem 1]{li2022transfer}) may be compromised by the q-aggregation procedure \citep{dai2012deviation}, ultimately degrading the convergence rate of Trans-Lasso (\citet[Theorem 3]{li2022transfer}).
Overall, we categorize the challenges of sufficient transfer learning without the prior knowledge of transferable set into four key aspects.
First, since any model is a simplified representation of the real world, working model for the target may be misspecified, making transfer learning more challenging.
Second, even under correct specification, inherent and unknowable real-world dissimilarities between sources and target may still result in negative transfer, leading to higher prediction risks or larger estimation errors.
Third, the smaller bias induced by randomness may mislead the selection of sufficient informative domain, leading to spurious sufficiency.
Forth, only valuable and limited target samples are used to evaluated the performances of candidate estimators or predictions, limiting the effectiveness of transfer learning. 
\note{(因此, 本文的目标在于回答以下问题: 在充分利用信息集的前提下, 如何设计出考虑 domain uncertainty 的 domain aggregation 算法?)}
Facing these challenges, we aim to address the following question. 
\begin{itemize}
  \item[(Q2)] \textit{How to design a novel transfer learning framework to make sufficient use of auxiliary informative knowledge when the transferable set is unknown?} \label{Q2}
\end{itemize}

\note{现有的文献已经有了一些域集成方法, 如针对个体相似性开发的 model selection aggregation 和 transferable source detection(引一些文章), 和针对组合相似性开发的 approximate-linear assembly 和 model averaging.}
\note{比如在个体相似性这类工作中, \citet{li2022transfer} 为了使每个源有更高的估计精度, 根据 source 的参数相似性嵌套地构造了候选的域, 使得每一个候选域具有更多的样本, 然而嵌套带来的迁移学习优势有可能会因为 q-aggregation 而消失}
\note{ (这是因为: 目标参数估计的 convergence rate 可以分为三项:  
  (1) 共享参数相对于信息源总样本量的复杂性
  (2) 目标特定参数相对于目标数据集样本量的复杂性
  (3) 与选择信息源相关的损失(\cite{li2022transfer}). 
  前两项的提升可能会因为第三项而消失, 受限于target的样本量n0. 
)} 
\note{总之, 如何提升 domain aggregation 的效果以更好的保持在先验知道什么信息可以迁移时的迁移学习方法的效果是个重要的问题, 在目前的文章并没有重点关注这一程序.}
\note{具体而言, domain aggregation 包含 3 个技术挑战.}

\note{(为了回答以上 2 个问题, 我们提出了新的的用于处理域不确定性的domain aggregation充分原则迁移学习框架, 在两种相似性下.)}
For (Q1) and (Q2), we propose an innovative sufficiency-principled transfer learning framework under both individual-similarity and combinatorial-similarity when the transferable set is unknown. 
\note{(在个体相似性下, 我们提出了新颖的乐观主义的迁移学习方法, 在domain aggregation过程中引入了一个乐观的惩罚函数, trade-off了模型选择和模型平均. 在该算法中, 乐观惩罚函数通过加大对较小样本量信息集的惩罚, 从而使得在实际中可以以更快的速度选出sufficient informative集. )}
For individual-similarity, we propose a novel sufficiency-principled weight selection criterion based on q-aggregation and a sufficient penalty, 
which guarantees transfer sufficiency while balances model selection with model averaging.
\note{(乐观惩罚函数通过加大对较小样本量信息集的惩罚来处理域不确定性, 从而使得在实际中可以以更快的速度选出 sufficient informative 集.)}
For combinatorial-similarity, benefiting from the sufficiency under individual-similarity, we propose an optimal weight selection criterion on the selected sufficient informative domain. 
Our sufficiency-principled transfer learning framework is broadly applicable across diverse contexts. 
Specifically, focusing on its application to linear regression models, we highlight four compelling advantages. 

\note{(这种充分原则迁移学习框架适用于广泛的背景, 具体而言我们研究线性模型, 具有如下的优点)}

\begin{itemize}
  \item[(i)]
\emph{Sufficiency principle} 
improves the efficiency of target estimation and prediction, which aims to simultaneously utilize the auxiliary informative samples with potential knowledge as much as possible and avoid negative transfer phenomenon in transfer learning. 
Under individual-similarity conditions, sufficiency guarantees weight convergence that asymptotically assigns zero weight to insufficient informative domains during aggregation, thereby enabling more effective utilization of informative samples.
\note{[我们提出的算法具有这种理想的充分性, 它可以有效地缓解 domain aggregation 引起的虚假充分信息性.]}
Under combinatorial-similarity, sufficiency guides us to perform a two-step transfer learning procedure on the sufficient informative domain rather than directly on the target, which differs from existing methods under combinatorial-similarity \citep{hu2023optimal, zhang2024prediction, lin2024profiled}.
\note{(因为迁移学习的收敛速度受target样本量的限制, 所以充分原则能潜在的提高迁移学习的效率)}
\note{[充分性的理论有什么?]}
\note{[实验也说明我们的充分性方法好]}

\item[(ii)]
\note{[别的迁移学习文章, 渐近分布的局限性]}
\emph{Asymptotic normality} of our model averaging estimator is derived, making inference feasible.
Asymptotic distribution of transfer learning estimators has been established in \cite{tian2023transfer} and \cite{li2024estimation} under a predetermined transferable set.
\note{[之前 ma 文章做迁移学习的局限性]}
Moreover, the asymptotic distribution of model-averaged estimators has not received much attention, with exceptions include the works of \cite{hjort2003frequentist, liu2015distribution,zhang2011focused} and \cite{zhang2020parsimonious}. 
Regarding foundational works on transfer learning with model averaging, the asymptotic distribution theory remains undeveloped in \cite{hu2023optimal} and \cite{zhang2024prediction}.
\note{[描述我们的渐近分布]}
Our work bridges this gap by establishing the asymptotic normality of our transfer learning estimator without prior knowledge of the transferable set, simplifying inference.

\item[(iii)]
\emph{Robustness to negative transfer} constitutes a notable advantage of our proposed transfer learning framework under both individual-similarity and combinatorial-similarity. 
This robustness is particularly valuable since transferring adversarial auxiliary samples may harm performance, arising negative transfer \citep{pan2009survey, olivas2009handbook}.
\note{[个体相似性时, stma在两种情况下都有一定的避免负迁移特性]}
Under individual-similarity, our proposed method does not require prior knowledge of the informative auxiliary set and theoretically ensures that the parameter transfer asymptotically occurs in potential sufficient informative domain, attempting to address negative transfer from a new perspective.
Under combinatorial-similarity, we provide high-probability optimality, achieving the lowest possible squared prediction error to avoid negative transfer.

\item[(iv)]
\emph{Privacy protection} is ensured, which is particularly important in the context of transfer learning since data from different sources often cannot be shared in full \citep{gao2019privacy, zhang2022data, wang2024differential, hu2023optimal}. 
By transmitting only summary statistics information and employing a regression cube technique \citep{chen2006regression,lin2011aggregated,schifano2016online}, our transfer learning framework provides a feasible strategy to effectively protect the privacy of individual data. 
\end{itemize}

Beyond primarily advancing transfer learning research, this article also contributes to model averaging.
In contrast to traditional frequentist model averaging, our proposed transfer learning procedure innovatively combines frequentist model averaging with q-aggregation, bridging the existing gap in statistical inference within transfer learning based on model averaging technique. 
While the asymptotic optimality of frequentist model averaging is generally studied typically when all candidate models are misspecified, we establish the high-probability optimality, without relying on the assumption of misspecification, providing more insights for model averaging research.

\note{[文章其他部分的构成]}
The rest of the paper is organized as follows. 
In Section~\ref{model}, we introduce our transfer learning framework via model averaging.
Section~\ref{theory} provides the theoretical properties of our approach, including under individual-similarity and combinatorial-similarity. 
Extensive simulations and a real-data analysis are investigated in Sections~\ref{sim} and~\ref{realdata}. 
Conclusions and future works are discussed in Section~\ref{discuss}. 
All technical details are presented in the Supplement~A and additional numerical results are provided in Supplement~B. 

Throughout this paper, we use the following notations. 
Define vectors $\bm{1}_{p} = (1,\cdots,1)\T \in \mathbb{R}^{p}$ and $\bm{0}_{p} = (0,\cdots,0)\T \in \mathbb{R}^{p}$.
For a positive integer $M$, $[M]$ denotes the set $\{0,1,2,\ldots,M\}$.
For a vector $\bm{v}$, $\|\bm{v}\|$ denotes $\ell_2$-norm, $\|\bm{v}\|_q$ denotes $\ell_q$-norm for $q \in [0,1]$ and $\|\bm{v}\|_{\infty}$ denotes $\ell_{\infty}$-norm.
For a vector $\bm{v}$ and a set of vectors $\mathcal{V}$, define $d(\bm{v},\mathcal{V}) = \inf_{\tilde{\bm{v}} \in \mathcal{\bm{V}}} \|\bm{v} - \tilde{\bm{v}}\|$.  
For an $m \times m$ dimensional matrix $\bA = (a_{ij})$, $\mathrm{tr}(\bA)$ denotes the trace of $\bA$, and $\bA_{i \cdot}$ denotes the $i^\mathrm{th}$ row of $\bA$. 
We use $\lambda_{\min}(\bA)$ and $\lambda_{\max}(\bA)$ to denote the minimum and maximum eigenvalues of $\bA$, respectively; $\|\bA\|$ for the spectral norm of $\bA$; and  
$\|\bA\|_{F}$ for the Frobenius norm of $\bA$. 
For a sub-Gaussian random variable $\xi \in \mathbb{R}$, its sub-Gaussian norm is defined by $\|\xi\|_{\psi_2} = \inf\{t > 0 : \mathbb{E}[\exp(\xi^{2}/t^2)] \le 2\}$. 
We write $a_n = O(b_n)$ if $\lvert a_n / b_n \rvert \le c$ for some constant $c< \infty$ and sufficiently large $n$. 
Similarly, $a_n = O_p(b_n)$ indicates that $\Pr(\lvert a_n / b_n \rvert \le c) \to 1$ for some constant $c < \infty$ as $n\to\infty$, and $a_n = o_p(b_n)$ indicates that $\Pr(\lvert a_n / b_n \rvert > c) \to 0$ for every $c > 0$ as $n\to\infty$. 
The notations $\xrightarrow{d}$ and $\xrightarrow{p}$ stand for convergence in distribution and convergence in probability, respectively. 
We write $\mathcal{N}(\bmu, \bOmega)$ for the multivariate normal distribution with mean vector $\bmu$ and covariance matrix $\bOmega$ and $\chi^{2}(k)$ for the chi-squared distribution with $k$ degrees of freedom. 
For real numbers $a$ and $b$, $a \vee b = \max\{a, b\}$ and $a \wedge b = \min\{a, b\}$. 
The symbols $\underline{\lim}$ and $\overline{\lim}$ denote the limit inferior and limit superior, respectively. 
Finally, for a set $\mathcal{A}$, $|\mathcal{A}|$ denotes its cardinality, and for sets $\mathcal{A}$ and $\mathcal{B}$, $\mathcal{A} \setminus \mathcal{B}$ is the set of elements in $\mathcal{A}$ that are not in $\mathcal{B}$.

%% file: body/model.tex
\section{Data Generating Process and Transfer Learning Algorithm via Model Averaging}\label{model}
This section outlines the data generation process and introduces our transfer learning algorithms employing model averaging techniques for both individual-similarity and combinatorial-similarity scenarios. 
Specifically, in Section~\ref{subsec:Model framework}, we propose a multiple linear regression framework for transfer learning. 
Building on this foundation, Section~\ref{subsec:candidate} details the construction of candidate domains via contrast threshold variation, 
and then introduce an innovative sufficiency-principled transfer learning algorithm, called \emph{Trans-MAI} as detailed in Section~\ref{subsec:sufficiency-transfer}. 
Furthermore, we extend this approach to strict combinatorial-similarity and combinatorial-similarity scenarios by developing \emph{Trans-MACs} and \emph{Trans-MAC} in Section~\ref{subsec:combinatorial} respectively, to improve the efficiency of leveraging auxiliary knowledge.


\subsection{Model framework}\label{subsec:Model framework}
As \cite{maulud2020review} assert in their review, perhaps one of the most common and comprehensive statistical and machine learning algorithms is linear regression. 
Linear regression characterizes linear relationships between response variables and their predictors, finding widespread applications across diverse disciplines including machine learning \citep{hastie2009elements, murphy2022probabilistic}, economics and finance \citep{campbell1998econometrics, wooldridge2016introductory}, social sciences \citep{gelman2007data, fox2015applied}, medical research \citep{casson2014understanding, schober2021linear}, among others. 
Inspired by this foundation, we investigate transfer learning in the context of multiple linear regression models.
Formally, we consider the target dataset $\mathcal{D}_0 = \{(y_i^{(0)}, \bm{x}_i^{(0)})\}_{i=1}^{n_0}$, consisting of $n_0$ independent and identically distributed (\textit{i.i.d.}) samples drawn from the following linear regression model
\begin{align*} 
  y_{i}^{(0)} 
  = \mu_{i}^{(0)} + \varepsilon_{i}^{(0)} 
  = (\bm{x}_{i}^{(0)})\T \bm{\beta}^{(0)} + \varepsilon_{i}^{(0)}, \quad i = 1,\ldots,n_0,
\end{align*}
where $\mu_i^{(0)} = \mathbb{E}[y_i^{(0)} \mid \bm{x}_i^{(0)}]$ is the conditional expectation, 
$\bm{x}_i^{(0)} = (x_{i1}^{(0)}, \ldots, x_{ip}^{(0)})^\top \in \mathbb{R}^p$ represents the $p$-dimensional covariate, 
$\varepsilon_i^{(0)}$ denotes the random noise satisfying $\mathbb{E}[\varepsilon_i^{(0)} \mid \bm{x}_i^{(0)}] = 0$ and $\mathbb{E}[(\varepsilon_i^{(0)})^2 \mid \bm{x}_i^{(0)}] = \sigma_{(0)}^2$,
and $\bm{\beta}^{(0)} \in \mathbb{R}^p$ is the coefficient vector of interest. 

In the framework of transfer learning, our primary objective is to enhance the estimation accuracy and predictive performance for target by leveraging information from $M$ auxiliary source studies. 
We consider $M$ source datasets $\{\mathcal{D}_m\}_{m =1}^{M}$, where each source $\mathcal{D}_m = \{(y_i^{(m)}, \bm{x}_i^{(m)})\}_{i=1}^{n_m}$ contains $n_m$ \iid observations. 
Similarly, we assume the samples in $\mathcal{D}_m$ are generated from the following linear regression model
\begin{align*}
  y_{i}^{(m)} 
  = \mu_i^{(m)} + \varepsilon_{i}^{(m)} 
  = (\bm{x}_{i}^{(m)})\T \bm{\beta}^{(m)} + \varepsilon_{i}^{(m)}, \quad i = 1,\ldots,n_m,
\end{align*}
where $\mu_i^{(m)} = \mathbb{E}[y_i^{(m)} \mid \bm{x}_i^{(m)}]$ is the conditional expectation, 
$\bm{x}_i^{(m)} = (x_{i1}^{(m)}, \ldots, x_{ip}^{(m)})^\top \in \mathbb{R}^p$ represents the $p$-dimensional covariate, 
$\varepsilon_i^{(m)}$ denotes the random noise satisfying $\mathbb{E}[\varepsilon_i^{(m)} \mid \bm{x}_i^{(m)}] = 0$ and $\mathbb{E}[(\varepsilon_i^{(m)})^2 \mid \bm{x}_i^{(m)}] = \sigma_{(m)}^2$,
and $\bm{\beta}^{(m)} \in \mathbb{R}^{p}$ is the regression coefficient for the $m^{\text{th}}$ source model. 
For convenience, we introduce the following matrix notations for response, conditional expectation, design matrix, error and Gram matrix as
$\bm{y}^{(m)} = (y_{1}^{(m)}, \ldots, y_{n_m}^{(m)})\T \in \mathbb{R}^{n_m}$, $\bm{\mu}^{(m)} = (\mu_{1}^{(m)}, \ldots, \mu_{n_m}^{(m)})\T \in \mathbb{R}^{n_m}$, $\bm{\varepsilon}^{(m)} = (\varepsilon_{1}^{(m)}, \ldots, \varepsilon_{n_m}^{(m)})\T \in \mathbb{R}^{n_m}$, $\bm{X}^{(m)} = (\bm{x}_{1}^{(m)}, \ldots, \bm{x}_{n_m}^{(m)})\T \in \mathbb{R}^{n_m \times p}$, and $\bm{G}^{(m)} = \bm{X}^{(m)\T}\bm{X}^{(m)} \in \mathbb{R}^{p \times p}$ for any $m \in [M]$, and the covariate dimension $p$ is allowed to diverge with the sample size $n_0$.

Since the regression parameters $\bm{\beta}^{(0)}$ and $\{\bm{\beta}^{(m)}\}_{m \in [M]}$ may be identical or different, depending on how similar they are, it is essential to measure the parameter similarity. 
The concept of \emph{individual-similarity} has been well-established in transfer learning literature \citep{bastani2021predicting,cai2021transfer,li2022transfer,lin2022transfer,tian2023transfer,gu2024robust,zhao2023residual}, 
which measures pairwise similarity between each source and target through parameter distance metrics. 
We first discuss the individual-similarity here and then extend to another one, which we refer to \emph{combinatorial-similarity}, in Section~\ref{subsec:combinatorial}.

Let $\bm{\delta}^{(m)} = \bm{\beta}^{(m)} - \bm{\beta}^{(0)}$ denote the contrast vector between the $m^\text{th}$ source and target parameter. 
We quantify the similarity or transferability by the $\ell_2$-norm $\|\bm{\delta}^{(m)}\|$, as it naturally corresponds to the performance metric under quadratic loss in transfer learning\footnote{Alternative metrics such as $\ell_q$-norms ($q \in [0, 1]$) have been explored in prior work \citep{li2022transfer,tian2023transfer}.}. 
Given a similarity threshold $h \geq 0$, we formally partition informative set and non-informative set as follows:

\begin{equation*}
\begin{aligned}
\mathcal{\tA}_{h} &= \{m \in [M] \mid \|\bm{\delta}^{(m)}\| \leq h\} \quad \text{(informative set),} \\
\mathcal{\tA}^c_{h} &= \{m \in [M] \mid \|\bm{\delta}^{(m)}\| > h\} \quad \text{(non-informative set).}
\end{aligned}
\end{equation*}


Threshold $h$ governs the allowable discrepancy between $\bm{\beta}^{(0)}$ and $\{\bm{\beta}^{(m)}\}_{m \in [M]}$. 
The informative set $\mathcal{\tA}_{h}$ contains the indices of relevant studies whose parameter contrast magnitudes $\|\bm{\delta}^{(m)}\|$ fall within tolerance $h$. 
We leverage samples from the informative set $\mathcal{\tA}_h$ to enhance the statistical efficiency of target prediction or estimation, while systematically excluding non-informative set $\mathcal{\tA}_h^c$. 
This selective transfer approach, as established in Section~\ref{subsec:sufficiency-transfer}, ensures that only relevant auxiliary data contributes to the target task while mitigating potential negative transfer effects.
Intuitively, the larger cardinality of informative set $|\mathcal{\tA}_h|$ is, the more auxiliary informative samples are available for the target statistical inference, improving estimation or prediction precision.
The composition of the informative set $\mathcal{\tA}_h$ is entirely determined by the contrast threshold $h$, which controls the similarity between source and target parameters and is data-dependently unknown in practice.
Smaller $ h $ yields high-quality transfer (smaller $\|\bm{\delta}^{(m)}\|$) but reduces $|\mathcal{\tA}_h|$, limiting available transfer information.
Larger $h$ increases $|\mathcal{\tA}_h|$ but risks including irrelevant sources ($\|\bm{\delta}^{(m)}\| \gg 0$), potentially causing negative transfer.
Therefore, $ h $ governs a trade-off between the quantity and quality of informative set $\mathcal{\tA}_h$. 
Remarkably, we emphasize that our proposed method in Section~\ref{model} does not require the explicit input of $h$, which coincides with \citet[Remark 7]{tian2023transfer} and the contrast magnitude $h$ will be further theoretically discussed in Section~\ref{theory}.

\subsection{Construction of candidate domains}\label{subsec:candidate}
To illustrate the motivation behind constructing candidate domains, consider an ideal scenario where all sources are identical to the target (\ie $\bm{\delta}^{(m)} = \bm{0}$ for all $m \in [M]$). 
In this setting, one may construct two ordinary least squares (OLS) estimators for $\bm{\beta}^{(0)}$: 
(a) pooled estimator using all available data with sample size $N_M = \sum_{m \in [M]}n_{m}$, and
(b) single-source estimator using only $\mathcal{D}_m$ with sample size $n_m$.
The pooled estimator achieves a superior convergence rate of $\|\widehat{\bm{\beta}}_{\text{pooled}} - \bm{\beta}^{(0)}\| = O_p(p^{1/2}N_M^{-1/2})$, compared to the single-source convergence rate of $O_p(p^{1/2}n_m^{-1/2})$. 
This demonstrates that when sources are similar to target (i.e., $\|\bm{\delta}^{(m)}\| \leq h$ for some threshold $h > 0$), pooling data from informative auxiliary studies can yield improved statistical efficiency gains and enhanced estimation precision through variance reduction while maintaining consistency.
However, pooling data from heterogeneous auxiliary studies may introduce bias to pooled estimator, leading to negative transfer.
Thus, a core challenge in data pooling lies in the uncertainty in identifying the truly informative set $\mathcal{\tA}_h$ \citep{li2022transfer,tian2023transfer}. 
Our framework formalizes this data pooling process as follows. 
\begin{itemize}
  \item[(i)] \emph{Threshold space construction}: construct the threshold space $\mathcal{H} = \{\|\bm{\delta}^{(m)}\| : m \in [M]\}$ containing all observed contrast magnitudes between sources and target. 
  \item[(ii)]  \emph{Candidate domain generation}:
  for each $h \in \mathcal{H}$ (ordered increasingly), generate candidate domains $\{\mathcal{\tA}_h\}_{h\in\mathcal{H}}$ satisfying
  $\mathcal{\tA}_{h_1} \subseteq \mathcal{\tA}_{h_2}$
  for $h_1 \leq h_2$.
  \item[(iii)]  \emph{Candidate estimation/prediction construction}: 
  for each $h \in \mathcal{H}$, construct estimations or predictions using data from $\mathcal{\tA}_h$. 
\end{itemize}

\begin{figure}[ht]
  \centering
  \includegraphics[width=0.7\linewidth]{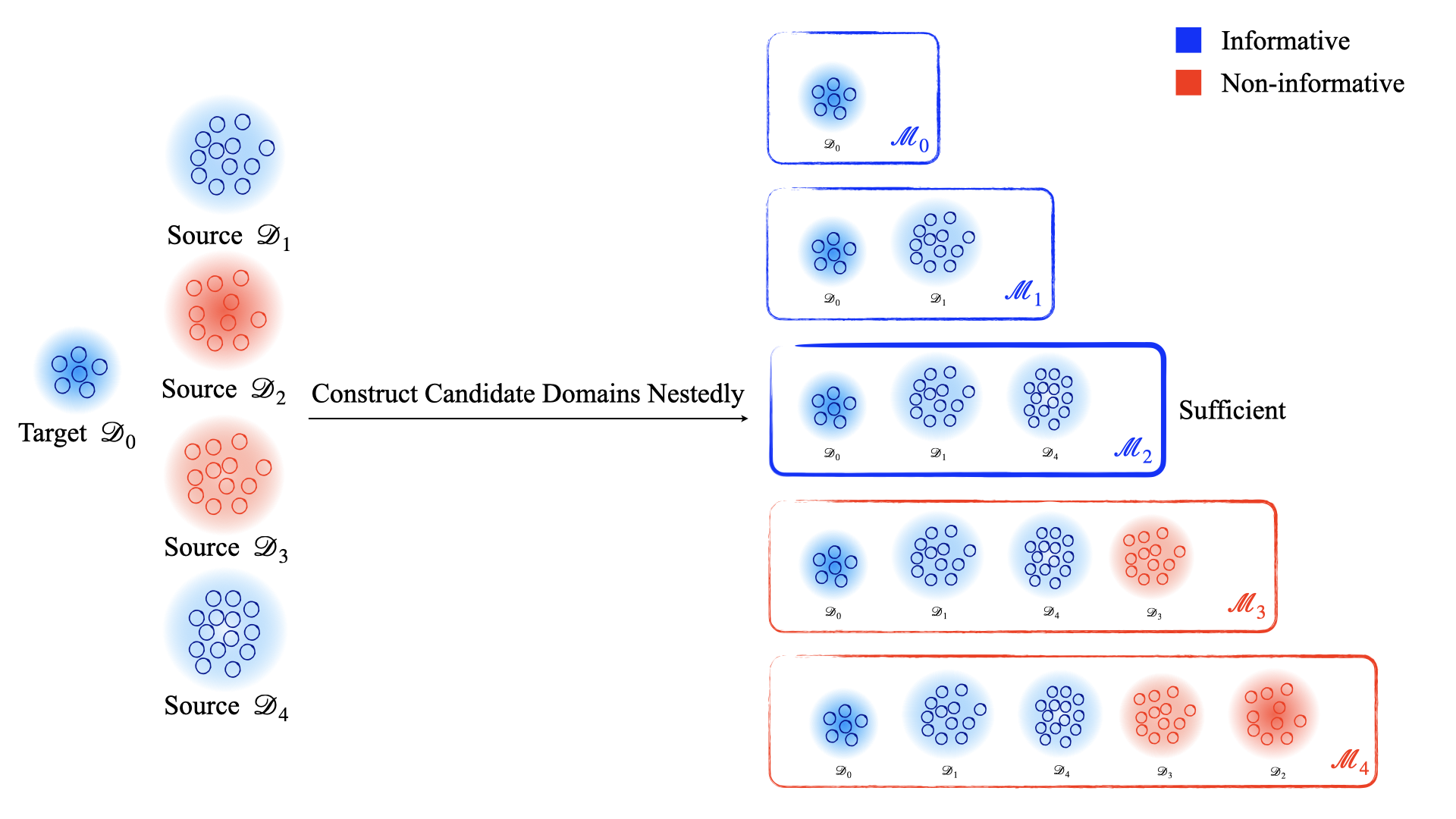}
  \caption{Illustration for constructing candidate domains.}
  \label{sort and construct}
\end{figure}

In practice, the contrast magnitude $\bm{\delta}^{(m)}$ is unknown, and thus we estimate it by an unbiased estimator $\hat{\bm{\delta}}^{(m)} = \hat{\bm{\beta}}^{(m)} - \hat{\bm{\beta}}^{(0)}$, where $\hat{\bm{\beta}}^{(m)} = (\bm{X}^{(m)\top}\bm{X}^{(m)})^{-1}\bm{X}^{(m)\top}\bm{y}^{(m)}$ for $m \in [M]$. 
Then, we nestedly construct candidate domains $\{\mathcal{M}_m\}_{m \in [M]}$ in increasing order of $\|\hat{\bm{\delta}}^{(m)}\|$ as follows:
\begin{equation} \label{eq:candidate_model}
    \mathcal{I}_m = \left\{ j \in [M] : \sum_{k \in [M]} I(\| \hat{\bm{\delta}}^{(k)} \| \leq \| \hat{\bm{\delta}}^{(j)} \| ) \leq m + 1\right\},  \mathcal{M}_m = \bigcup_{j\in\mathcal{I}_m} \mathcal{D}_j,
\end{equation}
where $\mathcal{I}_m$ is the index set of studies comprising candidate domain $\mathcal{M}_m$.
Following Eq.~\eqref{eq:candidate_model}, a series of data-driven, nested candidate domains $\mathcal{M}_m$ are obtained, ordered by their in-sample contrast magnitudes. 
For each $\mathcal{M}_m$ and $\mathcal{I}_{m}$, we denote the total sample size, pooled response vector, conditional expectation, design matrix, error vector and Gram matrix 
by $N_m = \sum_{j\in\mathcal{I}_m} n_j$,
$\bm{y}^{[m]} = [\bm{y}^{(j)\T}]\T_{j \in \mathcal{I}_m} \in\mathbb{R}^{N_m}$, 
$\bm{\mu}^{[m]} = [\bm{\mu}^{(j)\T}]\T_{j \in \mathcal{I}_m} \in\mathbb{R}^{N_m}$, 
$\bm{X}^{[m]} = [\bm{X}^{(j)\T}]\T_{j \in \mathcal{I}_m} \in \mathbb{R}^{N_m\times p}$, 
$\bm{\varepsilon}^{[m]} = [\bm{\varepsilon}^{(j)\T}]\T_{j \in \mathcal{I}_m} \in\mathbb{R}^{N_m}$ 
and $\bm{G}^{[m]} = \bm{X}^{[m]\top}\bm{X}^{[m]} \in \mathbb{R}^{p \times p}$, respectively.
This nested construction not only potentially accelerates the convergence of the parameter estimators, but also greatly simplifies computation compared to considering all possible combinations of auxiliary sources. 
Additionally, the residual vector for target is computed as $\hat{\bm{\varepsilon}}^{(0)} = \bm{y}^{(0)} - \bm{X}^{(0)}\hat{\bm{\beta}}^{(0)}$ with the corresponding residual variance estimator given by $\hat{\sigma}_{(0)}^{2} = n_0^{-1}\|\hat{\bm{\varepsilon}}^{(0)}\|^2$.
The complete procedure for constructing candidate domains is illustrated in Figure~\ref{sort and construct}.

After constructing candidate domains $\{\mathcal{M}_m\}_{m \in [M]}$, we utilize $\hat{\bm{\beta}}^{[m]} = (\bm{X}^{[m]\top}\bm{X}^{[m]})^{-1}\allowbreak \bm{X}^{[m]\top}\bm{y}^{[m]}$ as candidate estimate for each $\mathcal{M}_m$. 
Furthermore, we define the quasi-true parameter value for $\mathcal{M}_m$ as $\bm{\beta}^{[m]} = \mathbb{E}[\hat{\bm{\beta}}^{[m]} \mid \bm{X}^{[m]}]$, with the associated contrast vector $\bm{\delta}^{[m]} = \bm{\beta}^{[m]} - \bm{\beta}^{(0)}$ quantifying the similarity from target parameter. 
As our transfer learning method processes each candidate domain $\mathcal{M}_m$, it is necessary to define the informative and non-informative sets on the candidate domains.
For a similarity threshold $h \geq 0$, we formally partition informative set and non-informative set as follows:
\begin{equation*}
\begin{aligned}
\mathcal{A}_{h} &= \{m \in [M] \mid \|\bm{\delta}^{[m]}\| \leq h\} \quad \text{(informative set),} \\
\mathcal{A}^c_{h} &= \{m \in [M] \mid \|\bm{\delta}^{[m]}\| > h\} \quad \text{(non-informative set).}
\end{aligned}
\end{equation*}

\subsection{Sufficiency-principled model averaging transfer learning algorithm under individual-similarity}\label{subsec:sufficiency-transfer}
In this section, we propose a sufficiency-principled model averaging transfer learning algorithm under individual-similarity when the transferable set is unknown, named \emph{Trans-MAI} (\textbf{Trans}fer Learning via \textbf{M}odel \textbf{A}veraging under \textbf{I}ndividual-Similarly).
Trans-MAI constructs an aggregated estimator for the target parameter by combining estimators from candidate domains by a novel model-averaging based weight selection criterion, which incorporates both predictive performance and domain sufficiency.
In general, Trans-MAI serves two key advantages. 
First, the proposed procedure is asymptotically optimal in the sense that its objective squared prediction and risk function are asymptotically identical to those of the best but infeasible model averaging estimator.
Second, by integrating a sufficiency penalty, Trans-MAI can effectively achieve consistency in sufficient informative domain selection, yielding lower aggregation error. 

Given candidate estimators $\{\hat{\bm{\beta}}^{[m]}\}_{m \in [M]}$, model averaging technique guides the aggregated estimator to take form $\hat{\bm{\beta}}^{(0)}(\bm{w}) = \sum_{m=0}^{M} w_m \hat{\bm{\beta}}^{[m]}$, 
where $w_m$ represents the weight assigned to candidate domain $\mathcal{M}_m$. 
The weight vector $\bm{w} = (w_0, w_1, \dots, w_M)\T$ belongs to the probabilistic simplex
\begin{align*}
  \mathcal{W} = \left\{ \bm{w} \in [0, 1]^{M+1} : \sum_{m=0}^{M} w_m = 1 \right\}.
\end{align*}
The prediction for target is given by $\hat{\bm{\mu}}^{(0)}(\bm{w}) = \bm{X}^{(0)} \hat{\bm{\beta}}^{(0)}(\bm{w}) = \sum_{m=0}^M w_m \hat{\bm{\mu}}^{[m]}$, where $\hat{\bm{\mu}}^{[m]} = \bm{X}^{(0)} \hat{\bm{\beta}}^{[m]}$ is the predictor provided by the $m^\mathrm{th}$ candidate domain. 

To mathematically explain our weight selection criterion, we define the objective loss $L_{v}(\bw)$ and conditional risk function $R_v(\bw)$ on the weight space $\mathcal{W}$ as
\begin{align*} 
  L_{v}(\bw)= (1-v)\|\hat{\bm{\mu}}^{(0)}(\bm{w}) - \bm{\mu}^{(0)}\|^{2} + v \sum_{m=0}^{M} w_m\|\hat{\bm{\mu}}^{[m]} - \bm{\mu}^{(0)}\|^{2} , 
\end{align*}
and $R_{v}(\bw) = \mathbb{E} \bigl[ L_{v}(\bw) \mid \bm{X}^{[M]} \bigr]$ for a constant $v \in [0,1]$, respectively. The corresponding empirical objective loss $\hat{L}_v(\bm{w})$ is constructed by replacing the unobservable $\bm{\mu}^{(0)}$ with the response vector $\bm{y}^{(0)}$,
\begin{align*}
  \hat{L}_v(\bm{w}) = (1-v)\,\|\hat{\bm{\mu}}^{(0)}(\bm{w}) - \bm{y}^{(0)}\|^{2} + v\,\sum_{m=0}^{M} w_m\,\|\hat{\bm{\mu}}^{[m]} - \bm{y}^{(0)}\|^{2}. 
\end{align*}
Through careful derivation (see Supplement~A.2 for details), we establish the following key relationship under conditioning on $\bm{X}^{[M]}$,
\begin{equation} 
  \label{conditional_expectation}
  \mathbb{E}\Bigl[\hat{L}_v(\bm{w}) \mid \bm{X}^{[M]} \Bigr] = R_{v}(\bw) + n_0\sigma_{(0)}^{2} - 2\sigma_{(0)}^{2}\,\sum_{m=0}^{M} w_m\,\mathrm{tr} \Bigl(\bm{G}^{[m]^{-1}}\bm{G}^{(0)}\Bigr).
\end{equation}
This result reveals that $\hat{L}_v(\bm{w})$ becomes an unbiased estimator of $R_{v}(\bw)$ up to a constant term $n_0\sigma_{(0)}^2$, when augmented with the bias-correction term $2\sigma_{(0)}^{2}\sum_{m=0}^{M} w_m\,\mathrm{tr}(\bm{G}^{[m]^{-1}} \bm{G}^{(0)})$. 
Replacing $\sigma_{(0)}^{2}$ by its estimator $\hat{\sigma}_{(0)}^{2}$ and introducing a tuning parameter $\phi$ to the term $\sigma_{(0)}^{2}\sum_{m=0}^{M} w_m\,\mathrm{tr}(\bm{G}^{[m]^{-1}} \bm{G}^{(0)})$ ultimately yield the implementable criterion specified in Eq.~\eqref{eq:CMAv}.

Formally, we choose the weight vector by minimizing the sufficiency-principled weight selection  criterion
\begin{align}\label{eq:CMAv}
    \mathcal{C}_{v,\phi}(\bm{w})
    & = \underbrace{(1-v)\|\bm{y}^{(0)} - \hat{\bm{\mu}}^{(0)}(\bm{w})\|^{2}}_{\textbf{(i) loss of weighted predictor}} 
    + \underbrace{v \ \sum_{m=0}^{M} w_m \|\bm{y}^{(0)} - \hat{\bm{\mu}}^{[m]}\|^{2}}_{\textbf{(ii) weighted loss of candidates}} \nonumber \\ 
    & \qquad \qquad+ \underbrace{\phi \, \hat{\sigma}_{(0)}^{2} \sum_{m=0}^{M} w_{m} \, \mathrm{tr}(\bm{G}^{[m]^{-1}} \bm{G}^{[0]})}_{\textbf{(iii) sufficiency penalty}} , 
\end{align}
where $v \in [0,1]$ and $\phi >0$ are tuning parameters.

Term (i) in Eq.~\eqref{eq:CMAv} is the squared prediction loss of the aggregated predictor. 
In contrast, term (ii) in Eq.~\eqref{eq:CMAv} is a weighted average of the individual candidate prediction losses, which enhances the ability to distinguish between informative and non-informative domains. 
While model selection and averaging demonstrate asymptotic equivalence in certain scenarios \citep{peng2022improvability,xu2022model},
our theoretical analysis (detailed in Section~\ref{theory}) reveals that incorporating $v > 0$ provides a non-trivial improvement: 
it not only accelerates the convergence rate for weights on non-informative domains, but also assigns weight one on the sufficient informative domain, thereby mitigating negative transfer, achieving sufficiency and enhancing the transfer learning efficiency of the aggregated estimator.
Term (iii) in Eq.~\eqref{eq:CMAv} is a sufficiency penalty that adjusts for domain coverage. 
Letting $\bm{\Sigma}^{[m]}$ denote the sample covariance matrix for candidate domain $\mathcal{M}_m$, this penalty can be reexpressed as $\phi \, \hat{\sigma}_{(0)}^{2} n_0\sum_{m = 0}^{M} w_{m} N_{m}^{-1} \mathrm{tr} \bigl(\bm{\Sigma}^{[m]^{-1}} \bm{\Sigma}^{[0]}\bigr)$, 
so that candidate domains with larger sample sizes or better coverage of the target distribution (quantified through the trace term $\mathrm{tr}(\bm{\Sigma}^{[m]^{-1}}\bm{\Sigma}^{[0]}$) incur lower penalties. 
The combined effect adaptively favors domains that either contain more sample size or better represent the target population characteristics.
Our weight vector $\hat{\bw}$ is obtained by solving the constrained optimization problem
$$\hat{\bm{w}} = \argmin_{\bm{w} \in \mathcal{W}} \mathcal{C}_{v,\phi}(\bm{w}),$$ 
and the final aggregated estimator is given by $\hat{\bm{\beta}}^{(0)}(\hat{\bm{w}}) = \sum_{m=0}^{M} \hat{w}_m \hat{\bm{\beta}}^{[m]}$. 

Our proposed weight selection criterion is closely related to q-aggregation technique \citep{dai2012deviation,lecue2014optimal}, but Trans-MAI distinguishes that by incorporating a sufficiency penalty. 
The sufficiency penalty term is essential in transfer learning settings, as it not only asymptotically excludes non-informative domains (\ie the sum of their weights asymptotically converges to zero) and, but also, assigns negligible weights to insufficient informative domain (\ie informative but not sufficient). 
Additionally, although the sufficiency penalty is ideologically similar to parsimonious model averaging \citep{zhang2020parsimonious}, they consider different type of uncertainty. 

Another crucial practical consideration is \emph{data privacy}, which has become increasingly important in the era of transfer learning applications \citep{gao2019privacy,zhang2022data,wang2024differential}. 
In numerous real-world scenarios, raw data from different sources cannot be directly shared due to stringent privacy constraints \citep{hu2023optimal}. 
Our proposed algorithm effectively addresses this challenge through a \emph{regression cube} technique \citep{chen2006regression,lin2011aggregated,schifano2016online}, which enables privacy-preserving distributed computation. 
Mathematically, the algorithm computes each candidate domain estimator $\{\hat{\bbeta}^{[m]}\}_{m \in [M]}$ by
\begin{align} \label{eq:beta_aggregation}
  \hat{\bm{\beta}}^{[m]} = (\bm{X}^{[m]\T} \bm{X}^{[m]})^{-1} \bm{X}^{[m]\T} \bm{y}^{[m]} 
  = \Bigl(\sum_{j \in \mathcal{I}_m} \bm{G}^{(j)}\Bigr)^{-1} \sum_{j \in \mathcal{I}_m} \bm{G}^{(j)} \hat{\bm{\beta}}^{(j)},
\end{align}
which only requires transferring local statistics in each domain. 
Specifically, source $m$ provides $\bm{G}^{(m)}$ and $\hat{\bm{\beta}}^{(m)}$, and transmits them to the target, which ensures that the total data transmission is limited to $Mp(p+1)$ floating-point values, while rigorously preserving the privacy of individual-level data. 
The overall procedure for the sufficiency-principled model averaging transfer learning algorithm under individual-similarity concerning data privacy is summarized in Algorithm~\ref{alg:STMA}.

\begin{algorithm}[H]
  \SetKwInOut{Input}{Input}
  \SetKwInOut{Output}{Output}
  
  \Input{
    Target $\mathcal{D}_0$; 
    Sources $\{\mathcal{D}_m\}_{m=1}^{M}$; 
    $v \in [0,1]$ and $\phi$\;
  }
  \Output{
    Weight estimator $\hat{\bm{w}}$ and $\hat{\bm{\beta}}^{(0)}(\hat{\bm{w}})$\;
  }
  
  \BlankLine
  \textbf{Step 1. Local estimators collection}\;
  \For{$m \in [M]$}{
    Compute $\bm{G}^{(m)} = \bm{X}^{(m)\T} \bm{X}^{(m)}$\;
    Compute $\hat{\bm{\beta}}^{(m)} = \bm{G}^{(m)-1} \bm{X}^{(m)\T} \bm{y}^{(m)}$\;
    Transmit $\bm{G}^{(m)}, \hat{\bm{\beta}}^{(m)}$ to target\;
  }
  Compute $\hat{\bm{\varepsilon}}^{(0)} = \bm{y}^{(0)} - \bm{X}^{(0)}\hat{\bm{\beta}}^{(0)}$\;
  Compute $\hat{\sigma}_{(0)}^{2} = \|\hat{\bm{\varepsilon}}^{(0)}\|^2 /n_0$\;
  
  \BlankLine
  \textbf{Step 2. Candidate domains construction}\;
  \For{$m \in [M]$}{
    Compute $\hat{\bm{\delta}}^{(m)} = \hat{\bm{\beta}}^{(m)} - \hat{\bm{\beta}}^{(0)}$\;
  }
  \For{$m \in [M]$}{
  Compute $\mathcal{I}_m = \left\{ j \in [M] : \sum_{k \in [M]} I(\| \hat{\bm{\delta}}^{(k)} \| \leq \| \hat{\bm{\delta}}^{(j)} \| ) \leq m + 1\right\}$\;
  }
  \For{$m \in [M]$}{
    Compute $\bm{G}^{[m]} = \sum_{j\in \mathcal{I}_m}\bm{G}^{(j)}$\;
    Compute $\hat{\bm{\beta}}^{[m]} = \bm{G}^{[m]-1} \sum_{j\in \mathcal{I}_m}\bm{G}^{(j)} \hat{\bm{\beta}}^{(j)}$\;
  }
  
  \BlankLine
  \textbf{Step 3. Weight selection}\;
  Obtain $\hat{\bm{w}} = \argmin_{\bm{w} \in \mathcal{W}} \mathcal{C}_{v,\phi}(\bm{w})$, where $\mathcal{C}_{v,\phi}(\bm{w})$ is defined in Eq.~\eqref{eq:CMAv}\;
  Compute $\hat{\bm{\beta}}^{(0)}(\hat{\bm{w}}) = \sum_{m=0}^{M}\hat{w}_{m}\hat{\bm{\beta}}^{[m]}$\;
  
  \caption{Trans-MAI algorithm}\label{alg:STMA}
\end{algorithm}

We conclude Section~\ref{subsec:sufficiency-transfer} by the contributions of Trans-MAI algorithm. 
Trans-MAI effectively addresses the challenges in transfer learning without the prior knowledge of transferable set by 
(i) aggregating candidate estimators via a principled weighting criterion that integrates both empirical objective risk and domain sufficiency, 
(ii) automatically excluding non-informative domains to avoid negative transfer, and 
(iii) preserving data privacy by requiring only the transmission of summary statistics, rather than raw data. 
The theoretical analysis in Section~\ref{theory} further demonstrates that Trans-MAI achieves promising convergence properties and robust performance.

\subsection{Extensions for combinatorial-similarity}\label{subsec:combinatorial}
Although Trans-MAI aims to make sufficient use of the informative domains which is defined by the pair-wise contrast between each candidate and target, 
such a transfer learning framework under individual-similarity may be overly restrictive and overlook valuable knowledge within the non-informative domains,
since strategic combinations of non-informative domains might potentially provide a good approximation to the target. 
To address this limitation, we extend our individual-similarity framework to incorporate combinatorial-similarity\textemdash a more flexible similarity paradigm that allows the potential domain combinations to approximate the target, permitting each domain to be arbitrarily distant from the target,
named \emph{Trans-MACs} (\textbf{Trans}fer Learning via \textbf{M}odel \textbf{A}veraging under \textbf{s}trict \textbf{C}ombinatorial-Similarly) and \emph{Trans-MAC} (\textbf{Trans}fer Learning via \textbf{M}odel \textbf{A}veraging under \textbf{C}ombinatorial-Similarly).
This generalized similarity framework not only subsumes individual-similarity a special case, but also accommodates more complex transfer learning scenarios, which integrates knowledge across informative and non-informative domains, yielding superior performance as evidenced by our numerical results in Section~\ref{sim} and Section~\ref{realdata}.

We assume that the target parameter $\bm{\beta}^{(0)}$ can be approximated by a weighted combination of the candidate parameters, imposing the following strict combinatorial-similarity when $|\mathcal{A}_{h}| \neq M$,
\begin{align} \label{approximate linear}
  \bm{\beta}^{(0)} 
  = \sum_{m \in \mathcal{A}_{h}^c} \rho_m \bm{\beta}^{[m]} + \bm{\delta}_{\bm{\rho}},
\end{align}
where weight $\rho_m$ forms a vector in the $|\mathcal{A}_{h}^c|$-dimensional simplex $\mathcal{W}^{|\mathcal{A}_{h}^c|}$ and $\bm{\delta}_{\bm{\rho}} \in \mathbb{R}^p$ is a residual term. 
Combinatorial-similarity requires $\bm{\delta}_{\bm{\rho}}$ to be small, meaning that $\|\bm{\delta}_{\bm{\rho}}\| \leq h$. 
However, the vector $\bm{\rho}$ may not be uniquely identifiable without constraints \citep{lin2024profiled}. 
In fact, there may exist a vector $\bm{g} \in \mathbb{R}^{|\mathcal{A}_{h}^c|}$, satisfying $\sum_{m \in \mathcal{A}_{h}^c} g_m = 0$, such that
\begin{align*}
  \sum_{m \in \mathcal{A}_{h}^c } \rho_m \bm{\beta}^{[m]} + \bm{\delta}_{\bm{\rho}} 
  = \sum_{m \in \mathcal{A}_{h}^c } (\rho_m - g_m) \bm{\beta}^{[m]} + \bm{\delta}_{\bm{\rho}} + \sum_{m \in \mathcal{A}_{h}^c } g_m \bm{\beta}^{[m]} = \sum_{m \in \mathcal{A}_{h}^c } \tilde{\rho}_m \bm{\beta}^{[m]} + \bm{\delta}_{\bm{\tilde{\rho}}},
\end{align*}
where $\tilde{\rho}_{m} = \rho_m - g_m$ satisfies $\tilde{\brho} \in \mathcal{W}^{|\mathcal{A}_{h}^c|}$, and $\bm{\delta}_{\bm{\tilde{\rho}}} = \bm{\delta}_{\bm{\rho}} + \sum_{m \in \mathcal{A}_{h}^c } g_m \bm{\beta}^{[m]}$ satisfies $\|\bm{\delta}_{\bm{\tilde{\rho}}}\| \leq h$.
For identifiability, some orthogonality assumptions are considered in the literature \citep{zhang2020flexible,lin2024profiled}.
Formally, our strict combinatorial-similarity formally requires $\mathcal{S}_h \neq \varnothing$, 
where $\mathcal{S}_h = \{ \brho \in \mathcal{W}^{|\mathcal{A}_{h}^c|} \mid \|\bm{\beta}^{(0)} - \sum_{m \in \mathcal{A}_{h}^c} \rho_m \bm{\beta}^{[m]} \| \leq h \}$.

Motivated by the ideal sufficiency achieved by Trans-MAI under individual-similarity, we propose our weight selection criterion under strict combinatorial-similarity in detail.
Let $m_s = \arg\max_{m \in \mathcal{A}_h } N_{m}$, which is likely the index of the sufficient informative domain and $\mathcal{M}_{m_s}$ consists $\{\mathcal{D}_j\}_{j \in \mathcal{I}_{m_s}}$.
For any $\brho \in \mathcal{W}^{|\mathcal{A}_h^c|}$ and weighted prediction $\hat{\bm{\mu}}^{[m_s]}(\bm{\rho}) = \sum_{m \in \mathcal{A}_{h}^{c}} \rho_m \bm{X}^{[m_s]} \hat{\bm{\beta}}^{[m]}$, 
we define the squared prediction loss and risk function as $L'(\bm{\rho}) = \|\hat{\bm{\mu}}^{[m_s]}(\bm{\rho}) - \bm{\mu}^{[m_s]}\|^2$ and $R' (\bm{\rho}) = \E\left[ L'(\bm{\rho}) \mid \bm{X}^{[M]} \right]$, respectively. 
Moreover, define $\xi' = \inf_{\brho \in \mathcal{W}^{|\mathcal{A}_h^c|}} R'(\brho)$.
Letting $\hat{L}' (\bm{\rho}) = \|\hat{\bm{\mu}}^{[m_s]}(\bm{\rho}) - \bm{y}^{[m_s]}\|^2$,  it is straightforward to verify that
\begin{align*}
  \E \left[ \hat{L}' (\bm{\rho}) \mid \bm{X}^{[M]} \right] 
  = R' (\bm{\rho}) + \sum_{j \in \mathcal{I}_{m_s}} n_j \sigma_{(j)}^2 - 2 \sum_{m \in \mathcal{A}_{h}^c} \rho_m  \left(\sum_{j \in \mathcal{I}_{m_s}} \sigma_{(j)}^2 \operatorname{tr}(\bm{G}^{(j)} \bm{G}^{[m]^{-1}}) \right). 
\end{align*} 
Thus, $\hat{L}' (\bm{\rho}) + 2 \sum_{m \in \mathcal{A}_{h}^c} \rho_m  \left(\sum_{j \in \mathcal{I}_{m_s}} \sigma_{(j)}^2 \operatorname{tr}(\bm{G}^{(j)} \bm{G}^{[m]^{-1}}) \right)$ is an unbiased risk estimator of $R' (\bm{\rho})$ up to a constant term $\sum_{j \in \mathcal{I}_{m_s}} n_j \sigma_{(j)}^2$.
Plugging in the estimators $\{\hat{\sigma}_{(j)}^2\}_{j \in \mathcal{I}_{m_s}}$ yields our Trans-MACs criterion, formulated as
\begin{align*}
  \mathcal{P}(\bm{\rho}) 
  = \| \bm{y}^{[m_s]} - \hat{\bm{\mu}}^{[m_s]}(\bm{\rho})\|^2 + 2 \sum_{m \in \mathcal{A}_h^c} \sum_{j \in \mathcal{I}_{m_s}} \rho_m \hat{\sigma}_{(j)}^2 \operatorname{tr}(\bm{G}^{(j)} \bm{G}^{[m]^{-1}}) , 
\end{align*}
and the weight vector is chosen by $\hat{\bm{\rho}} = \argmin_{\bm{\rho} \in \mathcal{W}^{|\mathcal{A}_{h}^c|}} \mathcal{P}(\bm{\rho})$.  
The corresponding transfer learning estimator for the target parameter $\bbeta^{(0)}$ is $\hat{\bm{\beta}}_{\text{Trans-MACs}}(\hat{\bm{\rho}}) = \sum_{m \in \mathcal{A}_{h}^c} \hat{\rho}_m \hat{\bm{\beta}}^{[m]}$. 

In the literature, transfer learning under combinatorial-similarity has been previously investigated, with notable contributions including the profiled transfer learning approach via linear combinations of source domains proposed by \citet{lin2024profiled}, as well as the convex combination methods analyzed by \citet{zhang2024prediction} and \citet{hu2023optimal}. 
Unlike those works, the key innovation of our proposal lies in incorporating sufficiency principle to guide combinatorial weights, leveraging the sufficient informative domain instead of the target to construct the weight selection criterion and evaluate performances of candidate domains.
Given that the lack of target samples may limit the efficiency of transfer learning, as evidenced by \citet[Theorem 1]{lin2024profiled}, this distinction is promising for improving transfer learning effectiveness.

However, satisfying the strict combinatorial-similarity in Eq.~\eqref{approximate linear} may be challenging in practice. 
To address this, we propose a combinatorial-similarity requiring only $\mathcal{S}_h^{'} \neq \varnothing$, where
$\mathcal{S}_h^{'} = \{ \brho \in \mathcal{W}^{|\mathcal{A}_{h}^c|+1} \mid \|\bm{\beta}^{(0)} - \sum_{m \in \mathcal{A}_{h}^c \cup \{m_s\}} \rho_m \bm{\beta}^{[m]} \| \leq h \}$.
This combinatorial-similarity is more satisfiable, since $m_s \in \mathcal{A}_h$ guarantees that at least $\bm{\beta}^{[m_s]}$ is close to $\bm{\beta}^{(0)}$, making it weaker than the strict one.

Similar to Trans-MACs, we propose our Trans-MAC criterion
\begin{align} 
  \label{eq:mma} 
  \mathcal{Q}(\bm{\varphi}) 
  = \| \bm{y}^{[m_s]} - \hat{\bm{\mu}}^{[m_s]}(\bm{\varphi})\|^{2} + 2 \sum_{m \in \mathcal{A}_h^c \cup \{m_s\}} \sum_{j \in \mathcal{I}_{m_s}} \varphi_m \hat{\sigma}_{(j)}^{2} \operatorname{tr} (\bm{G}^{(j)} \bm{G}^{[m]^{-1}}) ,
\end{align}
where $\hat{\bm{\mu}}^{[m_s]}(\bm{\varphi}) = \sum_{m \in \mathcal{A}_h^c \cup \{m_s\}} \varphi_m \bm{X}^{[m_s]} \hat{\bm{\beta}}^{[m]}$. 
The weight estimator $\hat{\bm{\varphi}}$ is chosen by minimizing the criterion, \ie $\hat{\bm{\varphi}} = \argmin_{\bm{\varphi} \in \mathcal{W}^{|\mathcal{A}_h^c| + 1}}\mathcal{Q}(\bm{\varphi})$
and we can obtain the transfer learning estimator $\hat{\bm{\beta}}_\text{Trans-MAC}(\hat{\bm{\varphi}}) = \sum_{m \in \mathcal{A}_h^{c} \cup \{m_s\} } \hat{\varphi}_m \hat{\bm{\beta}}^{[m]}$. 
Note that if choosing $m_s = 0$, then the Trans-MAC criterion in Eq.~\eqref{eq:mma} degenerates to Trans-MAI criterion in Eq.~\eqref{eq:CMAv} with $\phi = 2$ and $v = 0$. 

\begin{remark}
  Although the combinatorial-similarity framework enables more comprehensive knowledge transfer, this enhanced capability comes at the cost of reduced privacy protection. 
  The estimators $\hat{\bm{\beta}}^{[m]}$ can still be obtained through regression cube technique, but the computation of $\mathcal{P}(\bm{\rho})$ and $\mathcal{Q}(\bm{\varphi})$ requires raw data in $\mathcal{M}_{m_s}$, implying a trade-off between knowledge exploitation and data privacy. 
  While sufficiency justifies selecting $m_s$ as $\arg\max_{m \in [M]}\hat{w}_m$ (see Proposition~\ref{selection consistency} in Section~\ref{theory}),
  privacy considerations may necessitate a more flexible choice of $m_s$ in practice.
\end{remark}

%% file: body/theory.tex
\section{Theoretical Analysis}\label{theory}

In this section, we analyze the theoretical properties of our proposed estimators in Sections~\ref{subsec:sufficiency-transfer} and \ref{subsec:combinatorial}. 
We note that all limiting processes discussed in this paper are as $n_0 \to \infty$ unless otherwise specified, which implies that $N_m \to \infty$ for $m \in [M]$ and dimension $p$ can also be diverging with $n_0$.
Our theoretical analysis mainly consists of two sections. 
The first one discusses the theoretical properties of Trans-MAI under individual-similarity and 
the second one focuses on the theoretical properties of Trans-MAC and Trans-MACs under combinatorial-similarity. 

For convenience, we introduce the notation $\eta = \min_{m \in \mathcal{A}_{h}^c} \|\bm{\delta}^{[m]}\|$ which is the minimum contrast between the target and the non-informative candidate domains. 
Let $s_m = \underline{\lim}_{n_0\to \infty} n_0/N_m$ and $s^{m} = \overline{\lim}_{n_0\to \infty} n_0/N_m$ for $m \in [M]$. 

\subsection{Theoretical analysis under individual-similarity}\label{subsec:theory-individual}
In this section, we consider the theoretical properties of Trans-MAI for individual-similarity under two scenarios: 
when all candidate models are correctly specified and when the target model is misspecified. 
Note that a correctly specified model in our settings means that $\mu_{i}^{(m)}$ is actually a linear function of the covariates $\bm{x}_i^{(0)}$.
When all candidates are correctly specified, we mainly discuss the 
weight convergence, estimation/prediction convergence rate and asymptotic distribution. 
When the target model is misspecified, we focus on the asymptotic optimality.

\subsubsection{Theoretical analysis for Trans-MAI under correctly specified candidate models}
Under correctly specified candidate models, we first consider the convergence of the weights assigned to non-informative candidates, demonstrating that Trans-MAI can effectively mitigate the negative transfer and some technical conditions are outlined below.

\begin{condition} \label{C.4}
    (i) For $m \in [M]$, there exist positive constants $\tau_ 1$ and $\tau_2$, such that 
    \begin{align*}
       0 < \tau_1 
       \leq \lambda_{\min}(\bm{\Sigma}^{[m]}) 
       \leq \lambda_{\max}(\bm{\Sigma}^{[m]}) 
       \leq \tau_2 < \infty . 
    \end{align*}

    \noindent(ii) For $m \in [M]$, $\lim_{N_m \to \infty} \max_{i=1,\dots,N_m} \|{\bm{X}}^{[m]}_{i\cdot}\|^{2}/N_m = 0$.

    \noindent(iii) $\sup_{m \in [M]} \max_{i=1, \dots, n_m} \E \bigl[|\bm{\epsilon}_i^{(m)}|^{2+\zeta}| \bm{X}^{[m]} \bigr] < \infty$, for some $\zeta > 0$. 
\end{condition}


\begin{condition} \label{finitebeta}
    $
    \sup_{m \in [M]} \|\bm{\beta}^{[m]}\|_\infty = O(1)
    $.
\end{condition}

\begin{condition} \label{C.1}
    There exist positive constants $c_1$ and $c_2$ such that
    \begin{align*}
        \Pr (c_1 \leq \hat{\sigma}_{(0)}^{2} \sigma_{(0)}^{-2} \leq c_2) \to 1.
    \end{align*}
\end{condition}

\begin{condition} \label{C.9weak}
    For $v \in (0,1/2]$, there exists a positive constant $\kappa_v > 0$, such that
    \begin{align*}
    (1-v) \bm{\delta}^{[m]\T } \bm{\Sigma}^{(0)} \bm{\delta}^{[m']} + 2^{-1} v \Bigl(\bm{\delta}^{[m]\T }\bm{\Sigma}^{(0)} \bm{\delta}^{[m]} + \bm{\delta}^{[m']\T } \bm{\Sigma}^{(0)} \bm{\delta}^{[m']} \Bigr) 
    \geq \kappa_v \|\bm{\delta}^{[m]}\| \|\bm{\delta}^{[m']}\| , 
    \end{align*}
    for any $m, m' \in \mathcal{A}_{h}^{c}$. 
\end{condition}

\begin{condition} \label{C.9}
    There exists a positive constant $\kappa > 0$ such that
    \begin{align*}
        \bm{\delta}^{[m]\T } \bm{\Sigma}^{(0)} \bm{\delta}^{[m']} \geq \kappa \|\bm{\delta}^{[m]}\| \|\bm{\delta}^{[m']}\|,
    \end{align*}
    for any $m, m' \in \mathcal{A}_{h}^{c}$. 
\end{condition}

Condition~\ref{C.4} assumes that the covariates is well-behaved, 
Condition~\ref{C.4}(i) implies the covariance matrices of covariates have bounded eigenvalues, 
Condition~\ref{C.4}(ii) requires the boundedness of the norm of the covariates, 
and Condition~\ref{C.4}(iii) bounds the conditional moment of order $2 + \zeta$ for $\bm{\epsilon}_i^{(m)}$, where $\zeta$ is a positive constant. 
Condition~\ref{C.4}(i) is equivalent to condition (F) in \citet{fan2004nonconcave}. 
Conditions~\ref{C.4} is generalized form of Assumptions (A1), (A2), and (A3) in \citet{zou2009adaptive}, and similar conditions have been considered in \citet{portnoy1984asymptotic} and \citet{zhang2020parsimonious}. 
Condition~\ref{finitebeta} requires that each element of $\{\bm{\beta}^{[m]}\}_{m \in [M]}$ are uniformly bounded. 
Condition~\ref{C.1} does not require that $\hat{\sigma}_{(0)}^{2}$ is consistent, thus making it easy to be satisfied, as specified in \citet[Condition 2]{yang2003regression}, \citet[Condition 2]{yuan2005combining} and  \citet[Condition C.1]{zhang2020parsimonious}. 

\begin{figure}[ht]
    \centering
    \includegraphics[width=0.8\linewidth]{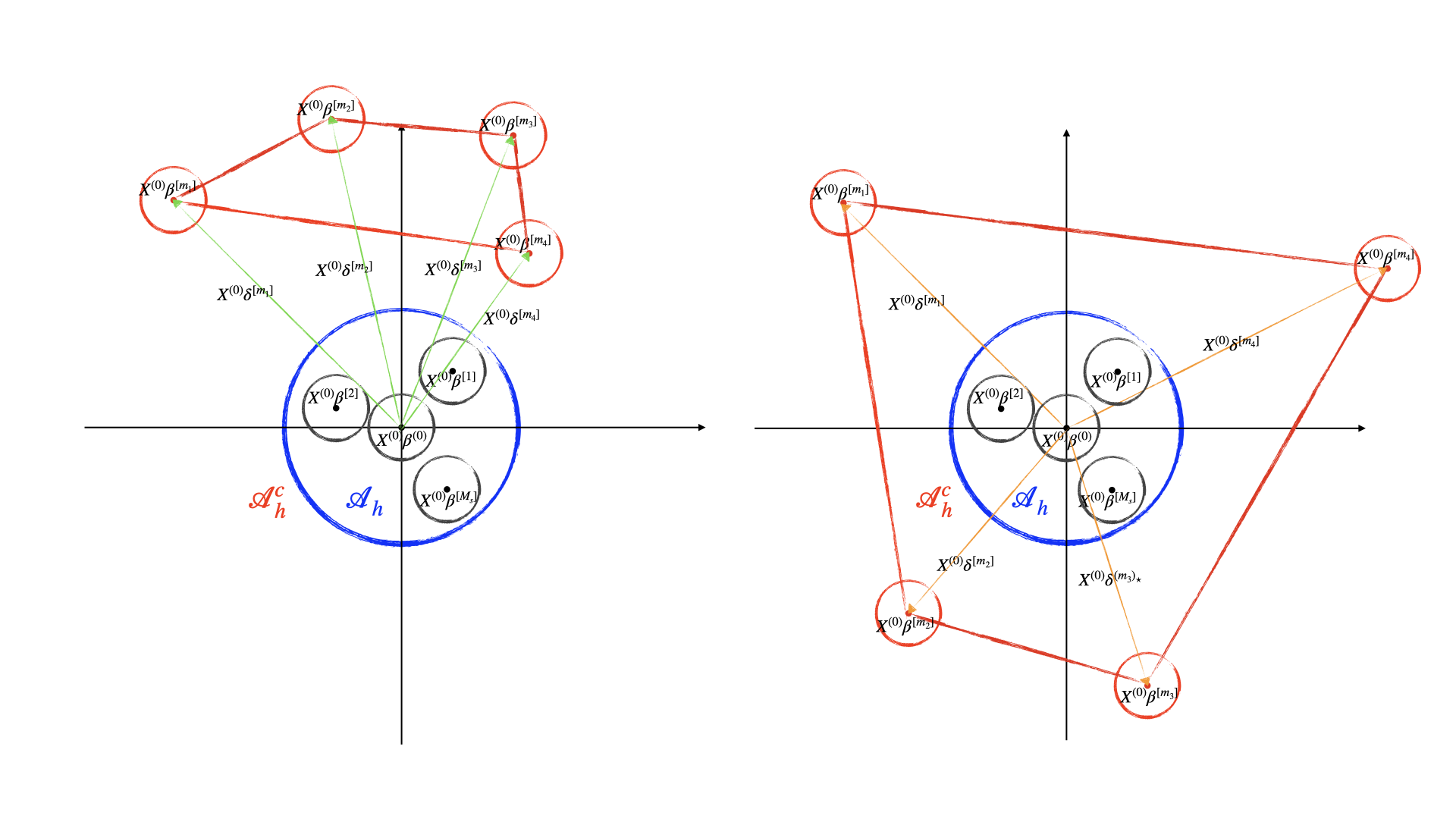}
    \caption{
        Geometric interpretation of the Condition~\ref{C.9}. 
        For $m \in [M]$, the points $\bm{X}^{(0)}\bm{\beta}^{[m]}$ are represented by dots, and the volatility of each estimate is indicated by circles. 
        The point $\bm{X}^{(0)}\bm{\beta}^{(0)}$ lies at the origin. 
        The blue circle represents the range of the informative domains. 
        Informative domains, indicated by black dots, lie within the blue circle, while non-informative domains, indicated by red dots, are located outside the blue circle. 
        The red polygon represents the simplex formed by $\bm{X}^{(0)} \bm{\beta}^{[m]}$ for $m \in \mathcal{A}_{h}^c$. 
        The green and orange arrows represent vectors $\bm{X}^{(0)}\bdelta^{[m]}$.
        The left case satisfies Condition~\ref{C.9}, while the right case violates Condition~\ref{C.9}.
    }
    \label{fig:Geometric interpretation of the Condition}
\end{figure}

Condition~\ref{C.9} requires that the inner product of $\bm{X}^{(0)} \bm{\delta}^{[m]} / \|\bm{\delta}^{[m]}\|$ of any two non-informative candidate models is positive, and bounded away from some constant. 
This implies that the target is not in the linear space spanned by the non-informative domains. 
To facilitate comprehension, we present the geometrical interpretation for Condition~\ref{C.9} in Fig.~\ref{fig:Geometric interpretation of the Condition}. 
On the left-hand side in Fig.~\ref{fig:Geometric interpretation of the Condition}, the green vectors denote $\bm{X}^{(0)}\bm{\delta}^{[m]}$ for $m \in \mathcal{A}_{h}^c$. 
Any pair of green vectors forms an acute angle, ensuring that Condition~\ref{C.9} is satisfied. 
In this case, the red polygon is far from the blue circle, indicating that the non-informative domains do not exhibit combinatorial-similarity. 
On the right side in Fig.~\ref{fig:Geometric interpretation of the Condition}, orange vectors denote $\bm{X}^{(0)}\bm{\delta}^{[m]}$ for $m \in \mathcal{A}_{h}^c$. 
Here, there exist pairs of orange vectors forming obtuse angles, causing Condition~\ref{C.9} to be violated. 
In this scenario, the red simplex encloses the blue circle, suggesting that the non-informative domains may exhibit potential combinatorial-similarity. 
In summary, Condition~\ref{C.9} ensures that the convex linear combinations of the non-informative domains do not coincide with the informative domains. 
Since that $\bm{\delta}^{[m]\T } \bm{\Sigma}^{(0)} \bm{\delta}^{[m]} + \bm{\delta}^{[m']\T } \bm{\Sigma}^{(0)} \bm{\delta}^{[m']} \geq 2 \bm{\delta}^{[m]\T } \bm{\Sigma}^{(0)} \bm{\delta}^{[m']}$, Condition~\ref{C.9weak} is a weaker version than Condition~\ref{C.9}.

\begin{lemma}[Weight convergence for non-informative candidates] \label{Convergence of weight new} 
    (i) When $v = 1$, if Conditions~\ref{C.4},~\ref{finitebeta},~\ref{C.1}, and $\phi n_0^{-1} p \eta^{-2} |\mathcal{A}_{h}^c|^{1/2} \to 0$ hold, 
    then the weight vector assigned to non-informative candidates, $\hat{\bm{w}}_{\mathcal{A}_h^c}$, satisfies 
    \begin{align*}
        \|\hat{\bm{w}}_{\mathcal{A}_h^c}\| = O_p\bigl(\phi n_0^{-1} p \eta^{-2}|\mathcal{A}_{h}^c|^{1/2}\bigr) . 
    \end{align*}
    (ii) When $v \in (1/2,1)$, if Conditions~\ref{C.4},~\ref{finitebeta},~\ref{C.1}, $h \eta^{-1} = o(1)$, and $\phi n_0^{-1} p \eta^{-2} |\mathcal{A}_{h}^c|^{1/2} \to 0$ hold, 
    \begin{align*}
        \|\hat{\bm{w}}_{\mathcal{A}_h^c}\| = O_p\bigl(\phi n_0^{-1} p \eta^{-2}|\mathcal{A}_{h}^c|^{1/2}\bigr) . 
    \end{align*}
    (iii) When $v \in (0,1/2]$, if Conditions~\ref{C.4},~\ref{finitebeta},~\ref{C.1},~\ref{C.9weak}, $h \eta^{-1} = o(1)$, and $\phi n_0^{-1} p \eta^{-2}|\mathcal{A}_{h}^c|^{1/2} \to 0$ hold,
    \begin{align*}
        \|\hat{\bm{w}}_{\mathcal{A}_h^c}\| = O_p\bigl(\phi n_0^{-1} p \eta^{-2}|\mathcal{A}_{h}^c|^{1/2}\bigr) .
    \end{align*}
    (iv) When $v = 0$, if Conditions~\ref{C.4},~\ref{finitebeta},~\ref{C.1},~\ref{C.9}, $h p^{1/2} \eta^{-2}|\mathcal{A}_{h}^c|^{1/2} = o(1)$, $\phi n_0^{-1} p \eta^{-2} |\mathcal{A}_{h}^c|^{1/2} \to 0$ and $n_0^{-1/2}p \eta^{-2}|\mathcal{A}_{h}^c|^{1/2} \to 0$ hold, 
    \begin{align*}
        \|\hat{\bm{w}}_{\mathcal{A}_h^c}\| 
        = O_p \bigl( \phi n_0^{-1} p \eta^{-2} |\mathcal{A}_{h}^c|^{1/2} + n_0^{-1/2}p \eta^{-2}|\mathcal{A}_{h}^c|^{1/2} + 
        h p^{1/ 2} \eta^{-2}|\mathcal{A}_{h}^c|^{1/2} \bigr) . 
    \end{align*}
\end{lemma}

Lemma~\ref{Convergence of weight new} implies that the weights assigned to non-informative domains converges to $0$ in probability as $n_0 \to \infty$ when $h$ and $\phi $ are appropriately chosen. 
The bounds in Lemma~\ref{Convergence of weight new} are used to further prove Lemma~\ref{Convergence of weight2}, Theorems~\ref{Convergence Rate of Trans-MA} and \ref{Oracle property}. 
Note that as $v$ increases, the conditions for weights assigned to non-informative domains asymptotically converging to $0$ gradually relax. 
Intuitively, increasing $v$ enlarges the penalty for each non-informative domain, thereby placing greater emphasis on model selection and mitigating the potential effects of combinatorial-similarity between the non-informative domains and the target.
Moreover, the minimal signal difference between the non-informative candidates and the target, $\eta$, influences the convergence rate of the non-informative candidates. 
Larger $\eta$, faster convergence rate of $\hat{\bm{w}}_{\mathcal{A}_{h}^c}$. 
The condition $h \eta^{-1} = o(1)$ implies that the contrast between the informative candidates and the target is negligible relative to that of the non-informative candidates. 
In the case of $v = 0$, $\|\hat{\bm{w}}_{\mathcal{A}_{h}^c}\|$ is dominated by $n_0^{-1/2} p \eta^{-2} |\mathcal{A}_{h}^c|^{1/2}$ even under Condition~\ref{C.9}, which suggests that only using standard model averaging (\ie $v=0$) is hard to distinguish between informative and non-informative domains compared to using a criterion mixing model averaging and model selection (\ie $v > 0$). 
Moreover, if Condition~\ref{C.9} is violated, the weights assigned to the non-informative domains $\hat{\bm{w}}_{\mathcal{A}_{h}^c}$ fail to converge to zero. 
A concrete numerical illustration of this phenomenon is provided in Figure~\ref{fig:exp2v} (Section~\ref{sim}).


Next, we examine the weight convergence for informative domains, which demonstrates transfer learning sufficiency. 
Specifically, under certain regularity conditions, we show that as $n_0 \to \infty$, the weights asymptotically concentrate on the sufficient informative domain.

\begin{condition}[Gap of sample size for informative candidates] \label{gap of sample size}
    There exists a positive constant $c_3 > 0$ such that 
	\begin{align*} 
    \tau_1^{2} \inf_{m \in \mathcal{A}_h \backslash \{m_s\}} s_m - \tau_2^2 s^{m_s} \geq c_3, 
    \end{align*}
    where $\tau_1$ and $\tau_2$ are positive constants defined in Condition~\ref{C.4}. 
\end{condition} 

Condition~\ref{gap of sample size} imposes a constraint on the sample size of informative candidate domains to bound $\operatorname{tr}(\bm{G}^{[m]^{-1}}\bm{G}^{(0)}) - \operatorname{tr}(\bm{G}^{[m_s]^{-1}} \bm{G}^{(0)})$ for $m \in \mathcal{A}_h \backslash \{m_s\}$, according to Ruhe's trace inequality \citep{ruhe1970perturbation}, implying $\mathcal{A}_{h} \backslash \{m_s\} \neq \varnothing$. 
For facilitating comprehension, we provide two cases for Condition~\ref{gap of sample size},
(i) $N_m = q^{m} n_0$, $m \in [M]$, where positive constant $q > \tau_{2}^{2}\tau_{1}^{-2}$, and 
(ii) $N_m = f(m) n_0$, $m \in [M]$, where $\inf_{m \in \mathcal{A}_h \backslash \{m_s\} } f(m_s)f(m)^{-1} > \tau_{2}^{2}\tau_{1}^{-2}$.
Condition~\ref{gap of sample size} ensures that Trans-MAI can automatically identify the sufficient informative domain, as established in Lemma~\ref{Convergence of weight2}. 


\note{ 
    For the case that $\min_{0 \leq j \leq |\mathcal{A}| -1}\varrho_j = 0$, one potential solution is to identify a suitable function, designated as $f:\mathbf{R} \to \mathbf{R}$, satisfied that $\min_{0 \leq j \leq |\mathcal{A}| -1}\varrho_j^{'} >0 $, where $\varrho_j^{'} = \overline{\lim}_{n_0\to \infty, N_{j} \to \infty}\frac{f(n_0)}{f(N_j)}$ and select the optimal weight by minimizing the following modified criterion as follows
    \begin{align}
        \mathcal{C}_{MA}^{'}(\bm{w}) = (1-v)\|\by^{(0)} - \hat{\bmu}(\bm{w})\|^{2} + v \sum_{j=0}^{M}w_j\|\by^{(0)} - \hat{\bmu}^{(j)}\|^{2} + \phi \hat{\sigma}_{(0)}^{2} \sum_{j = 0}^{M}w_{j} \frac{f(n_0)}{f(N_j)}\mathrm{tr}(\bm{\Sigma}^{(j)^{-1}}\bm{\Sigma}^{(0)}) . \label{CMAv'}
    \end{align}
}

\begin{lemma}[Sufficiency for informative domains] \label{Convergence of weight2}
    Under the same conditions as Lemma~\ref{Convergence of weight new} and Condition~\ref{gap of sample size},
    (i) when $v=1$, if $h = O\bigl(n_0^{-1/2}\bigr)$ and $\phi \to \infty$, then we have for $m \in \mathcal{A}_{h} \backslash \{m_s\}$, 
    \begin{equation*}
        \hat{w}_m = O_p\bigl(\phi ^{-1}\bigr).  
    \end{equation*}
    (ii) when $v \in (0,1)$, if $h\eta^{-1} = o(1), h = O\bigl(n_0^{-1/2}\bigr)$, $\phi \to \infty$ and $\phi p n_0^{-1/2}\eta^{-2}|\mathcal{A}_{h}^c| \to 0$, then we have for $m \in \mathcal{A}_{h} \backslash \{m_s\}$,
    \begin{equation*}
        \hat{w}_m = O_p\bigl(\phi ^{-1}\bigr). 
    \end{equation*}
\end{lemma}

Lemma~\ref{Convergence of weight2} implies that the weights assigned to the insufficient informative domains are bounded by $O_p(\phi ^{-1})$. 
As the sufficiency penalty term $\phi \to \infty$, Trans-MAI will asymptotically assign zero weight to insufficient informative domains, ensuring sufficiency. 
We note that the convergence rate is closely related to \citet[Theorem 5]{zhang2019inference}, which focuses on multi-model aggregation in single-domain learning. 
Additionally, for the tuning parameter $\phi$, we recommend using $\log{n_0}$, which corresponds to the penalty term in the Bayesian information criterion and satisfies the technical assumptions. 

Furthermore, denote $\hat{m}_s = \argmax_{m \in [M]} \hat{w}_m$ as the index of candidate domain with the largest estimated weight, which corresponds to the estimate of sufficient informative domain $\mathcal{M}_{\hat{m}_s}$. 
According to the convergence rates established in Lemmas~\ref{Convergence of weight new} and~\ref{Convergence of weight2}, we derive the consistency in sufficient informative domain selection as follows. 
\begin{proposition}[Consistency in sufficient informative domain selection] \label{selection consistency}
    For $v \in (0,1]$, under the same conditions as Lemmas~\ref{Convergence of weight new} and~\ref{Convergence of weight2}, $\Pr ({\hat{m}_s} = m_s ) \to 1$.
\end{proposition}
Proposition~\ref{selection consistency} establishes the consistency in sufficient informative domain selection for proposed Trans-MAI. 
Specifically, Proposition~\ref{selection consistency} tells us that  under certain conditions, the sufficient informative domain $\mathcal{M}_{\hat{m}_s}$ suggested by Trans-MAI will be $\mathcal{M}_{m_s}$ for some $h$.
It is necessary to emphasize that Trans-MAI does not require the explicit input of $h$ and similar discussions can be found in \citet[Remark 7]{tian2023transfer}.
Such a consistency is not only valuable for avoiding negative transfer, but also improving the efficiency of estimation/prediction for transfer learning, as established in Theorem~\ref{Convergence Rate of Trans-MA}. 



\begin{theorem}[Convergence rate of Trans-MAI] \label{Convergence Rate of Trans-MA}
    For $v \in (0,1]$, under the same conditions as Lemmas~\ref{Convergence of weight new} and~\ref{Convergence of weight2},
    \begin{align*} 
        &\|\hat{\bm{\beta}}^{(0)}(\hat{\bm{w}}) - \bm{\beta}^{(0)}\| \vee n_0^{-1}\|\bm{X}^{(0)}(\hat{\bm{\beta}}^{(0)}(\hat{\bm{w}})- \bm{\beta}^{(0)})\| \\
      = & O_p \bigl( \phi ^{-1}p^{1/2}n_0^{-1/2}(|\mathcal{A}_h|-1) \bigr) + O_p\bigl(p^{1/2}N_{m_s}^{-1/2}\bigr) + O(h) +  O_p\bigl(\phi n_0^{-1} p^{3/2} \eta^{-2} |\mathcal{A}_{h}^c|\bigr) . 
    \end{align*}
\end{theorem}
The convergence rate in Theorem~\ref{Convergence Rate of Trans-MA} is composed of four terms. 
The first term, $O_p\bigl(\phi ^{-1}p^{1/2} \allowbreak n_0^{-1/2}(|\mathcal{A}_h|-1)\bigr)$, represents the estimation error from the insufficient informative candidate domains. 
The second term, $O_p(p^{1/2} N_{m_s}^{-1/2})$, captures the estimation error arising from the sufficient informative domain. 
If $N_{m_s} \gg n_0$, this term converges faster than $O_p(p^{1/2}n_0^{-1/2})$, the error using only the target data.
Next term, $O(h)$, reflects the contrast magnitude among the informative candidate domains and is negligible when $h$ is small enough.
The final term, $O_p \bigl(\phi n_0^{-1} p^{3/2} \eta^{-2} |\mathcal{A}_h^c|\bigr)$, accounts for the non-transferability of the non-informative domains. 
If $\phi ^{-1} (|\mathcal{A}_h| -1 ) = o(1)$, $h = o(p^{1/2}n_0^{-1/2})$ and $\phi n_0^{-1/2}p \eta^{-2}|\mathcal{A}_h^c| = o(1)$, we have $\|\hat{\bm{\beta}}^{(0)}(\hat{\bm{w}}) -\bm{\beta}^{(0)}\| = o_p(p^{1/2}n_0^{-1/2})$, 
which clarifies the significance of transfer learning. 
Then, we briefly compare the convergence rate of Trans-MAI with Trans-Lasso (\citet[Corollary 1]{li2022transfer}) based on q-aggregation and recall some notations defined in \cite{li2022transfer}.
For some constant $c_1<1$, $\tilde{\mathcal{A}}^{o}=\left\{k \in \mathcal{A}:\left\|\bSigma^{(k)} \bw^{(k)}- \bSigma^{(0)} \bbeta\right\|_2^2 <c_1 \min _{k \in \mathcal{A}^c} \sum_{j \in H_k}\left|\bSigma_{j,}^{(k)} \bw^{(k)}-\bSigma_{j, .}^{(0)} \bbeta\right|^2\right\}$,
where $H_k=\left\{1 \leq j \leq p:\left|\bSigma_{j, .}^{(k)} \bw^{(k)}-\bSigma_{j,}^{(0)} \bbeta\right|>n_*^{-\kappa}, \kappa<\alpha / 2\right\}$.
Additionally, $C_{\bSigma}=1+\max _{j \leq p} \max _{k \in \mathcal{A}} \allowbreak \bigl\|e_j^{\top}\left(\bSigma^{(k)}-\bSigma^{(0)}\right)\left(\sum_{k \in \mathcal{A}} \alpha_k \bSigma^{(k)}\right)^{-1}\bigr\|_1$ and $\|\bbeta\|_0 \leq s$. 

\begin{corollary}[Comparisons of Trans-Lasso and Trans-MAI] \label{compare with Translasso}
    For $v \in (0,1]$, under the same conditions as Theorem~\ref{Convergence Rate of Trans-MA},
    if $\sqrt{\log{p}/n_0} = o(1)$, $h = o(n_0^{-1/2} (\log{M})^{1/2})$, $C_{\bSigma} = O(1)$,
    $n_0^{1/2} (n_{\tilde{\mathcal{A}}^{o}} + n_0)^{-1/2} = O(s^{-1/2} (\log{p})^{-1/2} (\log{M})^{1/2})$, 
    and $\phi^{-1} (|\mathcal{A}_h|-1) \vee n_0^{1/2} N_{m_s}^{-1/2} \vee \phi n_0^{-1/2}  |\mathcal{A}_h^c| = o(p^{-1/2} (\log{M})^{1/2})$ hold, 
    then we have 
    $\|\hat{\bm{\beta}}^{(0)} (\hat{\bm{w}}) - \bm{\beta}^{(0)}\| = o_p(\|\hat{\bm{\beta}}_{\text{Trans-Lasso}} - \bm{\beta}^{(0)}\|)$. 
\end{corollary}
Corollary~\ref{compare with Translasso} compares the error bound for Trans-MAI with Trans-Lasso, requiring some extra conditions compared with Theorem~\ref{Convergence Rate of Trans-MA}.
Notice that $C_{\bSigma}$ is a constant if $\max_{1 \leq j \leq p} \bigl\| \bm{e}_j\T \bigl(\bm{\Sigma}^{(k)}-\bm{\Sigma}^{(0)}\bigr) \bigr\|_0 \leq C < \infty$ holds for all $k \in \mathcal{A}$ \citep{li2022transfer}, thus $C_{\bm{\Sigma}} = O(1)$ is reasonable.
The equalities (i) and (ii) places restrictions on $s$, $p$, $M$, $\phi$, $n_{\tilde{\mathcal{A}}^{o}}$, $N_{m_s}$ and $n_0$.
Under conditions in Corollary~\ref{compare with Translasso}, $\hat{\bm{\beta}}^{(0)} (\hat{\bm{w}})$ serves a faster convergence rate than $\hat{\bm{\beta}}_{\text{Trans-Lasso}}$,
because Trans-MAI achieves sufficiency, reducing the error for q-aggregation procedure to $o(n_0^{-1}\log{(M)})$, thus improving the efficiency of estimation. 
Then, we establish the asymptotic distribution of Trans-MAI estimator. 





\begin{theorem}[Asymptotic normality of Trans-MAI] \label{Oracle property}
    Let $\bpsi \in \mathbb{R}^{p}$ be an arbitrary vector of $\ell_2$-norm $1$. 
    For $v \in (0,1]$, under the same conditions as Lemmas~\ref{Convergence of weight new} and~\ref{Convergence of weight2},
    if $h = o(N_{m_s}^{-1})$, $\phi ^{-1}N_{m_s}^{1/2}n_0^{-1/2}p^{1/2} (|\mathcal{A}_h|-1)^{1/2} \to 0$, and $\phi N_{m_s}^{1/2} n_0^{-1} p^{3/2} \eta^{-2} |\mathcal{A}_h^c| \to 0$ hold, then we have 
    \begin{equation*}
        \frac{\sqrt{N_{\hat{m}_s}}\bpsi\T \bSigma^{[\hat{m}_s]^{1/2}}}
            { \sqrt{\sum_{k \in \mathcal{I}_{\hat{m}_s}}(\bpsi\T (\bX^{(k)\T }\bX^{(k)})\bG^{[\hat{m}_s]^{-1}} \bpsi)\sigma_{(k)}^{2}} }
            (\hat{\bm{\beta}}^{(0)}(\hat{\bm{w}}) - \bm{\beta}^{(0)} )\xrightarrow{d} \mathcal{N}(0, 1).
        \end{equation*}
\end{theorem}

Theorem~\ref{Oracle property} demonstrates the asymptotic normality of Trans-MAI estimator. 
Notably, this result does not require the prior knowledge for the transferable set, simplifying statistical inference. 
Then, we explore the high-probability optimality of Trans-MAI and introduce the following conditions.
\begin{condition}\label{sub-Gaussian-target}
    $\varepsilon_{i}^{(0)}$ are sub-Gaussians and there exist a positive constant $C_{\psi_2} > 0$ such that $\|\varepsilon_{i}^{(0)}\|_{\psi_2} \leq C_{\psi_2}$, for $i = 1,\cdots, n_0$.
\end{condition}

\begin{condition}\label{mu_star_new_all}
    $\displaystyle\sup_{m \in [M]} N_{m}^{-1}\|\bm{\mu}^{[m]}\|^{2} = O(1)$.
\end{condition}

\begin{theorem}[High-probability optimality of Trans-MAI] \label{mma finite ao}
    Assuming that 
    \begin{equation}
        \epsilon_{\sigma_{(0)}} = \Pr( |\hat{\sigma}_{(0)}^2 - \sigma_{(0)}^2| > \tau_1 \tau_2 ^{-1} \sigma_{(0)}^2 ) < 1, \label{sigma condition 0}
    \end{equation}
    under Conditions~\ref{C.4}(i),~\ref{sub-Gaussian-target} and~\ref{mu_star_new_all}, 
    we have for any $\epsilon \in (0,1)$, 
    \begin{equation}\label{mma finite ao1}
        \frac{1}{n_0}R_v(\hat{\bw}) \leq \frac{1}{n_0} \inf_{\bw \in \mathcal{W}} R_v(\bw) + u_{1},
    \end{equation}
    with probability at least $1- \epsilon -\epsilon_{\sigma_{(0)}} $, 
    where $u_1 = c (M+1) \max \bigl\{ n_0^{-1}\log{\left(2(M+1)\epsilon^{-1}\right)}, \allowbreak p^{1/2} n_0^{-1/2}[\log{\left(2(M+1)\epsilon^{-1} \right)}]^{1/2} \bigr\}  + (|\phi - 2| + \phi) \sigma_{(0)}^2 (M + 1)pn_0^{-1}$ and $c$ is a positive constant. 
\end{theorem}

The condition in Eq.~(\ref{sigma condition 0}) does not require that the estimator $\hat{\sigma}_{(0)}^2$ is consistent, and similar condition can be found in \citet[Theorem 6.2]{bellec2018optimal}.
In Theorem~\ref{mma finite ao}, we establish the high-probability optimality of the weight estimator $\hat{\bw}$ selected by Trans-MAI allowing variance misspecification, demonstrating that, with high probability, $\hat{\bw}$ achieves near-optimal performance in minimizing the objective risk.

We end this correctly specified seceario with discussions on constructing candidate domains, 
which implies that the estimated contrasts $\hat{\bm{\delta}}^{(m)}$'s are well separated between $m \in \mathcal{\tA}_{h}$ and $m \in \mathcal{\tA}_{h}^c$ with high probability. 
Define $\un = \min_{m \in [M]} n_{m}$ and $\ueta = \min_{m \in \mathcal{\tA}_h^c} \|\bm{\delta}^{(m)}\|$ and the following condition is a source version of Condition~\ref{C.4}. 

\begin{condition} \label{C.4new}
    For any $m \in [M]$, there exist positive constants $\tau_3$ and $\tau_4$ such that $\tau_3 \leq \lambda_{\min}(\bm{\Sigma}^{(m)}) \leq \lambda_{\max}(\bm{\Sigma}^{(m)}) \leq \tau_4$, and $\lim_{n_m \to \infty} \max_{i = 1, \cdots, n_m} \|\bX_{i \cdot}^{(m)}\|/{n_m} = 0$. 
\end{condition}

\begin{proposition} \label{proposition 1}
    Under Conditions~\ref{C.4}(iii) and \ref{C.4new}, if $\ueta > 2h$, and $p^{1/2} \un^{-1/2} \ueta^{-1} \to 0$ as $\un \to \infty$, then we have
    \begin{align*} 
        \Pr \left( \max_{m \in \mathcal{\tA}_h} \|\hat{\bm{\delta}}^{(m)}\| < \min_{m \in \mathcal{\tA}_h^c} \| \hat{\bm{\delta}}^{(m)} \| \right) \to 1, 
    \end{align*}
    as $\un \to \infty$.
\end{proposition}



\subsubsection{Theoretical analysis for Trans-MAI under misspecified target model} 
Under misspecified target model, we consider the asymptotic behavior of Trans-MAI, 
establishing the asymptotic optimality, in the sense of minimizing the objective loss and risk function. 
Different from the above analysis, the data generating process is simplified to 
\begin{equation*}
    y_{i}^{(m)} = \mu_{i}^{(m)} + \varepsilon_{i}^{(m)}, \quad i = 1, \ldots, n_m, \quad m \in [M],
\end{equation*}
that is, we do not need $\mu_{i}^{(m)}$ is a linear function of the covariates.
Let $\xi_{n_0;v} = \inf_{w \in \mathcal{W}} R_v(\bm{w})$ for any $v \in [0,1]$ and the following conditions are introduced to prove asymptotic optimality.

\begin{condition}\label{mu_star}
    $\displaystyle\sup_{m \in [M]} n_0^{-1}\|\bm{\mu}^{[m]}\|^{2} = O(1)$.
\end{condition}

\begin{condition}\label{C.5}
    For $v \in [0,1]$, $\xi_{n_0;v}^{-1} p^{1/2} n_0^{1/2} \to 0$. 
\end{condition}

Condition~\ref{mu_star} ensures that $n_0^{-1}\|\bm{\mu}^{[m]}\|^{2}$ are finite for all candidate domains, which is a weaker version of Condition (C.4) in \citet{zhu2022frequentist}. 
Condition~\ref{C.5} imposes restrictions on the rates at which the infimum risk $\xi_{n_0;v}$ and parameter dimension $p$ with respect to $n_0$, and requires that the target model is sufficiently misspecified. 
Similar assumptions can be found in standard model averaging literature, for instance, \citet[Eq.~(7)]{ando2014model}, \citet[Assumption 1(e)]{liu2020model}, \citet[Assumption 5]{zhang2024prediction}, and \citet[Condition 8]{hu2023optimal}. 
A detail explanation of target model misspecification under Condition~\ref{C.5} is provided in Supplement~A.3.7.

\begin{theorem}[Asymptotic optimality of Trans-MAI] \label{Asymptotic optimality}
    For $v \in [0,1]$, under Conditions~\ref{C.4}(i),~\ref{C.1},~\ref{mu_star} and~\ref{C.5}, if $ \xi_{n_0;v}^{-1}p \phi \to 0$, then
    \begin{align} \label{Asymptotic optimality1}
        \frac{L_v(\hat{\bm{w}})}{\inf_{\bm{w} \in \mathcal{W}} L_v(\bm{w})}  
        = 1 + O_{p}( \xi_{n_0;v}^{-1} \phi p + \xi_{n_0;v}^{-1} n_0^{1/2} p^{1/2}) = 1 + o_p(1). 
    \end{align}
    If, in addition, $\bigl( L_v(\hat{\bm{w}}) - \xi_{n_0;v}\bigr) \xi_{n_0;v}^{-1}$ is uniformly integrable, then
    \begin{align} \label{Asymptotic optimality2}
        \frac{\E\{L_v(\hat{\bm{w}})\}}{\inf_{\bm{w} \in \mathcal{W}} R_v(\bm{w})} \to 1. 
    \end{align}
\end{theorem}

Theorem~\ref{Asymptotic optimality} demonstrates that the transfer model averaging procedure is asymptotically optimal, implying that its objective loss and risk functions converge asymptotically to those of the infeasible but best possible transfer model averaging predictor. 
It is worth noting that our asymptotic optimality always holds regardless of whether the candidate models except target are correctly specified, which is reasonable because the target model is our main concern and naturally dominates the performance.
If $v = 0$, Theorem~\ref{Asymptotic optimality} degenerates to the standard asymptotic optimality in the sense of minimizing the squared prediction loss and corresponding prediction risk. 
Furthermore, we note that the proof of Eq.~(\ref{Asymptotic optimality1}) yields the following intermediate result
\begin{align} 
    \label{Asymptotic optimality3}
    \frac{R_v(\hat{\bm{w}})}{\inf_{w \in \mathcal{W}} R_v(\bm{w})} \xrightarrow{p} 1,
\end{align}
which has been extensively studied in \cite{hansen2012jackknife}, \cite{zhang2024prediction} and \cite{hu2023optimal}. 
However, Eq.~(\ref{Asymptotic optimality2}) is different from Eq.~(\ref{Asymptotic optimality3}), 
since $\hat{\bw}$ in Eq.~(\ref{Asymptotic optimality3}) is directly plugged in $R_v(\bw)$, 
while Eq.~(\ref{Asymptotic optimality2}) considers the randomness of $\hat{\bw}$, also discussed in \cite{zhang2020parsimonious}.

\subsection{Theoretical analysis for Trans-MACs and Trans-MAC under combinatorial-similarity}\label{subsec:theory-combinatorial}
We now turn to the theoretical properties of our proposed methods under combinatorial-similarity, Trans-MAC and Trans-MACs. 
For this purpose, we first introduce some notations and technical conditions. 
Define the matrix $\bV = (v_{m,m'}) \in \mathbb{R}^{(|\mathcal{A}_{h}^{c}|) \times (|\mathcal{A}_{h}^{c}|)}$ with entries
$$
v_{m,m'} =\|\bdelta^{[m]} \|^{-1}\|\bdelta^{[m']} \|^{-1} \bdelta^{[m]\T} \bX^{[m_s]\T} \bX^{[m_s]} \bdelta^{[m']}, \quad \forall m, m' \in \mathcal{A}_h^c.
$$
and its submatrix $\bV_{b} = (v_{m,m'}) \in \mathbb{R}^{(|\mathcal{A}^{c}|-1) \times (|\mathcal{A}^{c}|-1)}$ of $\bV$ with indices restricted to $m, m' \in \mathcal{A}_h^c \setminus \{b\}$.
Let $\bar{\sigma}^2 = \max_{i \in \mathcal{I}_{m_s}} \sigma_{(m)}^2$.

\subsubsection{Theoretical analysis for Trans-MACs}
In this section, we investigate the theoretical properties for Trans-MACs, including weight convergence, convenience rate for estimation/prediction, and high-probability optimality.
We introduce the following technical conditions. 

\begin{condition} \label{C.1'}
    $\max_{m \in \mathcal{I}_{m_s}} \hat{\sigma}_{(m)}^{2} / \sigma_{(m)}^{2} = O_p(1)$.
\end{condition}

\begin{condition} \label{evi}
There exist a vector $\brho \in \mathcal{S}_h$ and an index $b \in \{m \in \mathcal{A}_h^c | \rho_m > 0 \}$ such that 
 $N_{m_s}^{-1} \lambda_{\min} \left(\bm{V}_{b} \right) \geq \tau'$,
where $\tau'$ is a positive constant.
\end{condition}

Condition~\ref{C.1'} is a generalized version of Condition~\ref{C.4}, assuming the asymptotic behavior of the estimated variance $\hat{\sigma}_{(m)}^{2}$. 
In Condition~\ref{evi}, existence of $\brho$ ensures $\mathcal{S}_h \neq \varnothing$, thus implying that the strict combinatorial-similarity satisfies.
If $h = 0$, $\brho \in \mathcal{S}_h$ means that $\sum_{m \in \mathcal{A}_h^c} \rho_m \bbeta^{[m]} = \bbeta^{(0)}$, suggesting that $\bbeta^{[m]}$ for $m \in \mathcal{A}_h^c$ are linearly dependent and consequently $\bV$ is singular. 
When $\bbeta^{[b]}$ with index $b$ satisfying $\rho_b > 0$ is removed, 
Condition~\ref{evi} demonstrates that $\bbeta^{(0)}$ is not in the simplex of the remaining $\bbeta^{[m]}$, otherwise $\bV_b$ matrix is singular, violating Condition~\ref{evi}.
Therefore, Condition~\ref{evi} can be interpreted as a type of minimal linear dependence under the simplex constraint.

\begin{lemma}[Weight convergence for Trans-MACs] \label{Weight convergence under informative combination new}
    Under Conditions~\ref{C.4}, \ref{finitebeta}, \ref{C.1}, \ref{C.1'} and \ref{evi}, 
    if $hp^{1/2}(|\mathcal{A}_h^c| -1)^{1/2} \to 0$ and $N_{m_s}^{-1/2}p \eta^{-2} (|\mathcal{A}_h^c| - 1)^{1/2} \rightarrow 0$ hold, then
    \begin{align*} 
        d(\hat{\brho}, \mathcal{S}_h)
        = O_p\bigl( N_{m_s}^{-1/2} p \eta^{-2} (|\mathcal{A}_h^{c}| - 1)^{1/2} + h p^{1/2} \eta^{-2} (|\mathcal{A}_h^{c}| - 1)^{1/2} \bigr).
    \end{align*}
\end{lemma}

Lemma~\ref{Weight convergence under informative combination new} 
establishes the weight convergence for Trans-MACs  under combinatorial-similarity, providing a generalization of Lemma~\ref{Convergence of weight new}(iv) for $v = 0$. 
In particular, if $m_s = 0$, the error bound for $\hat{\brho}$ derived in Lemma~\ref{Weight convergence under informative combination new} exactly coincides with the corresponding bound for $\hat{\bw}_{\mathcal{A}_h^c}$ in Lemma~\ref{Convergence of weight new}(iv) under the setting $\phi = 2$. 
The technical condition $hp^{1/2}(|\mathcal{A}_h^c| -1)^{1/2} = o(1)$, which necessarily implies $h\eta^{-1} = o(1)$, ensures that the contrast term $\bdelta_{\brho}$ becomes asymptotically negligible when compared to the individual contrasts $\bdelta^{[m]}$ for $m \in \mathcal{A}_h^c$. 
Moreover, in the regime where $h = O\bigl(p^{1/2}N_{m_s}^{-1/2}\bigr)$, the convergence rate $d(\hat{\brho}, \mathcal{S}_h)$ is primarily determined by the dominated term $N_{m_s}^{-1/2} p \eta^{-2} (|\mathcal{A}_h^{c}| - 1)^{1/2}$.
Based on the weight convergence established in Lemma~\ref{Weight convergence under informative combination new}, we further discuss convergence rate of Trans-MACs. 

\begin{theorem}[Convergence rate of Trans-MACs] \label{stma2convergence} 
    Under the same conditions of Lemma~\ref{Weight convergence under informative combination new}, if $h p \eta^{-2} |\mathcal{A}_h^{c}| \to 0$ and $ N_{m_s}^{-1/2} p^{3/2} \eta^{-2} |\mathcal{A}_h^{c}| \to 0$ hold, then
    \begin{align*}
    	& \quad \|\hat{\bm{\beta}}_\text{Trans-MACs}(\hat{\brho}) - \bm{\beta}^{(0)}\| 
        \vee 
        N_{m_s}^{-1} \|\bm{X}^{[m_s]}(\hat{\bm{\beta}}_\text{Trans-MACs}(\hat{\brho}) - \bm{\beta}^{(0)})\| \\
        & = O_p \bigl( N_{m_s}^{-1/2} p^{3/2} \eta^{-2} |\mathcal{A}_h^{c}| + h p \eta^{-2} |\mathcal{A}_h^{c}| \bigr).
    \end{align*}
\end{theorem}

In Theorem~\ref{stma2convergence}, the estimation error as meassured by $\|\hat{\bm{\beta}}_\text{Trans-MACs}(\hat{\brho}) - \bm{\beta}^{(0)}\|$ can be decomposed into two interpretable terms that correspond to the weight convergence components identified in Lemma~\ref{Weight convergence under informative combination new}. 
The first term $N_{m_s}^{-1/2} p^{3/2} \eta^{-2} |\mathcal{A}_h^{c}|$ is from the estimation error of $\hat{\bbeta}^{[m]}$ and the second term $h p \eta^{-2} |\mathcal{A}_h^{c}|$ is from the strict combinatorial-similarity threshold $h$.

\begin{corollary}[Comparisons of Trans-Lasso and Trans-MACs] \label{compare with Translasso Trans-MACs}
    Under the same conditions as Theorem~\ref{stma2convergence},
    if $h = o(n_0^{-1/2} (\log{M})^{1/2} |\mathcal{A}_h^c|^{-1})$, $C_{\bSigma} = O(1)$, $n_0^{1/2} (n_{\Tilde{\mathcal{A}}^{0}} + n_0)^{-1/2} = O(s^{-1/2} (\log{p})^{-1/2} (\log{M})^{1/2})$ and $n_0^{1/2} N_{m_s}^{-1/2} = o(p^{-1/2} (\log{M})^{1/2}|\mathcal{A}_h^c|^{-1})$ hold, then we have 
    $\|\hat{\bm{\beta}}_\text{Trans-MACs}(\hat{\brho}) - \bm{\beta}^{(0)}\| = o_p(\|\hat{\bm{\beta}}_{\text{Trans-Lasso}} - \bm{\beta}^{(0)}\|)$. 
\end{corollary}

Corollary~\ref{compare with Translasso Trans-MACs} compares the estimation error bound for Trans-MACs with Trans-Lasso. 
It is reasonable that under some conditions $\hat{\bm{\beta}}_\text{Trans-MACs}(\hat{\brho})$ serves a faster convergence rate than $\hat{\bm{\beta}}_{\text{Trans-Lasso}}$, 
because Trans-MACs select the weight upon a informative domain instead of only target inspired by sufficiency and leverages more knowledge from strict combinatorial-similarity than Trans-Lasso.

\begin{condition}\label{mu_star_new_2}
    $\displaystyle\sup_{m \in \mathcal{A}_{h}^{c} } N_{m}^{-1}\|\bm{\mu}^{[m]}\|^{2} = O(1)$.
\end{condition}

\begin{condition}\label{sub-Gaussian}
    $\varepsilon_{i}^{(m)}$ are sub-Gaussians and there exist a positive constant $C_{\psi_2} > 0$ such that $\|\varepsilon_{i}^{(m)}\|_{\psi_2} \leq C_{\psi_2}$ for $i = 1,\cdots, n_m, m \in \mathcal{I}_{m_s}$.
\end{condition}
Condition~\ref{mu_star_new_2} ensures that the squared error $\|\bm{\mu}^{[m]}\|^2$ is uniformly bounded for $m \in \mathcal{A}_h^c$ and similar condition is established in \citet[Proposition 7.2]{bellec2018optimal}.
In Condition~\ref{sub-Gaussian}, we assume that the error terms $\varepsilon_i^{(m)}$ for $m \in \mathcal{I}_{m_s}$ are sub-Gaussian random variables. 
This assumption is widely used in the literature, as demonstrated in previous works such as \citet{li2022transfer}, \citet{zhu2018sparse} and \citet{gold2020inference}, among others.

\begin{theorem}[High-probability optimality of Trans-MACs] \label{mma ao transmacs}
    Assuming that 
    \begin{equation}
        \epsilon_{\sigma} = \Pr(\max_{m \in \mathcal{I}_{m_s}} |\hat{\sigma}_{(m)}^2 - \sigma_{(m)}^2| > 2^{-1} \tau_1 \tau_2 ^{-1} \bar{\sigma}^2 ) < 1, \label{sigma condition}
    \end{equation}
    under Conditions~\ref{C.4}(i),~\ref{mu_star_new_2} and~\ref{sub-Gaussian}, 
    we have for any $\epsilon \in (0,1)$, 
    \begin{align*}
        \frac{1}{N_{m_s}} R'(\hat{\brho}) \leq  \inf_{\bvarphi \in \mathcal{W}^{|\mathcal{A}_h^c| + 1}} \frac{1}{N_{m_s}} R'(\brho) + u_2,
    \end{align*}
    with probability at least $1-\epsilon_{\sigma} - \epsilon$, 
    where $u_2
      = c |\mathcal{A}_h^c| \max \bigl\{p^{1/2} N_{m_s}^{-1/2}[\log{\left(2|\mathcal{A}_h^c|\epsilon^{-1} \right)}]^{1/2},  \allowbreak N_{m_s}^{-1} \log{\left(2|\mathcal{A}_h^c|\epsilon^{-1}\right)} \bigr\} + \bar{\sigma}^2 |\mathcal{A}_h^c|N_{m_s}^{-1}$ and $c$ is a positive constant. 
\end{theorem}

The condition in Eq.~(\ref{sigma condition}) requires that the estimators $\{\hat{\sigma}_{(m)}^2 \}_{m \in \mathcal{I}_{m_s}}$ satisfy with high probability $\max_{m \in \mathcal{I}_{m_s}} |\hat{\sigma}_{(m)}^2 - \sigma_{(m)}^2| > 2^{-1} \tau_1 \tau_2 ^{-1} \bar{\sigma}^2$, which is weaker than consistency and similar condition can be found in \citet[Theorem 6.2]{bellec2018optimal}.
In Theorem~\ref{mma ao transmacs}, we establish the high-probability optimality of the weight estimator $\hat{\brho}$ selected by Trans-MACs allowing variance misspecification, demonstrating that, with high probability, $\hat{\brho}$ achieves near-optimal performance in minimizing the mean squared prediction risk. 

Next, under misspecified informative domain $\mathcal{M}_{m_s}$, we establish the asymptotic optimality of Trans-MACs, in the sense of minimizing the mean squared prediction loss. 

\begin{condition}\label{mu_star'}
    $\displaystyle\sup_{m \in \mathcal{A}_h^c } N_{m_s}^{-1}\|\bm{\mu}^{[m]}\|^{2} = O(1)$.
\end{condition}

\begin{condition}\label{C.5'}
    $\xi'^{-1} p^{1/2} N_{m_s}^{1/2} \to 0$. 
\end{condition}

\begin{theorem}[Asymptotic optimality of Trans-MACs] \label{Asymptotic optimality TransMAcs}
    Under Conditions~\ref{C.4}(i),~\ref{C.1'},~\ref{mu_star'} and~\ref{C.5'}, we have
    \begin{align*} 
        \frac{L'(\hat{\brho})}{\inf_{\brho \in \mathcal{W}^{|\mathcal{A}_h^c|}} L'(\brho)}  
        = 1 + O_{p}(\xi'^{-1} p^{1/2} N_{m_s}^{1/2}) = 1 + o_p(1). 
    \end{align*}
    If, in addition, $\bigl( L'(\hat{\brho}) - \xi'\bigr) \xi'^{-1}$ is uniformly integrable, we have
    \begin{align*} 
        \frac{\E\{L'(\hat{\brho})\}}{\inf_{\brho \in  \mathcal{W}^{|\mathcal{A}_h^c|} } R'(\brho)} \to 1. 
    \end{align*}
\end{theorem}

\subsubsection{Theoretical analysis for Trans-MAC}
In this section, we discuss the theoretical properties of Trans-MAC under combinatorial-similarity, 
including weight convergence, convenience rate for estimation/prediction, and high-probability optimality and introduce some additional notations and conditions.

For any $\bvarphi \in \mathcal{W}^{|\mathcal{A}_h^c| + 1}$, let $\hat{\bm{\mu}}^{[m_s]}(\bm{\varphi}) = \sum_{m \in \mathcal{A}_{h}^{c} \cup \{m_s\}} \rho_m \bm{X}^{[m_s]} \hat{\bm{\beta}}^{[m]}$
and we define the squared prediction loss and risk function $L''(\bm{\varphi}) = \|\hat{\bm{\mu}}^{[m_s]}(\bm{\varphi}) - \bm{\mu}^{[m_s]}\|^2$ and
$R''(\bm{\varphi})= \E\left[ L_{N_{m_s}}(\bm{\varphi}) \mid \bm{X}^{[M]} \right]$ respectively.
Furthermore, let $\xi'' = \inf_{\bvarphi \in \mathcal{W}^{|\mathcal{A}_h^c| + 1}} R''(\bvarphi)$.

\begin{condition} \label{evi new}
    There exists a positive constant $\tau''> 0$ such that $N_{m_s}^{-1} \lambda_{\min} \left(\bm{V} \right) \geq \tau''$.
\end{condition}

Condition~\ref{evi new} requires that $\bV$ is non-singular and positive definite, 
implying that the vectors $\bdelta^{[m]}$ for $m \in \mathcal{A}_h^c$ are linearly independent, 
thus $\{\beta^{[m]}\}_{m \in \mathcal{A}_h^c}$ cannot linearly combine to form $\bbeta^{(0)}$.
If Condition~\ref{evi new} is satisfied, it is obvious that $\mathcal{S}_h = \varnothing$.

\begin{lemma}[Weight convergence for Trans-MAC] \label{Weight convergence under informative combination new 2}
    Under Conditions~\ref{C.4}, \ref{finitebeta}, \ref{C.1}, \ref{C.1'} and \ref{evi}, 
    if $hp^{1/2}(|\mathcal{A}_h^c| -1)^{1/2} \to 0$ and $N_{m_s}^{-1/2}p \eta^{-2} (|\mathcal{A}_h^c| - 1)^{1/2} \rightarrow 0$ hold, then
    \begin{align*} 
        d(\hat{\bvarphi}, \mathcal{S}_h^{'})
        = O_p\bigl( N_{m_s}^{-1/2} p \eta^{-2} (|\mathcal{A}_h^{c}| - 1)^{1/2} + h p^{1/2} \eta^{-2} (|\mathcal{A}_h^{c}| - 1)^{1/2} \bigr).
    \end{align*}
    Under Conditions~\ref{C.4}, \ref{finitebeta}, \ref{C.1}, \ref{C.1'} and \ref{evi new}, if $hp^{1/2}|\mathcal{A}_h^c|^{1/2} \to 0$ and $N_{m_s}^{-1/2}p \eta^{-2} |\mathcal{A}_h^c|^{1/2} \rightarrow 0$ hold, then
    \begin{align*} 
        d(\hat{\bvarphi}, \mathcal{S}_h^{'})
        = O_p\bigl( N_{m_s}^{-1/2} p \eta^{-2} |\mathcal{A}_h^{c}|^{1/2} + h p^{1/2} \eta^{-2} |\mathcal{A}_h^{c}|^{1/2} \bigr).
    \end{align*}
\end{lemma}

Comparing Lemma~\ref{Weight convergence under informative combination new} with~\ref{Weight convergence under informative combination new 2}, 
we observe that Trans-MAC maintains well-behaved weight convergence even when Condition~\ref{evi new} holds ($\mathcal{S}_h = \varnothing$), 
demonstrating its superior feasibility over Trans-MACs.
Based on the weight convergence in Lemma~\ref{Weight convergence under informative combination new 2}, we discuss the convergence rate of Trans-MAC and further compare it with Trans-Lasso.

\begin{theorem}[Convergence rate of Trans-MAC] \label{stma3convergence} 
    Under the same conditions as Lemma~\ref{Weight convergence under informative combination new 2}, 
    if $h p \eta^{-2} |\mathcal{A}_h^{c}| \to 0$ and $ N_{m_s}^{-1/2} p^{3/2} \eta^{-2} |\mathcal{A}_h^{c}| \to 0$ hold, then
    \begin{align*}
    	& \quad \|\hat{\bm{\beta}}_\text{Trans-MAC}(\hat{\bvarphi}) - \bm{\beta}^{(0)}\| 
        \vee 
        N_{m_s}^{-1} \|\bm{X}^{[m_s]}(\hat{\bm{\beta}}_\text{Trans-MAC}(\hat{\bvarphi}) - \bm{\beta}^{(0)})\| \\
        & = O_p \bigl( N_{m_s}^{-1/2} p^{3/2} \eta^{-2} |\mathcal{A}_h^{c}| + h p \eta^{-2} |\mathcal{A}_h^{c}| \bigr).
    \end{align*}
\end{theorem}

\begin{corollary}[Comparisons of Trans-Lasso and Trans-MAC] \label{compare with Translasso Trans-MAC}
    Under the same conditions as Theorem~\ref{stma3convergence},
    if $h = o(n_0^{-1/2} (\log{M})^{1/2} |\mathcal{A}_h^c|^{-1})$, $C_{\bSigma} = O(1)$, $n_0^{1/2} (n_{\Tilde{\mathcal{A}}^{0}} + n_0)^{-1/2} = O(s^{-1/2} (\log{p})^{-1/2} (\log{M})^{1/2})$ and $n_0^{1/2} N_{m_s}^{-1/2} = o(p^{-1/2} (\log{M})^{1/2}|\mathcal{A}_h^c|^{-1})$ hold, then we have 
    $\|\hat{\bm{\beta}}_\text{Trans-MAC}(\hat{\bvarphi}) - \bm{\beta}^{(0)}\| = o_p(\|\hat{\bm{\beta}}_{\text{Trans-Lasso}} - \bm{\beta}^{(0)}\|)$. 
\end{corollary}
In Theorem~\ref{stma3convergence}, we establish the estimation/prediction error bound for Trans-MAC.
Moreover, Corollary~\ref{compare with Translasso Trans-MAC} compares the estimation error bound for Trans-MAC with Trans-Lasso,
demonstrating that under some conditions $\hat{\bm{\beta}}_\text{Trans-MAC}(\hat{\brho})$ serves a faster convergence rate than $\hat{\bm{\beta}}_{\text{Trans-Lasso}}$, 
since Trans-MAC select the weight upon a informative domain instead of only target inspired by sufficiency and leverages more knowledge from combinatorial-similarity than Trans-Lasso.
Next, we focus on the high-probability optimality of Trans-MAC without misspecification assumption, and introduce the following condition.


\begin{condition}\label{mu_star_new}
    $\displaystyle\sup_{m \in \mathcal{A}_{h}^{c} \cup \{m_s\}} N_{m}^{-1}\|\bm{\mu}^{[m]}\|^{2} = O(1)$.
\end{condition}

\begin{theorem}[High-probability optimality of Trans-MAC] \label{mma ao}
    Assuming that 
    \begin{equation*}
        \epsilon_{\sigma} = \Pr(\max_{m \in \mathcal{I}_{m_s}} |\hat{\sigma}_{(m)}^2 - \sigma_{(m)}^2| > 2^{-1} \tau_1 \tau_2 ^{-1} \bar{\sigma}^2 ) < 1, 
    \end{equation*}
    under Conditions~\ref{C.4}(i),~\ref{sub-Gaussian} and~\ref{mu_star_new}, 
    we have for any $\epsilon \in (0,1)$, 
    \begin{align}\label{mma ao1}
        \frac{1}{N_{m_s}} R''(\hat{\bvarphi}) \leq  \inf_{\bvarphi \in \mathcal{W}^{|\mathcal{A}_h^c| + 1}} \frac{1}{N_{m_s}} R''(\bvarphi) + u_3,
    \end{align}
    with probability at least $1-\epsilon_{\sigma} - \epsilon$, 
    where $u_3
      = c (|\mathcal{A}_h^c|+1) \max \bigl\{N_{m_s}^{-1}\log{\left(2(\mathcal{A}_h^c|+1)\epsilon^{-1}\right)},\allowbreak  p^{1/2} N_{m_s}^{-1/2}[\log{\left(2(|\mathcal{A}_h^c|+1)\epsilon^{-1} \right)}]^{1/2} \bigr\} + \bar{\sigma}^2 (|\mathcal{A}_h^c| + 1) N_{m_s}^{-1}$ and $c$ is a positive constant. 
\end{theorem}

Compared with Condition~\ref{mu_star_new_2}, Condition~\ref{mu_star_new} further requires that the squared error $\|\bm{\mu}^{[m]}\|^2$ is uniformly bounded for $m \in \mathcal{A}_h^c \cup \{m_s\}$.
In Theorem~\ref{mma ao}, we establish the high-probability optimality of the weight estimator $\hat{\bvarphi}$ selected by Trans-MAC allowing variance misspecification, demonstrating that, with high probability, $\hat{\bvarphi}$ achieves near-optimal performance in minimizing the mean squared prediction risk. 
Theorem~\ref{mma ao} implies that the performance of $\hat{\bm{\beta}}_{Trans-MAC}(\hat{\bvarphi})$ is only determined by the best convex linear combination, indicating that $\hat{\bm{\beta}}_{Trans-MAC}(\hat{\bvarphi})$ is only slightly worse than the best but infeasible convex linear combination. 
Formally, 'slightly worse' refers to the error term $u_3$, which can be seen as the cost of searching for the best convex linear combination. 
This term $u_3$ sometimes becomes negligible, for instance, when $|\mathcal{A}_h^c| = O(1), p^{1/2}N_{m_s}^{-1/2} = o(1)$, illustrating the robustness of Trans-MAC against adversarial auxiliary samples. 

Additionally, it is important to note that although Eq.~(\ref{mma ao1}) and the oracle inequality (e.g., \citet{gaiffas2011hyper}, \citet{rigollet2012kullback}, \citet{dai2012deviation}) are similar in form, they differ in their treatment of uncertainty. 
Specifically, Theorem~\ref{mma ao} accounts for the uncertainty of each candidate estimator $\hat{\bm{\beta}}^{(m)}$, while in oracle inequalities, the candidate predictions are assumed to be fixed, as seen in Theorem 4.1 of \citet{dai2012deviation}. 
This distinction allows Trans-MAC to avoid the need for a data-splitting step, as used in \citet{li2022transfer}, and to construct candidate estimators for $\hat{\bbeta}^{(0)}$ using more auxiliary samples.

Next, under misspecified informative domain $\mathcal{M}_{m_s}$, we discuss the asymptotic optimality of Trans-MAC, in the sense of minimizing the mean squared prediction loss. 

\begin{condition}\label{mu_star''}
    $\displaystyle\sup_{m \in \mathcal{A}_h^c \cup \{m_s\}} N_{m_s}^{-1}\|\bm{\mu}^{[m]}\|^{2} = O(1)$.
\end{condition}

\begin{condition}\label{C.5''}
    $\xi''^{-1} p^{1/2} N_{m_s}^{1/2} \to 0$. 
\end{condition}

\begin{theorem}[Asymptotic optimality of Trans-MAC] \label{Asymptotic optimality TransMAc}
    Under Conditions~\ref{C.4}(i),~\ref{C.1'},~\ref{mu_star''} and~\ref{C.5''}, we have
    \begin{align*} 
        \frac{L''(\bvarphi)}{\inf_{\bvarphi \in \mathcal{W}^{|\mathcal{A}_h^c| + 1}} L''(\bvarphi)}= 1 + O_p( \xi''^{-1}  N_{m_s}^{1/2}p^{1/2} ) = 1 + o_p(1).
    \end{align*}
    If, in addition, $\bigl(L''(\bm{\varphi}) - \xi''(\bm{\varphi})\bigr) \xi''^{-1}(\bm{\varphi})$ is uniformly integrable, we have
    \begin{align*} 
        \frac{\E\{L''(\bm{\varphi})\}}{\inf_{\bvarphi \in \mathcal{W}^{|\mathcal{A}_h^c|+1}} R''(\bvarphi
        )} \to 1. 
    \end{align*}
\end{theorem}


%% file: body/sim.tex
\section{Simulations}\label{sim}

In this section, we perform comprehensive simulation studies to evaluate the effectiveness of our proposed methodologies. 
First, we systematically compare the performance of our three proposed methods (Trans-MAI, Trans-MAC, and Trans-MACs) against several alternative approaches, examining their behaviors under varying experimental settings, including different cardinalities of the informative set and diverse sample sizes. 
Additionally, we conduct experimental validation of Trans-MAI's theoretical properties, specifically investigating how non-zero values of $v$ influence the weight convergence rate for non-informative domains and verifying its asymptotic normality.

\subsection{Method comparisons}
We evaluate the empirical performance of our proposed methods alongside several comparable methods through various numerical experiments. 
Specifically, we assess the following methods: 
\begin{itemize}
    \item[(a)] The “OLS-Tar” method represents OLS only using target data.
    \item[(b)] The “OLS-Pool” method represents OLS using pooled data, including all sources and target.
    \item[(c)] The “MAP” method represents the model-averaging based transfer learning method proposed by \cite{zhang2024prediction}.
    \item[(d)] The “Trans-Lasso” method represents the high-dimensional linear regression transfer learning method proposed by \cite{li2022transfer}.
    \item[(e)] The “PTL” method represents the high-dimensional linear regression transfer learning method proposed by \cite{lin2024profiled}.
    \item[(f)] The “Trans-MAI” method represents our proposed sufficient-principled transfer learning method via model averaging under individual-similarity with \(v = 0.5\) and $\phi = \log{n_0}$.
    \item[(g)] The “Trans-MACs” method represents our proposed sufficient-principled transfer learning method via model averaging under strict combinatorial-similarity with \(v = 0.5\), $\phi = \log{n_0}$ and $m_s = \arg\max_{m \in [M]}\hat{w}_m$ selected via Trans-MAI.
    \item[(h)] The “Trans-MAC” method represents our proposed sufficient-principled transfer learning method via model averaging under combinatorial-similarity with \(v = 0.5\) and $\phi = \log{n_0}$ and $m_s = \arg\max_{m \in [M]}\hat{w}_m$ selected via Trans-MAI.
\end{itemize}

Throughout the study, we fix $M = 10$ and $p = 20$. 
The source data are \iid observations generated from the model $y_i^{(m)} = \bx_i^{(m)\T}\bbeta^{(m)} + \varepsilon_{i}^{(m)}$, where $\bx_{i}^{(m)} \sim \mathcal{N}(\bm{0},\bOmega_{(m)})$ and $\varepsilon_{i}^{(m)} \sim \mathcal{N}(0,\sigma_{(m)}^{2})$ for $1 \leq m \leq M$. 
The target data are \iid generated from the model $y_i^{(0)} = \bx_i^{(0)\T}\bbeta^{(0)} + \varepsilon_{i}^{(0)}$ with $\bx_i^{(0)} \sim \mathcal{N}(\bm{0}, \bOmega_{(0)})$ and $\varepsilon_{i}^{(0)} \sim \mathcal{N}(0, \sigma_{(0)}^{2})$.
The parameters $\bbeta^{(m)}$, $\bbeta^{(0)}$, $\sigma_{(m)}^{2}$, $\sigma_{(0)}^{2}$, $\bOmega_{(m)}$ and $\bOmega_{(0)}$ will be specified later for different scenarios. 

The performance of the aforementioned methods is evaluated according to the prediction accuracy and estimation accuracy. 
The prediction accuracy is quantified using the empirical mean squared prediction error (MSPE) computed on $n_{test} = 10000$ test samples \iid generated from the target data distribution, \ie $n_{test}^{-1} \sum_{i=1}^{n_{test}} \{\hat{\bmu}(\bx_i) - \bmu(\bx_i)\}^{2}$. 
The estimation accuracy is measured by mean squared error (MSE), \ie $\|\hat{\bbeta}^{(0)} - \bbeta^{(0)}\|^2$, where $\hat{\bbeta}^{(0)}$ is an estimate of $\bbeta^{(0)}$.

\subsubsection{Experiment 1: homogenous design in the absence of combinatorial-similarity}\label{section exp 1.1}

In this simulation, we primarily investigate the role of informative sources in transfer learning. 
This experiment setup is revised from \cite{li2022transfer} and \cite{lin2024profiled}, which assume the existence of informative domains. 
Specifically, we consider $|\mathcal{A}| = \{0,1,2,3,4,5,6,7,8\}$ informative domains. 
Following \cite{tripuraneni2021provable}, we set $\bOmega_{m} = \bOmega_{0} = \bm{I}_{p}$ for all domains, and fix $\sigma_{(m)} = \sigma_{(0)} = 1.0$ for $m = 1,\cdots,M$. 
The sample sizes are fixed at $n_0 = 100$ for the target and $n_{m} = 200$ for each source. 

The parameter vector $\bbeta$ is generated through the following procedure.
First, for $0 \leq m \leq |\mathcal{A}|$, $\bbeta^{(m)} = \bbeta^{(0)}$ and the coefficients of $\bbeta^{(0)}$ are specified by \iid drawn from $\mathcal{N}(2,4)$. 
Next, for the regression coefficients in auxiliary samples, we consider two distinct configurations: (i) sparse $\boldsymbol{\delta}^{(m)}$ and (ii) non-sparse $\boldsymbol{\delta}^{(m)}$.

(i) For $1 \leq m \leq |\mathcal{A}|$, let $|\mathcal{V}^{(m)} |= v_{\delta}$ randomly sampled from $\{1,\cdots,p\}$ without replacement, where $v_{\delta} = 3$. 
For  each $j \in \mathcal{V}^{(m)}$, we generate $\delta_j^{(m)}$ independently from $\mathcal{N}(0,h^2)$, where $h = \{0, 0.04,0.08,0.12\}$ and for each $j \notin \mathcal{V}^{(m)}$, set $\delta_j^{(m)} = 0$. 
Then, the source parameter is assembled by $\bbeta^{(m)} = \bbeta^{(0)} + \bdelta^{(m)}$ for $1 \leq m \leq |\mathcal{A}|$.

(ii) For $1 \leq m \leq |\mathcal{A}|$,  $\delta_j^{(m)} \sim_{i.i.d.} \mathcal{N}(0,1)$ where $j \in \{1,\cdots,p\}$, and then $\bdelta^{(m)} = h\frac{\bdelta^{(m)}}{\|\bdelta^{(m)}\|}$ where $h = \{0, 0.04,0.08,0.12\}$. 
The source parameter is then assembled by $\bbeta^{(m)} = \bbeta^{(0)} + \bdelta^{(m)}$ for $1 \leq m \leq |\mathcal{A}|$.

In the meantime, for $ |\mathcal{A}| + 1 \leq m \leq M$, the coefficients of $\bbeta^{(m)} \neq \bbeta^{(0)}$ are specified by drawing \iid random numbers from $\mathcal{N}(-1,4)$.

\begin{figure}
    \centering
    \includegraphics[width=0.9\linewidth]{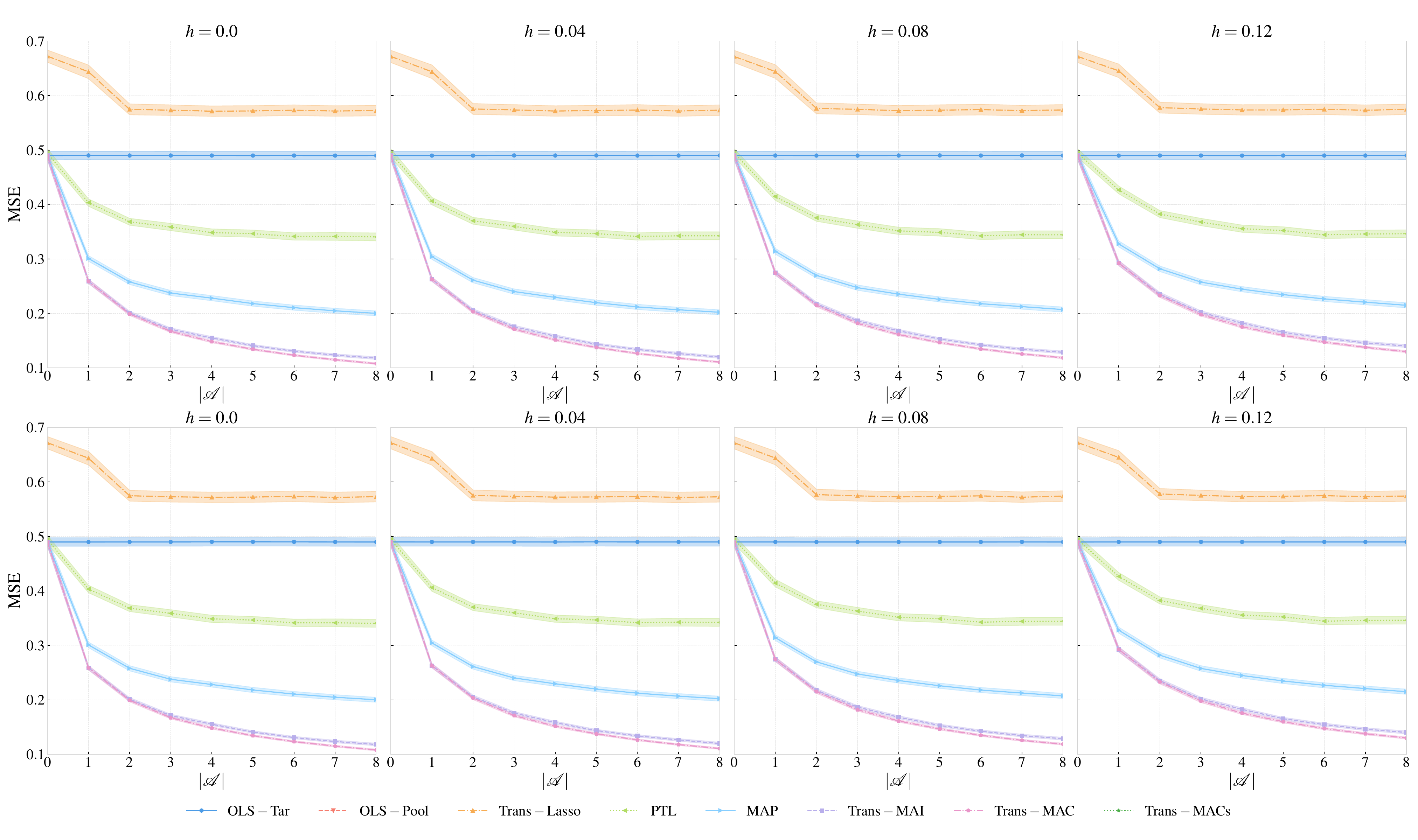}
    \caption{Estimation	performance of OLS-Tar, OLS-Pool, MAP, Trans-Lasso, PTL, Trans-MAI, Trans-MACs and Trans-MAC for Experiment~1 and the light areas indicate $95\%$ empirical confidence intervals. The	two	rows correspond	to	configurations (i) and	(ii) generating $\bbeta^{(m)}$ respectively for $1 \leq m \leq |\mathcal{A}|$. The y-axis	corresponds	to mean of squared estimation errors (MSE) for each estimator $\hat{\bbeta}^{(0)}$, $\|\hat{\bbeta}^{(0)} - \bbeta^{(0)}\|^{2}$. OLS-Pool and Trans-MACs methods behave off-axis.}
    \label{fig:exp1_est}
\end{figure}
\begin{figure}
    \centering
    \includegraphics[width=0.9\linewidth]{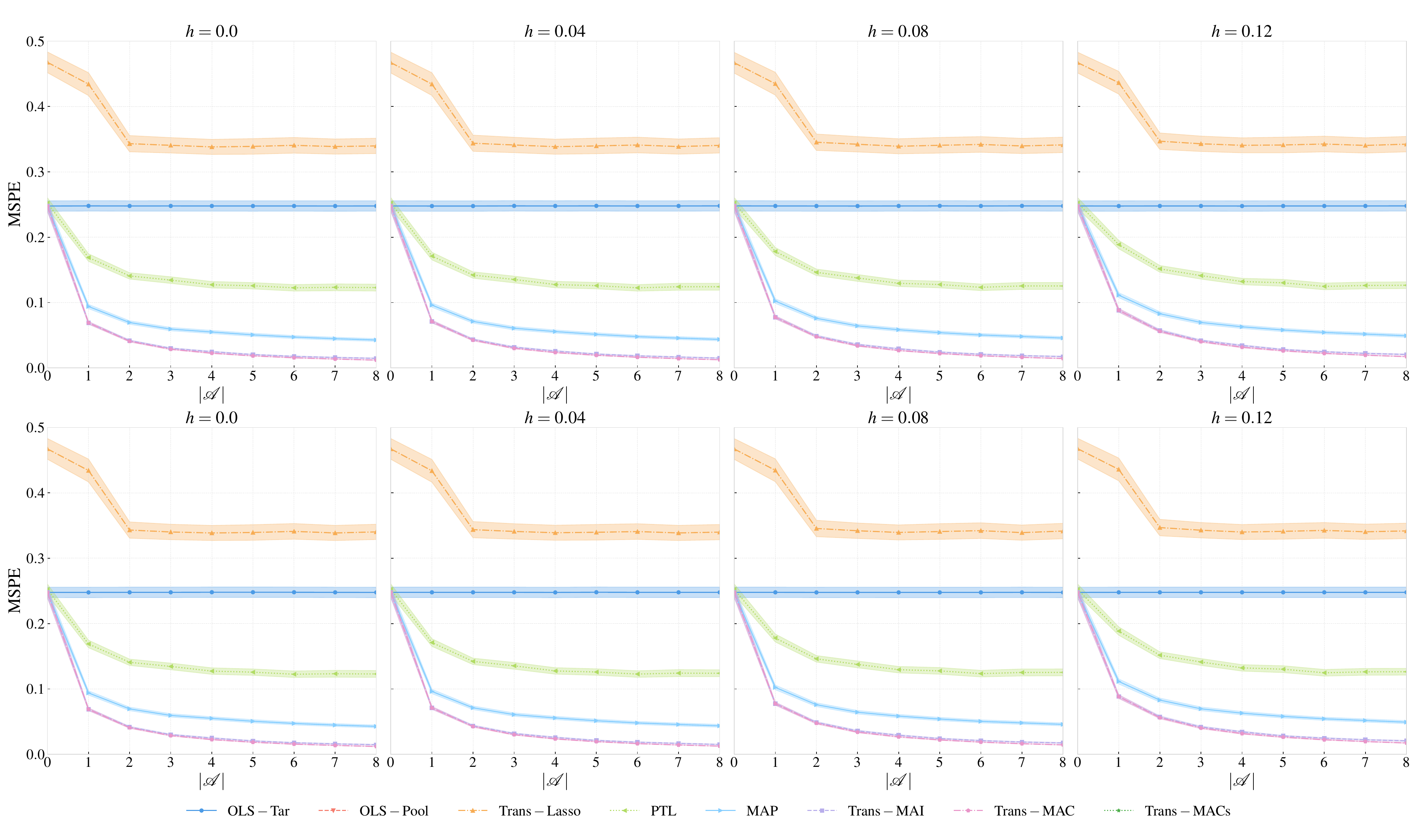}
    \caption{Prediction	performance of OLS-Tar, OLS-Pool, MAP, Trans-Lasso, PTL, Trans-MAI, Trans-MACs and Trans-MAC for Experiment~1 and the light areas indicate $95\%$ empirical confidence intervals.	The	two	rows correspond	to	configurations (i) and	(ii) generating $\bbeta^{(m)}$ respectively for $1 \leq m \leq |\mathcal{A}|$. The y-axis	corresponds	to mean
    squared prediction error (MSPE) $n_{test}^{-1} \sum_{i=1}^{n_{test}} (\hat{\bmu}(\bx_{i}) - \bmu(\bx_i))^{2}$ for some predictor $\hat{\bmu}(\bx_{i})$. OLS-Pool and Trans-MACs methods behave off-axis.}
    \label{fig:exp1_pred}
\end{figure}

In Figures~\ref{fig:exp1_est} and \ref{fig:exp1_pred}, we report the means of squared estimation and prediction errors for Experiment~1. 
Each point is summarized from $B = 500$ independent simulations, and the light areas indicate $95\%$ empirical confidence intervals. 
The two rows correspond to configurations (i) and (ii) generating $\bbeta^{(m)}$, respectively, for $1 \leq m \leq |\mathcal{A}|$. 
As expected, the performance of the OLS-Tar does not change as $|\mathcal{A}|$ increases, while all seven other transfer learning algorithms exhibit decreasing estimator and prediction errors as $|\mathcal{A}|$ increases. 
As $h$ increases, the problem becomes more challenging, and the estimation and prediction errors of all methods increase.  
In both settings (i) and (ii), Trans-MAC and Trans-MAI perform best and second best, respectively. 
The OLS-Pool performs poorly due to negative transfer from non-informative sources. 
Trans-MACs performs poorly because the strict combinatorial-similarity does not be satisfied, whereas the combinatorial-similarity allows Trans-MAC to perform better and more stably. 
The reason why Trans-Lasso performs worse than OLS-Tar is that Trans-Lasso does not consider the sufficiency and Lasso is not the best linear unbiased estimator compared with OLS in low-dimensional linear regression, leading negative transfer. 
As $|\mathcal{A}|$ increases, Trans-MAI outperforms both PTL and MAP because it makes more sufficient use of the informative domains under individual-similarity. 
Trans-MAC performs slightly better than Trans-MAI because it also leverages potential knowledge in non-informative domains under combinatorial-similarity.  It is also worth noting that due to the randomness of	the	parameter generation, our	definition	of $\mathcal{A}$ may not always	be	the	best subset	of	auxiliary	samples	that give the smallest estimation errors, demonstrated by \cite{li2022transfer}.


\subsubsection{Experiment 2: homogenous design under combinatorial-similarity}\label{exp2}
For this experiment, in addition to assuming the existence of informative domains, we also assume that  non-informative domains satisfy strict combinatorial-similarity assumption. 
Let $\bOmega_{0} = \bOmega_{m} = \bm{I}_p$ and  $\sigma_{(0)} = \sigma_{(m)} = 0.5$, $|\mathcal{A}| = \{0,1,2,3,4,5,6,7,8\}$. 

Then we generate the parameters $\{\bbeta\}_{m \in [M]}$ by the following steps.
First, generate the coefficients $\brho  = \frac{1}{M - |\mathcal{A}|}\bm{1}_{M-|\mathcal{A}|}\in \mathbb{R}^{M-|\mathcal{A}|}$. 
Next, for $ |\mathcal{A}| + 1 \leq m \leq M$, generate the coefficients $\bbeta^{(m)} \neq \bbeta^{(0)}$ \iid drawn from $\mathcal{N}(2,16)$.
Then, generate the target parameter by 
    \begin{equation*}
        \bbeta^{(0)} = \biggl(\bm{I}_p - \sum_{m = |\mathcal{A}|+1}^{M} \rho_m \bG^{[m]^{-1}} \bG^{[|\mathcal{A}|]} \biggr)^{-1} \biggl(\sum_{m = |\mathcal{A}| + 1}^{M} \rho_m \bG^{[m]^{-1}} \bigl(\sum_{j = |\mathcal{A}| + 1}^{m}\bX^{(j)^{T}} \bX^{(j)}\bbeta^{(j)}\bigr) \biggr).
    \end{equation*}
Additionally, for $1\leq m \leq |\mathcal{A}|$, generate $\bdelta^{(m)}$ by two configurations as the same as Experiment~1, and let $\bbeta^{(m)} = \bbeta^{(0)} + \bdelta^{(m)}$.
It is straightforward to verify that $\{\bbeta^{[m]}\}_{m \in [M]}$ satisfy strict combinatorial-similarity in Eq.~(\ref{approximate linear}) and combinatorial-similarity. 

The mean squared estimation errors and prediction errors for Experiment~2 are reported in Figure~B.1 and Figure~B.2 (Supplement~B.1), respectively. These results exhibit similar patterns to those observed in Experiment~3.
Consistent with previous findings, our proposed methods Trans-MAI, Trans-MAC, and Trans-MACs demonstrate superior performances compared to all alternative approaches.

\subsubsection{Experiment 3: heterogeneous design under combinatorial-similarity}
This experiment maintains the same parameter specifications as Experiment~2, with two key distinctions: 
the covariance matrices of covariates may differ between target and source datasets, and 
the error variances are allowed to vary across datasets, following \cite{li2022transfer} and \cite{lin2024profiled}.
The specific generation procedures for $\bOmega_{m}$, $\bOmega_{0}$, $\sigma_{(m)}^{2}$, and $\sigma_{(0)}^{2}$ are implemented as follows.

For $1 \leq m \leq M$, we construct the covariance matrix $\bOmega_{m}$ as a symmetric Toeplitz matrix with the first row is given by $(1, \bm{1}_{2m-1}\T/(m+1), \bm{0}_{p-2m}\T)$. 
For the target, we set $\bOmega_0 = \bm{I}_p$.
The error variances are specified under two distinct scenarios:
(a) Let the \(m^{\mathrm{th}}\) source \(\sigma_{(m)} = 1.0\) for \(1 \leq m \leq M\) and the target \(\sigma_{(0)} = 1.0\).
(b) Let the \(m^{\mathrm{th}}\) source \(\sigma_{(m)} = 0.2m\) for \(1 \leq m \leq M\) while maintaining $\sigma_{(0)} = 1.0$ for the target.

\begin{figure}
    \centering
    \includegraphics[width=0.9\linewidth]{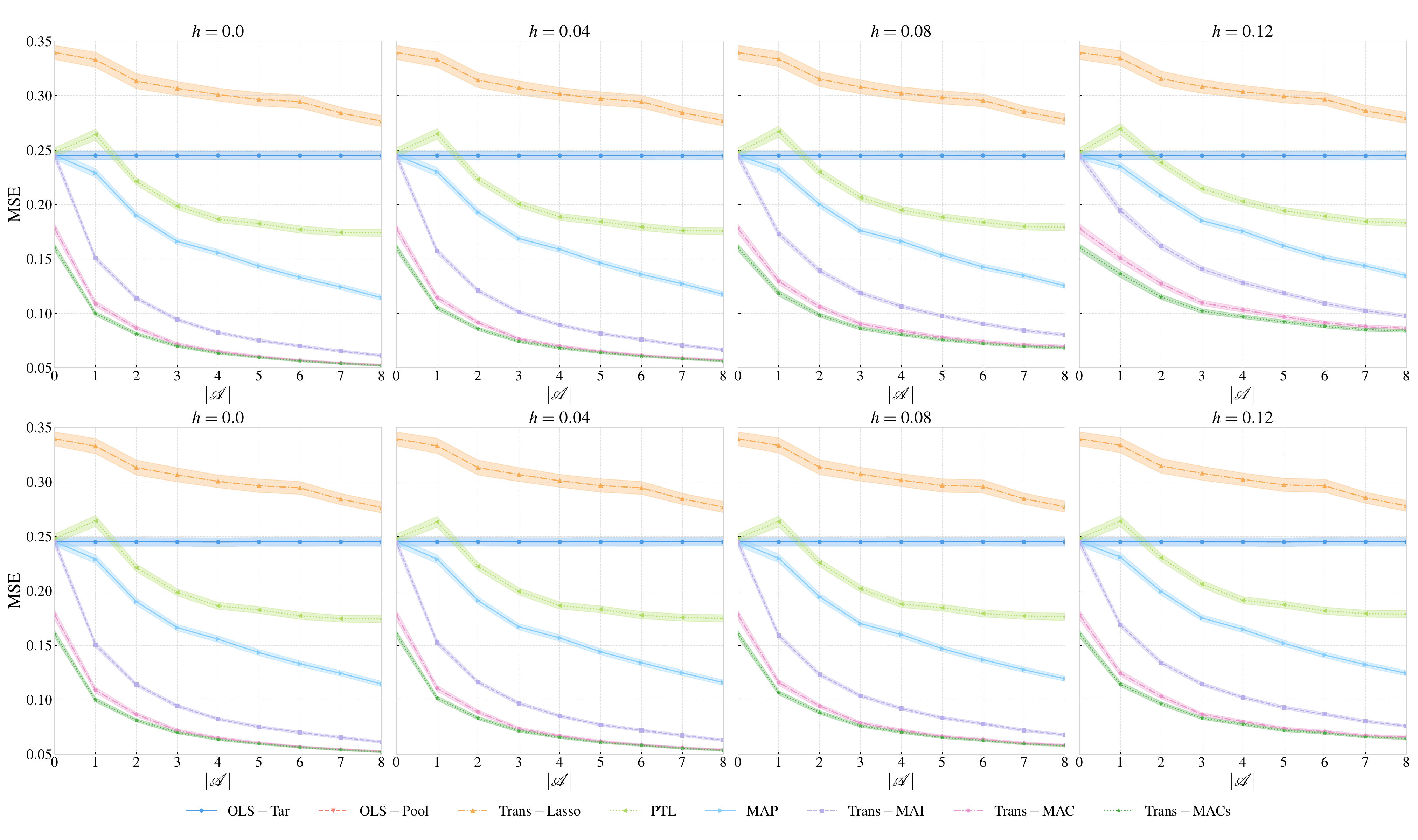}
    \caption{Estimation	performance of OLS-Tar, OLS-Pool, MAP, Trans-Lasso, PTL, Trans-MAI, Trans-MACs and Trans-MAC for Experiment~3 under the case that $\sigma_{(m)} = \sigma_{(0)}$ and the light areas indicate $95\%$ empirical confidence intervals.	The	two	rows correspond	to	configurations (i) and	(ii) generating $\bbeta^{(m)}$ respectively for $1 \leq m \leq |\mathcal{A}|$. The y-axis	corresponds	to mean of squared estimation errors (MSE) for each estimator $\hat{\bbeta}^{(0)}$, $\|\hat{\bbeta}^{(0)} - \bbeta^{(0)}\|^{2}$. OLS-Pool behaves off-axis.}
    \label{fig:exp3_est}
\end{figure}

\begin{figure}
    \centering
    \includegraphics[width=0.9\linewidth]{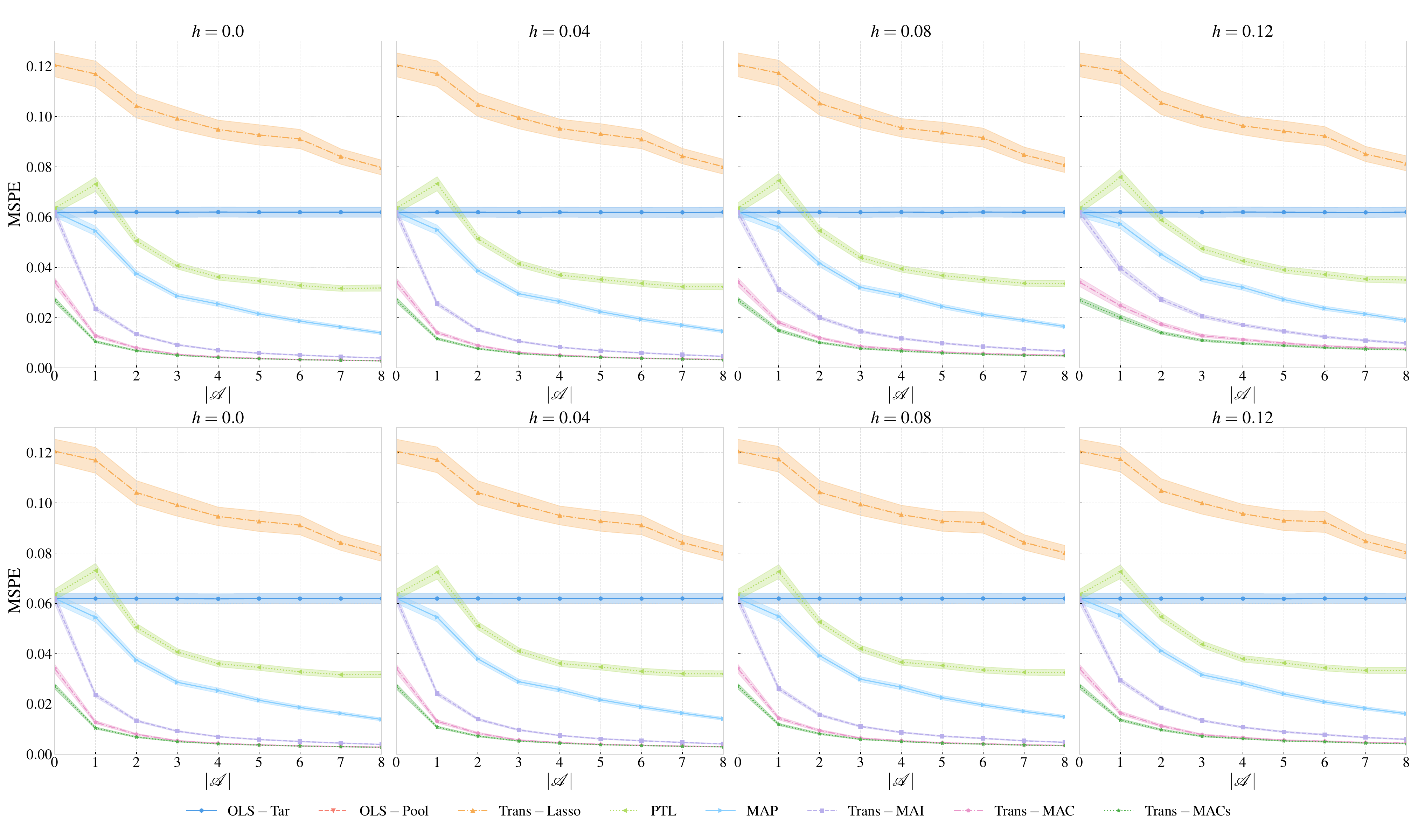}
    \caption{Prediction	performance of OLS-Tar, OLS-Pool, MAP, Trans-Lasso, PTL, Trans-MAI, Trans-MACs and Trans-MAC for Experiment~3 under the case that $\sigma_{(m)} = \sigma_{(0)}$ and the light areas indicate $95\%$ empirical confidence intervals.	The	two	rows correspond	to	configurations (i) and	(ii) generating $\bbeta^{(m)}$ respectively for $1 \leq m \leq |\mathcal{A}|$. The y-axis	corresponds	to $n_{test}^{-1} \sum_{i=1}^{n_{test}} (\hat{\bmu}(\bx_{i}) - \bmu(\bx_i))^{2}$ for some predictor $\hat{\bmu}(\bx_{i})$. OLS-Pool behaves off-axis.}
    \label{fig:exp3_pred}
\end{figure}

In Figure~\ref{fig:exp3_est} and \ref{fig:exp3_pred}, we report the mean squared estimation errors and squared prediction errors for Experiment~3 under the condition $\sigma_{(m)} = \sigma_{(0)}$. 
Consistent with the results of Experiment~1, the performances of all methods consistently decrease as $|\mathcal{A}|$ increases, except for OLS-Tar. 
OLS-Pool maintains the worst performance due to negative transfer effects from non-informative sources.
Our proposed methods maintain their performance ranking best, second best, and third best, respectively.
The effectiveness of Trans-MAI arises from its incorporation of sufficiency under individual-similarity, thereby enhancing transfer learning efficiency. 
Trans-MACs and Trans-MAC demonstrate superior performance compared to PTL and MAP because constructing candidate domains nestedly accelerates the convergence rate of each candidate, thereby improving the performance of transfer learning, particularly when $|\mathcal{A}| = 0$. 
As $|\mathcal{A}|$ increases, Trans-MAI under individual-similarity performs progressively closer to Trans-MACs and Trans-MAC under combinatorial-similarity, 
since fewer domains remain for informative linear combinations and the advantage of combinatorial-similarity diminishes progressively.
Trans-MACs outperforms Trans-MAC for more effective utilization of strict combinatorial-similarity due to faster convergence in candidate domains.
However, as $|\mathcal{A}|$ increases, Trans-MAC gradually approaches Trans-MACs because of the increasing performance of the sufficient informative domain in Trans-MAC.


Figure~B.3 and~B.4 in Supplement~B.2 present the mean squared estimation and rediction errors for Experiment~3 under the condition $\sigma_{(m)} \neq \sigma_{(0)}$. 
Consistent with previous findings, our proposed methods remain superior to other alternatives.

\subsubsection{Experiment 4: heterogeneous design under combinatorial-similarity}
For this experiment, the parameter specifications are the same as those in Experiment~2. 
The specific generation procedures for $\bOmega_{m}$, $\bOmega_{0}$, $\sigma_{(m)}^{2}$, and $\sigma_{(0)}^{2}$ are implemented as follows.

For $1 \leq m \leq M$, let the covariance matrix $\bOmega_{m}$ be a symmetric Toeplitz matrix with the first row given by $(1, \bm{1}_{2m-1}\T/(m+1), \bm{0}_{p-2m}\T)$, and let the covariance matrix $\bOmega_0$ for the target be $\bOmega_{0} = (0.5^{|i-j|})_{p \times p}$. 
The error variances are specified under two distinct scenarios:
(a) Let the $m^{\mathrm{th}}$ source $\sigma_{(m)} = 0.2m$ for $1 \leq m \leq M$, and the target $\sigma_{(0)} = 1.0$.
(b) Let the $m^{\mathrm{th}}$ source $\sigma_{(m)} = 1.2^{m-3}$ for $1 \leq m \leq M$, and $\sigma_{(0)} = 1.0$.

We report the mean squared estimation errors and squared prediction errors for Experiment~4 under the case of $\sigma_{(m)} = 0.2m$ in Figure~B.5 and B.6 in the Supplement~B.3, 
and for the case of $\sigma_{(m)} = 1.2^{m-3}$ in Figure~B.7 and B.8 in Supplement~B.3.
In both scenarios, our proposed methods maintain outperforming other alternatives.

\subsubsection{Experiment 5: effect of different sample sizes}
This experiment evaluates the estimation performance of different methods under varying sample sizes for the target and sources. 
Let $M = 10$, $|\mathcal{A}| = 4$, and $p = \{50, 80\}$. The sample size of target is $n_0 = \{150, 200, 250\}$, and the sample sizes of sources are $n_{m} = \{100, 150, 200\}$ for $1 \leq m \leq M$. 
We generate $\bdelta^{(m)}$ by (ii) with $h = 0.12$ for $1 \leq m \leq |\mathcal{A}|$.

\begin{figure}
    \centering
    \includegraphics[width=0.8\linewidth]{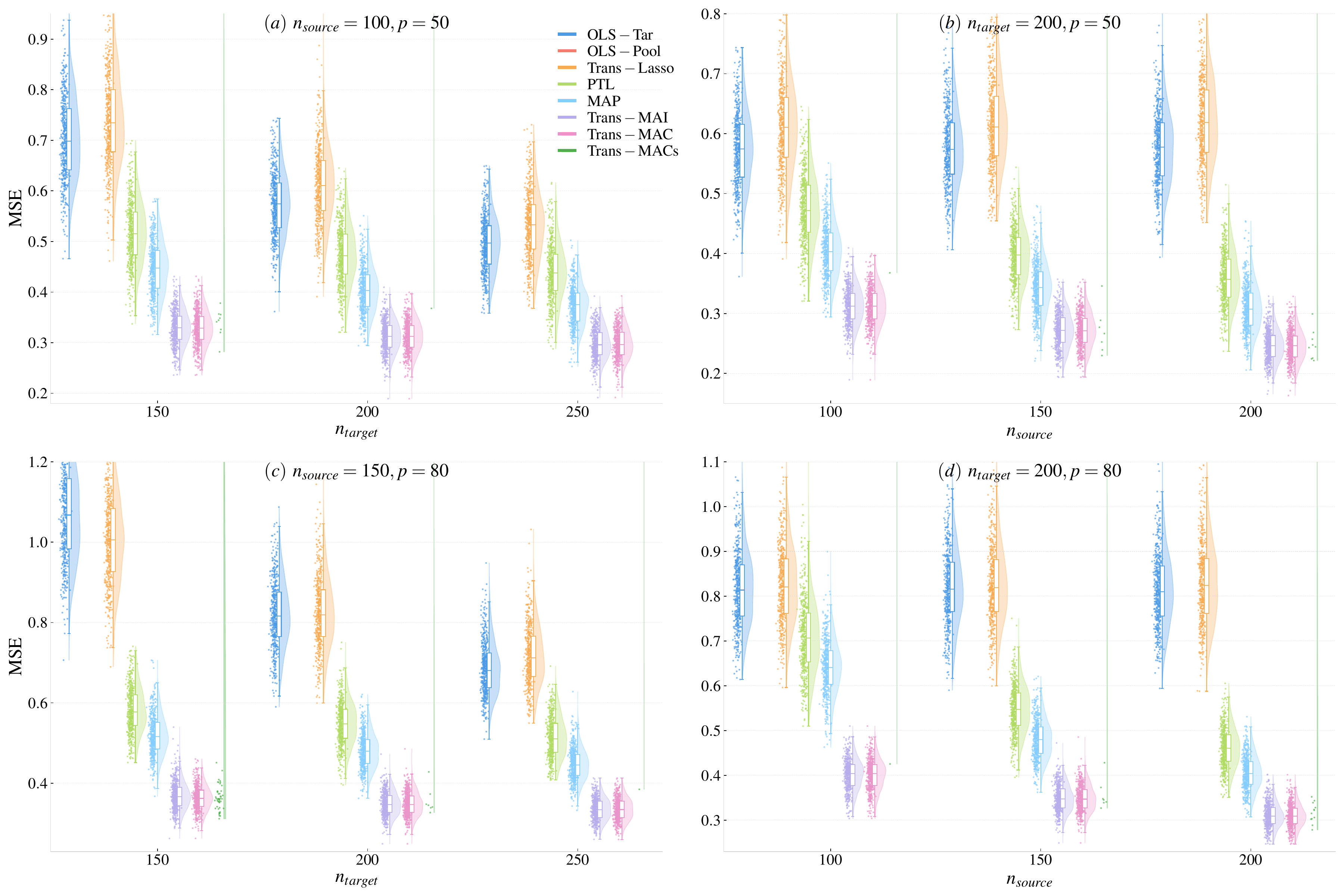}
    \caption{Estimation performances of MSE values for different estimators in Example~1 with scatterplots, boxplots and voilinplots. $\sigma_{(0)} = \sigma_{(m)} = 1.0$ for $1\leq m \leq M$. The first row presents the cases with $p = 50$ and the second row presents the cases with $p = 80$. The first and second columns study the target sample size $n_0$ and the source sample size $n_{m}$, repectively. OLS-Pool and Trans-MACs methods behave off-axis.} 
    \label{fig:aexp1nN}
\end{figure}

We replicate Experiment~1 $B = 500$ times and the resulting MSE values are visualized by scatter plots, box plots, and violin plots in Figure~\ref{fig:aexp1nN}.
Figure~\ref{fig:aexp1nN} demonstrates that Trans-MAC and Trans-MAI consistently achieve the best and second-best performance, respectively. 
In some cases, Trans-MAC slightly outperforms Trans-MAI due to its ability to simultaneously leverage both individual-similarity and potential combinatorial-similarity, while Trans-MACs underperforms due to violations of strict combinatorial-similarity assumptions. 
Furthermore, subplots (b) and (d) reveal a clear pattern where the MSE values of PTL, MAP, Trans-MAI, and Trans-MAC decrease with increasing source sample size $n_{m}$, indicating that larger sources generally improve target parameter estimation accuracy.
\begin{figure}
    \centering
    \includegraphics[width=0.8\linewidth]{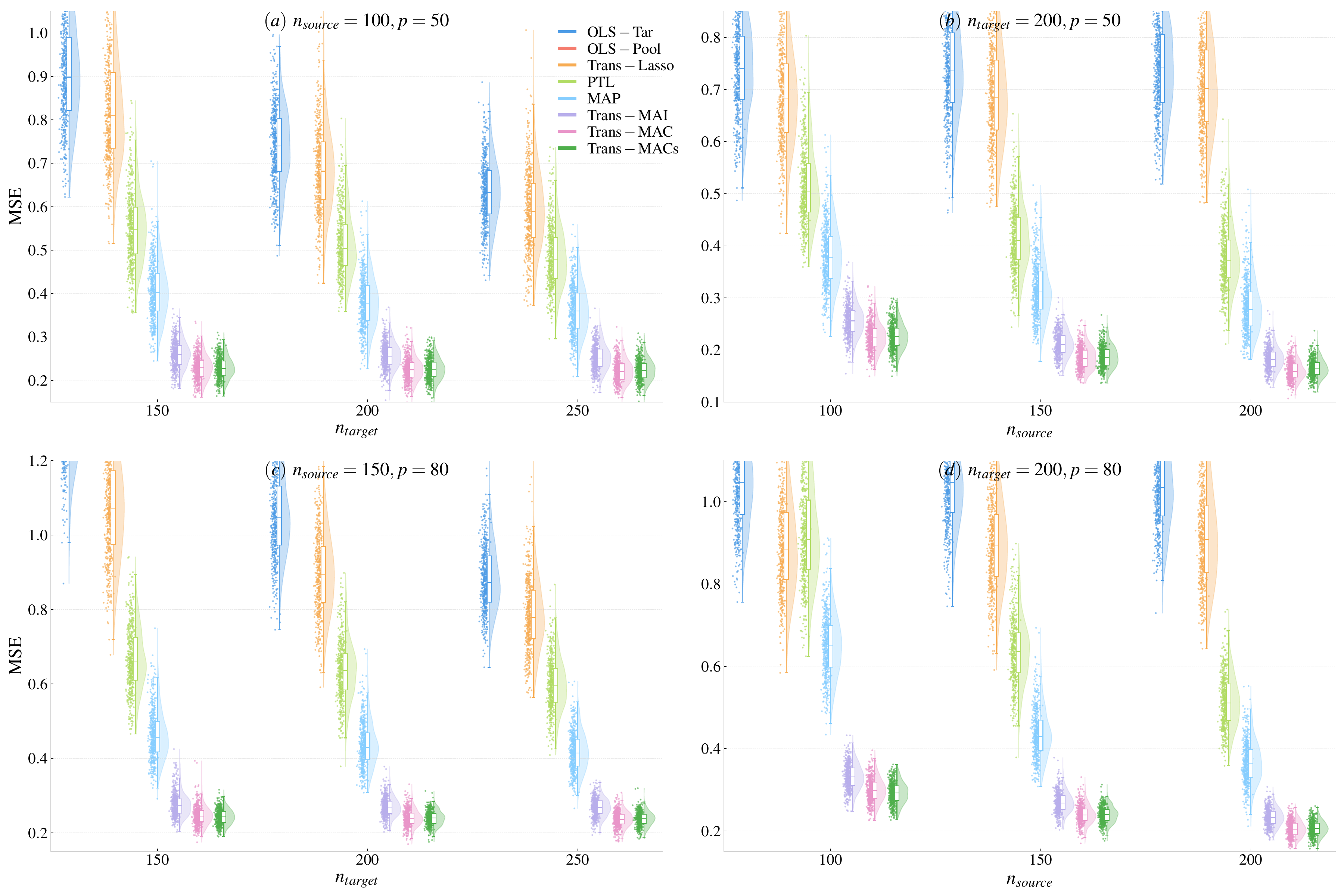}
    \caption{Estimation performances of MSE values for different estimators in Example~4 with scatterplots, boxplots and voilinplots. $\sigma_{(0)} = 1.0$ and $\sigma_{(m)} = 0.2 m$ for $1\leq m \leq M$. The first row presents the cases with $p = 50$ and the second row presents the cases with $p = 80$. The first and second columns study the target sample size $n_0$ and the source sample size $n_{m}$, repectively. OLS-Pool behaves off-axis.}
    \label{fig:aexp5nN_estq5080ns100150200}
\end{figure}

We replicate Experiment~4 $B = 500$ times with $\sigma_{(0)} = 1.0$ and $\sigma_{(m)} = 0.2m$ for $1 \leq m \leq M$,
and illustrates MSE values by  scatter plots, box plots, and violin plots in Figure~\ref{fig:aexp5nN_estq5080ns100150200}.
While most numerical findings in Figure~\ref{fig:aexp5nN_estq5080ns100150200} qualitatively align with those in Figure~\ref{fig:aexp1nN}, two crucial differences distinguish these experiments: (1) Example~4 employs a heterogeneous experimental setting compared to the homogeneous setup in Example~1, and (2) Example~4 satisfies combinatorial-similarity whereas Example~1 does not.

Unlike in Figure~\ref{fig:aexp1nN}, Trans-MACs and Trans-MAC perform significantly better than Trans-MAI in Figure~\ref{fig:aexp5nN_estq5080ns100150200}, 
since Trans-MACs and Trans-MAC leverage knowledge from individual-similarity, as well as combinatorial-similarity.
Although Trans-MACs achieves superior performance in certain scenarios under strict combinatorial-similarity, Trans-MAC exhibits greater stability across diverse scenarios due to its more flexible combinatorial-similarity. 
Consequently, we recommend Trans-MAC as the preferred choice for broader practical applications, given its robust performance across varying experimental settings.

\subsection{Weight convergence for Trans-MAI with varying $v$}
\begin{figure}
    \centering
    \includegraphics[width=0.8\linewidth]{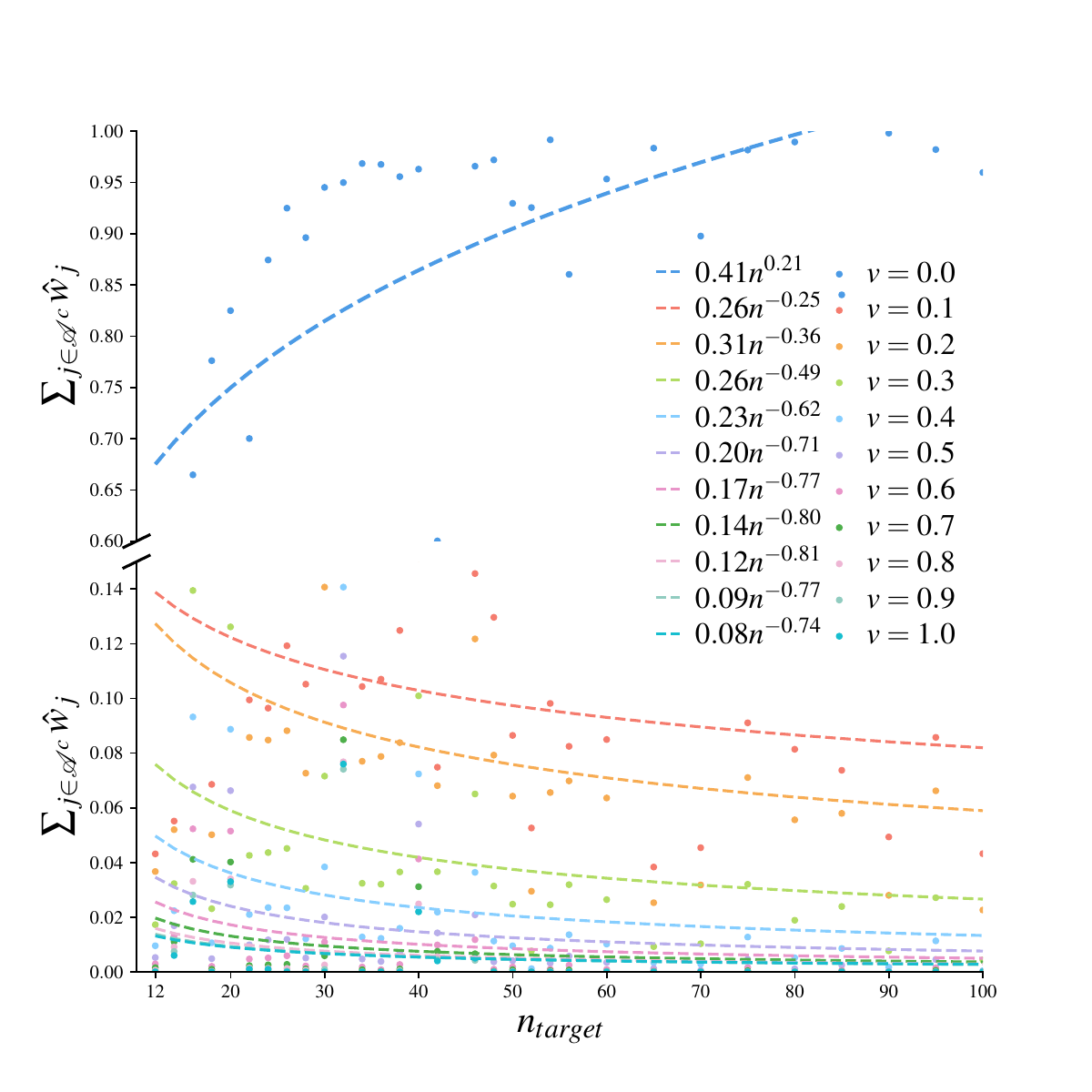}
    \caption{The sum of weights for non-informative domains for Trans-MAI in Experiment~2 with different $v$. For each value of $n_{target}$, Experiment~2 is repeated $B = 1000$ times.
    The mean of $\sum_{j \in \mathcal{A}^c} \hat{w}_j$ is scatterplotted.  For each $v$, the dashed line $c_v n^{-a_v}$ is plotted, where $(c_v,a_v) = \arg \min_{c,a} \sum_{n_{0,k}} \bigl[ \frac{1}{B} \sum_{b = 1}^{B}( \sum_{j \in \mathcal{A}^c} \hat{w}_j )_{[b], n_{0,k}}^{v} - c n_{0,k}^{-a} \bigr]^{2}$.
    }
    \label{fig:exp2v}
\end{figure}
In this experiment, we investigate how different values of $v$ affect weight convergence for non-informative domains in Trans-MAI. 
We investigate the effect of different $v$ values on the weight convergence for non-informative domains in Trans-MAI. 
The experimental setup fixes $p = 10$, $M = 10$, $h = 0.0$, and $|\mathcal{A}| = 3$, 
with source and target sample sizes $n_{m} = n_0 \in \{ 12, 14, 16, 18, 20, \allowbreak 22, 24, 26, 28, 30, 32, 34, 36, 38, 40, 42, 46, 48, 50, 52, 54, 56, 60, 65, 70, 75, 80, 85, 90, 95, 100 \}$
and all other parameters follow Experiment~2's specifications.
For each value of $n_0$, the experiment is repeated $B = 1000$ times, and the mean of $\sum_{j \in \mathcal{A}^c} \hat{w}_j$ is scatterplotted in Figure~\ref{fig:exp2v}. 
For each value of $v$, we fit these points with dashed curves of form $c_v n^{-a_v}$ in Figure~\ref{fig:exp2v}, where $(c_v,a_v) = \arg \min_{c,a} \sum_{n_{0,k}} \bigl[ \frac{1}{B} \sum_{b = 1}^{B}( \sum_{j \in \mathcal{A}^c} \hat{w}_j )_{[b], n_{0,k}}^{v} - c n_{0,k}^{-a} \bigr]^{2}$.

Firstly, from Figure~\ref{fig:exp2v}, we observe that when $v \neq 0$, the sum of weights for the non-informative domains $\sum_{j\in\mathcal{A}^c} \hat{w}_j$ gradually converges to $0$ as $n_{\text{target}}$ increases. 
However, when $v = 0$, the sum of weights for the non-informative domains $\sum_{j\in\mathcal{A}^c} \hat{w}_j$ 
fails to converge to $0$, instead remaining non-negligible even with increasing sample size $n_{\text{target}}$.
This behaviour is attributed to the fact that, in the parameter settings of Experiment~2 satisfying strict combinatorial-similarity, $\mathcal{C}_{v,\phi}(\bw)$ lacks the ability to distinguish between informative domains and non-informative domains.
Secondly, Figure~\ref{fig:exp2v} illustrates that the line corresponding to larger $v$ lies lower,
since as $v$ increases, the penalty for the non-informative domains becomes stronger, thus enhancing the ability of $\mathcal{C}_{v,\phi}(\bw)$ to exclude linear convex combinations of non-informative domains close to the target.
Thirdly, it is evident that when $v \geq 0.4$, $\sum_{j \in \mathcal{A}^c} \hat{w}_j$ behaves as $o_p(n_0^{-1/2})$, which aligns with Theorem~\ref{Convergence of weight new}.

In addition, the weight convergence for Trans-MAI with different $v$ values is investigated, according to the settings in Experiment~1 and the results are presented in Figure~B.9 in Supplement~B.4.


\subsection{Illustration for asymptotic normality of Trans-MAI}
We experimentally verify the asymptotic normality of Trans-MAI. 
Specifically, we set $M = 10$, $\bpsi = p^{-1/2}\bm{1}_p \in \mathbb{R}^{p}$, $p = 20$, $|\mathcal{A}| = 2$, $h = 0$, and $n_0 = 200$. 
Each experiment is repeated $B = 2000$ times. 
Table~\ref{table:normality} presents the results, including the average bias, standard deviation and an illustration of normality for 
$\bigl(\sum_{k \in \mathcal{I}_{\hat{m}_s}} \bpsi^{T}(\bX^{(k)^{T}}\bX^{(k)})\bG^{[\hat{m}_s]^{-1}} \bpsi \sigma_{(k)}^{2}\bigr)^{-1/2} N_{\hat{m}_s}^{1/2}\bpsi^{T}\bSigma^{[\hat{m}_s]^{1/2}} \allowbreak (\hat{\bbeta}^{(0)}(\hat{\bw}) - \bbeta^{(0)} )$.
The last column of Table~\ref{table:normality} displays the histograms of the estimation biases, along with the asymptotic normal distributions characterized by the estimated means and variances,
suggesting that the distribution is well approximated by a standard normal distribution $\mathcal{N}(0,1)$.
\begin{center}
\begin{table}[]
\resizebox{\textwidth}{!}{
\begin{tabular}{cccl}
\hline
\textbf{Design}                                                                                                                                                                         & \textbf{Average Bias} & \textbf{Standard Deviation} & \textbf{Normality} \\ \hline
\textbf{\begin{tabular}[c]{@{}c@{}}Experiment 1\\ $\Omega_{0} = \Omega_{(m)} = I_p$\\ $\sigma_{(0)} = \sigma_{(m)} = 1.0$\\ $n_m = 200$\end{tabular}}                                   & -0.0110               & 0.9854                      &     \begin{minipage}[b]{0.20\columnwidth}
		\centering
		\raisebox{-.5\height}{\includegraphics[width=\linewidth]{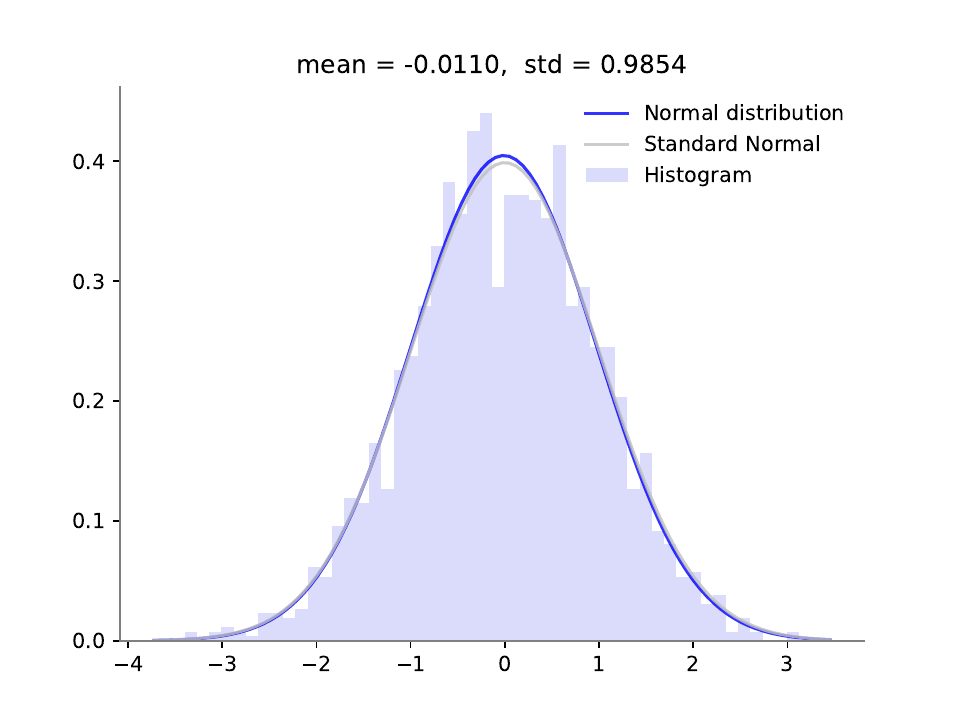}}\end{minipage}               \\
\textbf{\begin{tabular}[c]{@{}c@{}}Experiment 1\\ $\Omega_{0} = \Omega_{(m)} = I_p$\\ $\sigma_{(0)} = \sigma_{(m)} = 1.0$\\ $n_m = 500$\end{tabular}}                                   & -0.0122               & 1.0284                      &         \begin{minipage}[b]{0.20\columnwidth}
		\centering
		\raisebox{-.5\height}{\includegraphics[width=\linewidth]{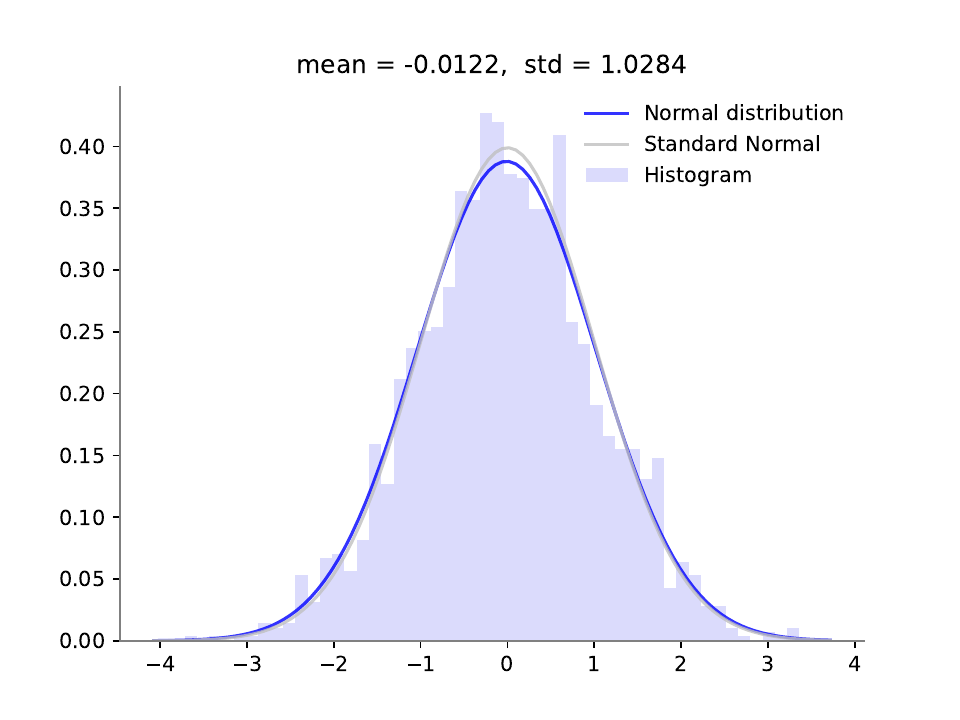}}\end{minipage}                               \\
\textbf{\begin{tabular}[c]{@{}c@{}}Experiment 2\\ $\Omega_{0} = \Omega_{(m)} = I_p$\\ $\sigma_{(0)} = \sigma_{(m)} = 1.0$\\ $n_m = 200$\end{tabular}}                                   & -0.0160               & 1.0167                      &      \begin{minipage}[b]{0.20\columnwidth}
		\centering
		\raisebox{-.5\height}{\includegraphics[width=\linewidth]{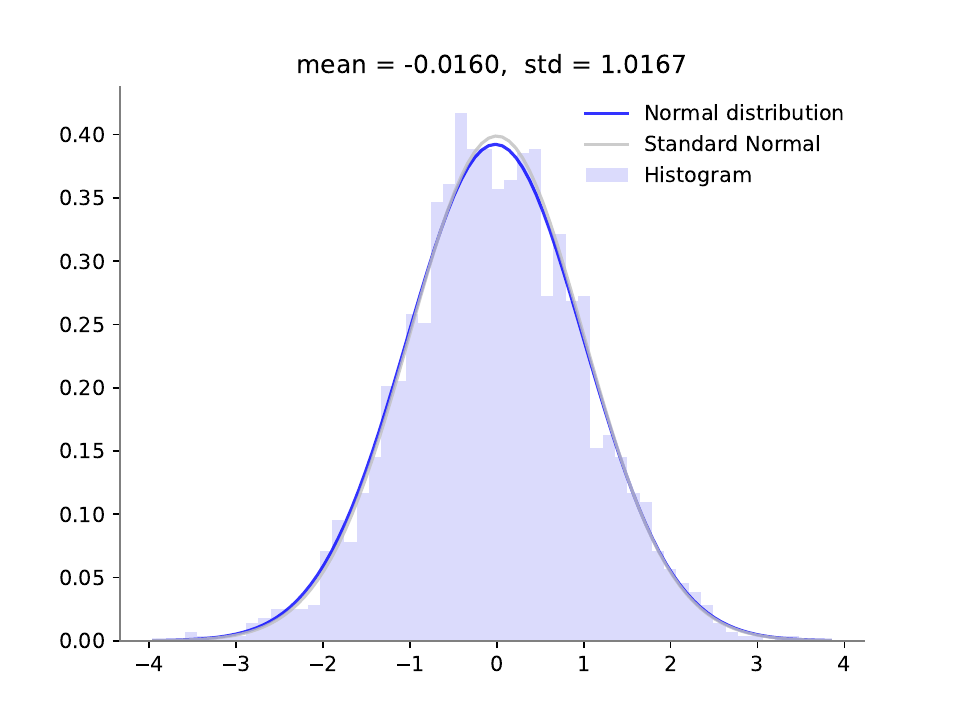}}\end{minipage}                                  \\
\textbf{\begin{tabular}[c]{@{}c@{}}Experiment 2\\ $\Omega_{0} = \Omega_{(m)} = I_p$\\ $\sigma_{(0)} = \sigma_{(m)} = 1.0$\\ $n_m = 500$\end{tabular}}                                   & -0.0246               & 0.9934                      &       \begin{minipage}[b]{0.20\columnwidth}
		\centering
		\raisebox{-.5\height}{\includegraphics[width=\linewidth]{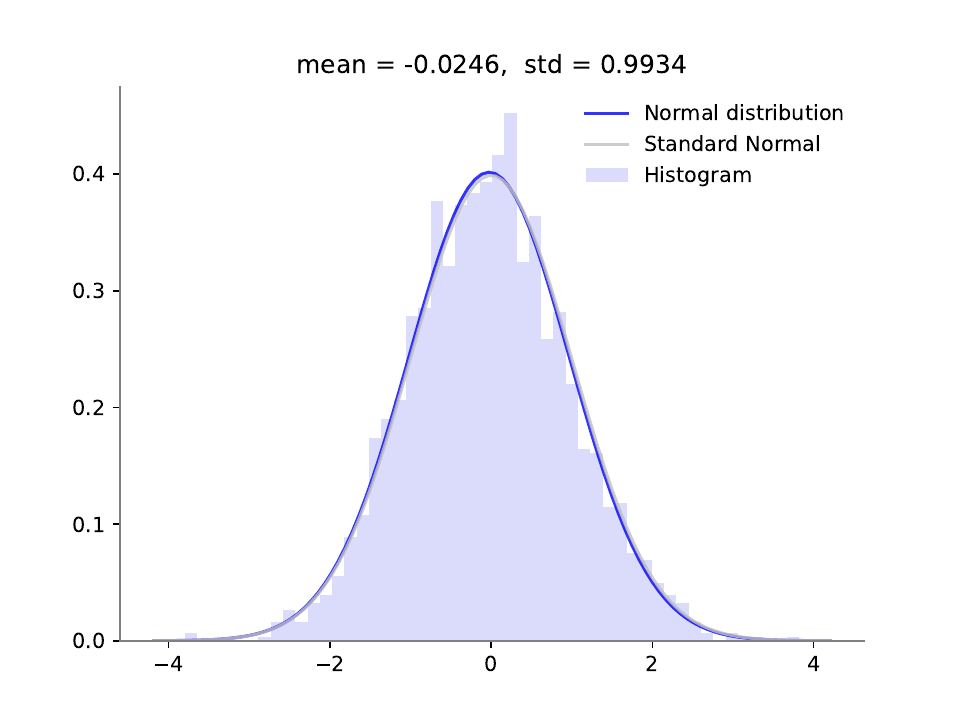}}\end{minipage}                                 \\
\textbf{\begin{tabular}[c]{@{}c@{}}Experiment 3\\ $\Omega_{(0)} = I_p, \Omega_{(m)}= \text{Toeplitz}(m)$\\ $\sigma_{(0)} = 1.0, \sigma_{(m)} = 0.2 m$\\ $n_m = 200$\end{tabular}}       & 0.0051                & 1.0065                      &      \begin{minipage}[b]{0.20\columnwidth}
		\centering
		\raisebox{-.5\height}{\includegraphics[width=\linewidth]{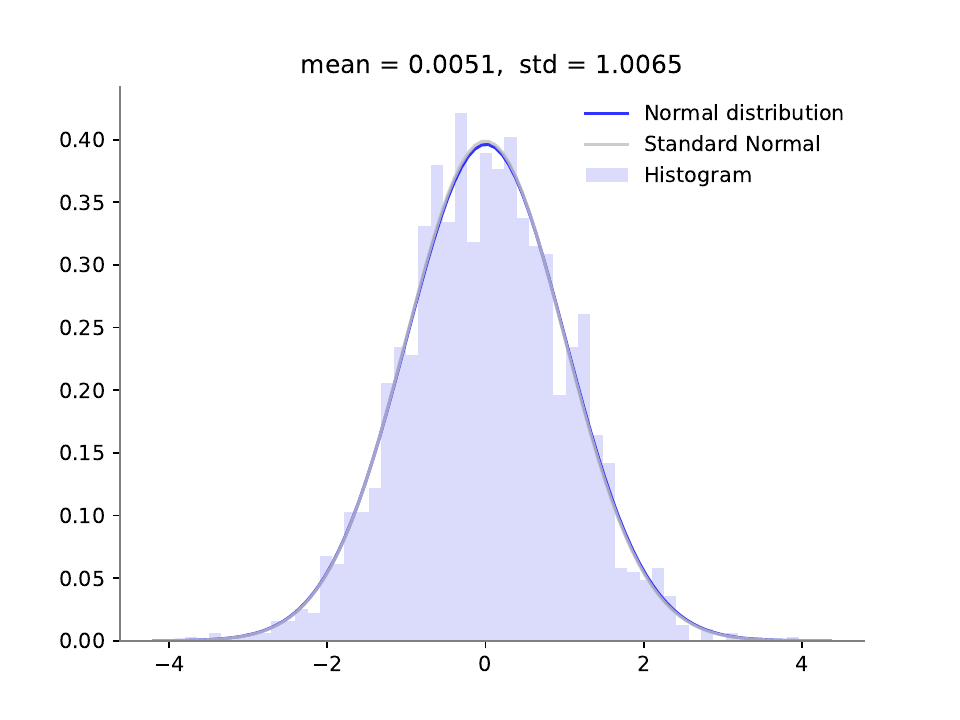}}\end{minipage}                                  \\
\textbf{\begin{tabular}[c]{@{}c@{}}Experiment 3\\ $\Omega_{(0)} = I_p, \Omega_{(m)}= \text{Toeplitz}(m)$\\ $\sigma_{(0)} = 1.0, \sigma_{(m)} = 0.2 m$\\ $n_m = 500$\end{tabular}}       & -0.0084               & 0.9899                      &         \begin{minipage}[b]{0.20\columnwidth}
		\centering
		\raisebox{-.5\height}{\includegraphics[width=\linewidth]{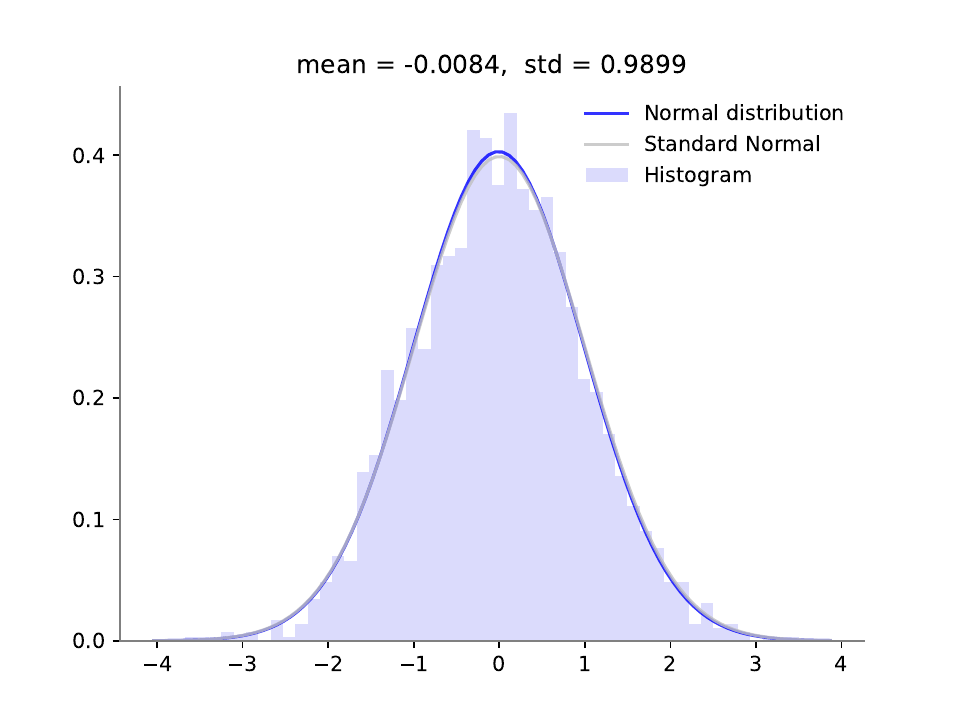}}\end{minipage}                               \\
\textbf{\begin{tabular}[c]{@{}c@{}}Experiment 4\\ $\Omega_{(0)} = (0.5), \Omega_{(m)}= \text{Toeplitz}(m)$\\ $\sigma_{(0)} = 1.0, \sigma_{(m)} = 1.2^{m-3}$\\ $n_m = 200$\end{tabular}} & 0.0104                & 1.0070                      &        \begin{minipage}[b]{0.20\columnwidth}
		\centering
		\raisebox{-.5\height}{\includegraphics[width=\linewidth]{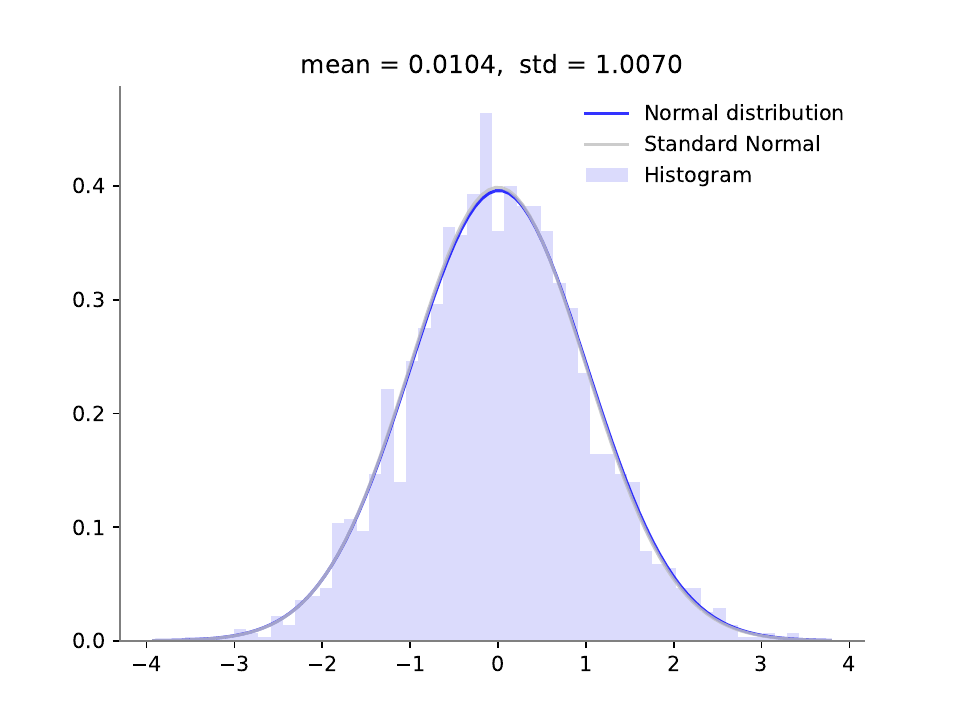}}\end{minipage}                                \\
\textbf{\begin{tabular}[c]{@{}c@{}}Experiment 4\\ $\Omega_{(0)} = (0.5), \Omega_{(m)}= \text{Toeplitz}(m)$\\ $\sigma_{(0)} = 1.0, \sigma_{(m)} = 1.2^{m-3}$\\ $n_m = 500$\end{tabular}} & 0.0109                & 0.9907                      &         \begin{minipage}[b]{0.20\columnwidth}
		\centering
		\raisebox{-.5\height}{\includegraphics[width=\linewidth]{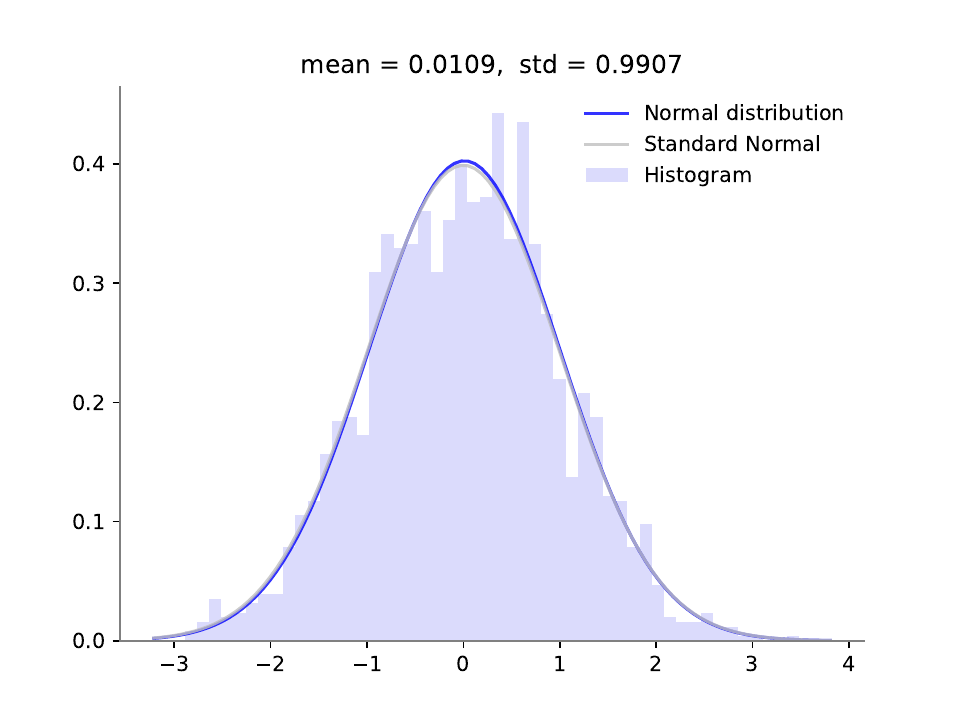}}\end{minipage}                               \\ \hline
\end{tabular}
}
\caption{Verification of the asymptotic normality for Trans-MAI under different experimental designs}
\label{table:normality}
\end{table}
\end{center}

%% file: body/realdata.tex
\section{Empirical Data Analysis}\label{realdata}
In this section, we implement our approaches to analyze Beijing housing rental data, sourced from the publicly available dataset at \url{http://www.idatascience.cn/dataset}. 
Our objective is to predict monthly rental prices, facilitating a deeper understanding of the housing rental market. 
Given similarities in geographical location, population structure, and rental demand, four adjacent districts in southern central Beijing are selected for analysis: Chaoyang, Shijingshan, Fengtai, and Xicheng.
The specific locations of rental houses in these four districts is visualized in Figure~B.10 (Supplement~B.5), providing a geographical representation of our data sources.
Overall, the dataset consists of 2,165 observations with 33 variables. 
To implement our transfer learning framework, we treat data from different districts as multi-source datasets. 
The sample sizes for four sources, Chaoyang, Shijingshan, Fengtai, and Xicheng are $(1241, 269, 347, 308)$, respectively. 
The response variable, denoted as $\by^{(m)}$, represents the natural logarithm of the monthly rent. 
Following preliminary variable selection to exclude irrelevant variables, 12 variables remain in our models, 
including the number of rooms ($\bx_{1}^{(m)}$), 
the number of restrooms ($\bx_{2}^{(m)}$), 
the number of living rooms ($\bx_{3}^{(m)}$), 
total area ($\bx_{4}^{(m)}$), 
whether a bed is present ($\bx_{5}^{(m)}$), 
whether a wardrobe is present ($\bx_{6}^{(m)}$), 
whether an air conditioner is present ($\bx_{7}^{(m)}$), 
whether fuel gas is available ($\bx_{8}^{(m)}$), 
floor level ($\bx_{9}^{(m)}$), 
total number of floors ($\bx_{10}^{(m)}$), 
the number of schools within 3 km ($\bx_{11}^{(m)}$), 
and the number of hospitals within 5 km ($\bx_{12}^{(m)}$) for the $m^{\mathrm{th}}$ dataset. 
Table~B.1 in Supplement~B.5 provides more details of covariates in our data analysis. 
All covariates have been properly transformed and scaled. 
To further validate the replicability of our proposal, each dataset is treated as the target, with the others serving as sources. 
Subsequently, we explore multiple combinations of target and source datasets, applying our procedure iteratively, as in \cite{lin2024profiled}, and \cite{zhang2024prediction}.

\begin{figure}
    \centering
    \includegraphics[width=1.0\linewidth]{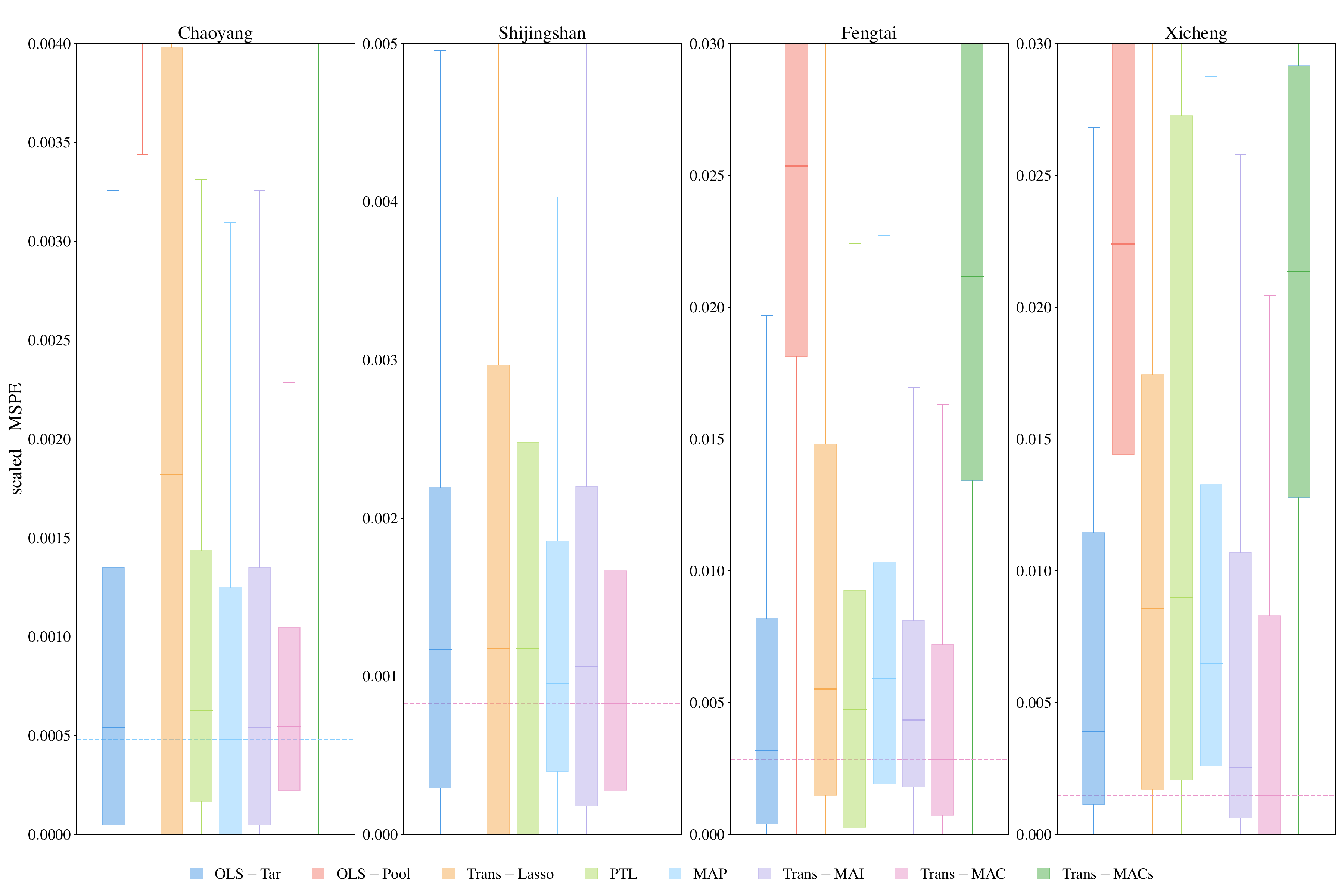}
    \caption{Boxplots of the scaled MSPE for different target districts in the housing rental information data analysis. 
    Each subgraph is labeled with the corresponding district used as the target. 
    The dashed line represents the median of the method with the smallest median, colored to correspond with the respective method.
    }
    \label{fig:housedata}
\end{figure}

To evaluate out-of-sample prediction risk, the target samples are randomly divided into two subgroups, with 70\% used for training and 30\% for testing and we repeat this process $B = 500$ times for each target.
The scaled mean squared prediction error (scaled MSPE) for method \( k \) is calculated as \( n_{m}^{-1} ( \|\by_{[b]}^{(m)_k} - \hat{\by}_{[b]}^{(m)_k}\|^{2} - \min_{\tilde{k}} \|\by_{[b]}^{(m)_{\tilde{k}}} - \hat{\by}_{[b]}^{(m)_{\tilde{k}}}\|^{2} ) \), where \([b]\) denotes the \(b^{\mathrm{th}}\) replication, \(n_m\) is the sample size for testing from the \(m^{\text{th}}\) target and \(m = 1, 2,3, 4\). 
By definition, the scaled MSPE is non-negative. 

Figure~\ref{fig:housedata} illustrates boxplots of the scaled MSPE values.
As shown in Figure~\ref{fig:housedata}, Trans-MAC outperforms alternative methods across all districts, achieving a lower median scaled $\mathrm{MSPE}$ compared to all other alternatives, except for the Chaoyang district, thereby demonstrating the effectiveness of our weight selection criterion. 
For the Chaoyang district as the target, although Trans-MAC has a slightly higher median compared to MAP, it exhibits smaller volatility, indicating more stable predictive performance. 
Among all methods, OLS-Pool yields the poorest performances due to negative transfer effects from incorporating non-informative domains. 
Trans-MACs underperforms as the datasets fail to satisfy the strict combinatorial-similarity specified in Eq.~\eqref{approximate linear}.
OLS-Tar shows poor performance due to its inability to utilize auxiliary information effectively. 
Trans-MAC performs better than Trans-MAI because it leverages auxiliary knowledge both from individual-similarity identified by Trans-MAI and  combinatorial-similarity. 
Trans-MAC consistently outperforms other alternatives (Trans-Lasso, PTL, and MAP) across districts, corroborating our simulation results. 
In summary, the empirical data analysis demonstrates the effectiveness of Trans-MAC in achieving superior scaled MSPE, highlighting its potential for predictive tasks in future applications.

%% file: body/discuss.tex
\section{Discussions}\label{discuss}

In this paper, we propose a novel sufficient-principled transfer learning framework when the transferable set is unknown, incorporating both individual-similarity and combinatorial-similarity. 
For individual-similarity, we propose an innovative weight selection criterion based on q-aggregation \citep{dai2012deviation} and parsimonious model averaging \citep{zhang2020parsimonious} to select the sufficient informative domain and derive theoretical properties, including weight convergence, $\ell_2$-estimation/prediction error bounds, selection consistency, and generalized asymptotic optimality under appropriate conditions. 
Additionally, leveraging the sufficiency achieved earlier, we extend our transfer learning framework to combinatorial-similarity for more auxiliary knowledge and provide upper bounds for weight convergence, $\ell_2$-estimation/prediction error bounds, high-probability optimality and asymptotic optimality under appropriate conditions. 
Numerical simulations and empirical data analysis support our theoretical findings and demonstrate the effectiveness of our proposed method in practical applications.

Several promising future directions merit further research.
First, our sufficiency-principled framework has the potential to extend beyond regression models and extending it to nonlinear, semiparametric, and nonparametric settings, such as varying coefficient models, single-index models, and their generalized versions in high-dimensional scenarios, is a compelling area for exploration, thus offering a generalizable paradigm for statistical transfer learning.
Second, while this paper explores individual-similarity and combinatorial-similarity through two-step weight selection criteria, designing a one-step criterion that simultaneously considers both similarities would be a meaningful advancement.
Third, whether $v = 1$ or $v \in (0,1)$ is preferable, presents an intriguing point of comparison, as it has the potential to offer a novel perspective on the relative merits of model selection and model averaging.
Forth, whether there exists optimal $\phi$ in sufficiency penalty remains a valuable but unknown question. 
Fifth, the current approach involves fitting a single model for each domain. Expanding this to allow for multiple models per domain presents an interesting avenue for future work.
